\documentclass[a4paper]{article}
\pdfoutput=1
\usepackage{jcappub}

\usepackage{graphicx,color,amsmath}
\usepackage{graphicx}
\usepackage{dcolumn}
\usepackage{multirow}
\usepackage{tabularx}
\usepackage{wrapfig}
\usepackage{bm}
\usepackage{amsmath}
\usepackage{leftindex}
\usepackage{amssymb}
\usepackage{comment}
\usepackage{setspace}
\usepackage{physics}
\usepackage[dvipsnames]{xcolor}
\usepackage{enumerate}
\usepackage{booktabs}
\graphicspath{{figs/}}
\usepackage{tabularx}
\usepackage{array}

\usepackage{caption}
\usepackage{subcaption}

\definecolor{mypurple}{RGB}{164,64,214}

\newcommand{\nhat}{\hat{n}}

\newcommand{\perimeter}{Perimeter Institute for Theoretical Physics, 31 Caroline St N, Waterloo, ON N2L 2Y5, Canada}

\newcommand{\york}{Department of Physics and Astronomy, York University, Toronto, ON M3J 1P3, Canada}

\newcommand{\USC}{Physics and Astronomy Department, University of Southern California, Los Angeles, CA, 90089, USA}

\begin{document}

\title{Constraints on the remote quadrupole field from the polarized Sunyaev Zel'dovich effect}

\author[a,b]{Jordan Krywonos}
\author[a,b]{, Matthew C. Johnson}
\author[c]{, Elena Pierpaoli}
\author[a,c]{, and Haoyu Wu}

\affiliation[a]{\perimeter}
\affiliation[b]{\york}
\affiliation[c]{\USC}

\emailAdd{jkrywonos@perimeterinstitute.ca}
\emailAdd{mjohnson@perimeterinstitute.ca}

\date{\today}

\abstract{The polarized Sunyaev Zel'dovich (pSZ) effect is a cosmic microwave background (CMB) polarization anisotropy induced by Thomson scattering from free-electrons in non-linear structure. The pSZ signal is determined by the distribution of ionized gas tracing the cosmic web and the CMB quadrupole at the location of free-electrons -- the remote quadrupole field. Measuring the pSZ effect provides a consistency check of the optical depth to reionization and sheds light on the anomalous nature of the large-scale CMB temperature anisotropies, such as the low observed CMB temperature quadrupole. In this paper, we demonstrate that a CMB-CMB-galaxy bispectrum summarizes several existing pSZ statistics, and that in our observable Universe the ideal galaxy sample to detect pSZ is at $z \sim 1-2$. We evaluate the bispectrum using CMB data from \textit{Planck} and ACT with galaxy density from the unWISE galaxy redshift catalog as well as \textit{Planck} cosmic infrared background (CIB) maps. We do not make a statistically significant detection of the pSZ effect, which is consistent with the expected $\mathcal{O}(1)$ signal-to-noise from this data combination. The measured amplitude of the pSZ bispectrum provides constraints on the optical depth bias associated with large-scale structure (the amplitude of the pSZ signal) of $\hat{b}_q=1.02 \pm 2.64$, the optical depth to reionization of $\hat{\tau}_{\rm rei} = -0.01 \pm 0.14$, and the tensor-to-scalar ratio $r$ of $\sigma_r \sim 150$ ($n_t = 0$) or $\sigma_r \sim 3$ ($n_t  = -1$). We forecast that future measurements could tighten the constraints on these quantities by roughly a factor of 3, which is sufficient to provide independent confirmation of the low CMB quadrupole and the optical depth to reionization.}

\maketitle

\section{Introduction}

Increasingly precise measurements of the cosmic microwave background (CMB) radiation~\cite{AtacamaCosmologyTelescope:2025blo, SPT-3G:2025bzu, SimonsObservatory:2025wwn} 
are ushering in a new era of discovery driven by measurements of the {\em secondary CMB anisotropies}. CMB secondaries are temperature and polarization anisotropies induced by the scattering of CMB photons from mass (CMB lensing) and charge (Sunyaev Zel'dovich effects) associated with large scale structure (LSS). Weak lensing of the CMB is a mature observable~\cite{ACT:2023dou,k5yr-3h6d}, contributing significantly to constraints on cosmological parameters; recently a new gravitational secondary, the moving lens effect, was detected for the first time~\cite{Hotinli:2026wly}. Both the thermal  Sunyaev Zel'dovich (tSZ) and kinetic  Sunyaev Zel'dovich (kSZ) effects induced by scattering of CMB photons from free electrons in the post-reionization Universe have been detected at high significance, and provide meaningful information about baryonic feedback in galaxies as well as cosmology \cite{HSC:2025lum,kclp-x5j1,ACT:2025llb, Krywonos:2024mpb,Chaussidon:2026vmn}. In this paper, we focus on the polarization anisotropies induced by scattering from charges in the post-reionization Universe, the polarized Sunyaev Zel'dovich (pSZ) effect~\cite{Sunyaev1980,Kamionkowski:1997na}, which has yet to be detected.

\begin{figure}
    \centering
\includegraphics[width=.75\textwidth]{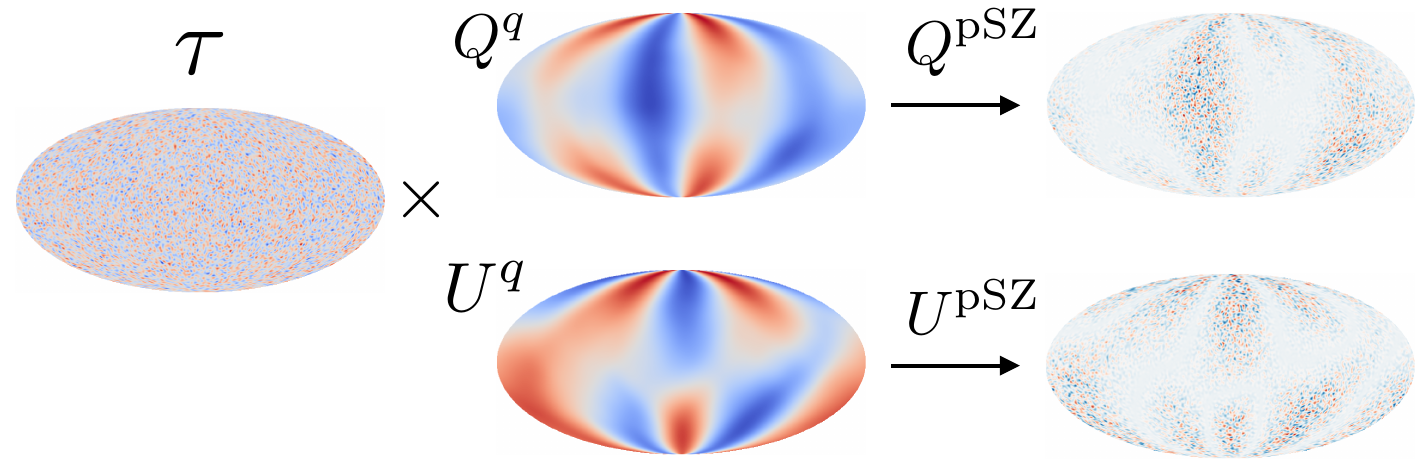}
    \caption{A depiction of the pSZ effect from a redshift slice at $z\sim 1$ of width $\Delta z \sim 0.5$. The pSZ effect in the Stokes parameters $Q^{\rm pSZ}$ and $U^{\rm pSZ}$ (right) is the product of the optical depth $\tau$ associated with ionized gas in the redshift slice (left) and the Stokes parameters $Q^q$ and $U^q$ of the remote quadrupole field in the redshift slice (center).} \label{fig:qusim}
\end{figure}

The pSZ effect is a line of sight integral over the inhomogeneous differential optical depth and the CMB quadrupole in the location of ionized gas - the remote quadrupole field. We show a schematic image of the expected signal from a redshift slice at $z\sim1$ in figure~(\ref{fig:qusim}), illustrating how the pSZ effect is the product of these two fields. The pSZ effect is one of a number of polarization anisotropies induced by Thomson scattering after recombination. CMB polarization from reionization (the `reionization bump') is sourced by the remote quadrupole field convolved with the mean optical depth; we classify this as part of the `primary CMB'. CMB polarization is also sourced by inhomogeneities in electron density produced during reionization, referred to as `patchy reionization'~\cite{Gruzinov_1998,Liu_2001} (see e.g.~\cite{Roy:2020cqn} for a recent assessment of the detectability of this signal). Finally, the motion of ionized gas induces the kinetic polarized Sunyaev Zel'dovich (kpSZ) effect~\cite{Audit_1999,Itoh:1997ks,Challinor:1999yz}; see e.g. \cite{Hotinli:2022wbk} for a recent assessment of the detectability of this signal. We do not study these additional effects here.

The magnitude of the pSZ effect is estimated by considering the typical optical depth fluctuations associated with galaxy clusters $\tau \sim 3 \times 10^{-3}$ (e.g. \cite{SPT-3G:2022zrq}) and the magnitude of the primary CMB quadrupole $q^{\pm} \sim \sqrt{C_2^{\rm obs}} \sim 14 \ {\rm \mu K}$, which yields an expected amplitude of polarization fluctuations of order $\sim 40 \ {\rm nK}$ in the direction of a cluster. This is an exceedingly small signal. Even the deepest existing CMB surveys - such as SPT-3G with a polarization map depth of $\sim 5 \ \mu{\rm K}$ \cite{Camphuis_2026} - fall short of the sensitivity required to resolve the pSZ signal for an individual object by an order of magnitude. Furthermore, pSZ is a blackbody anisotropy, making it difficult to distinguish at the map level from contributions from the (lensed) primary CMB polarization, which is itself orders of magnitude larger than the pSZ signal. Finally, adding to the challenge, the amplitude of the locally observed CMB quadrupole is anomalously low compared to the $\Lambda$CDM expectation -- $C_2^{\rm obs} = 219 \ {\mu K}^2$ vs $1060 \ {\mu K}^2$ -- implying that the amplitude of the pSZ signal in the local Universe is smaller than in a typical realization of $\Lambda$CDM.

\begin{figure}
    \centering
    \includegraphics[width=0.4\linewidth]{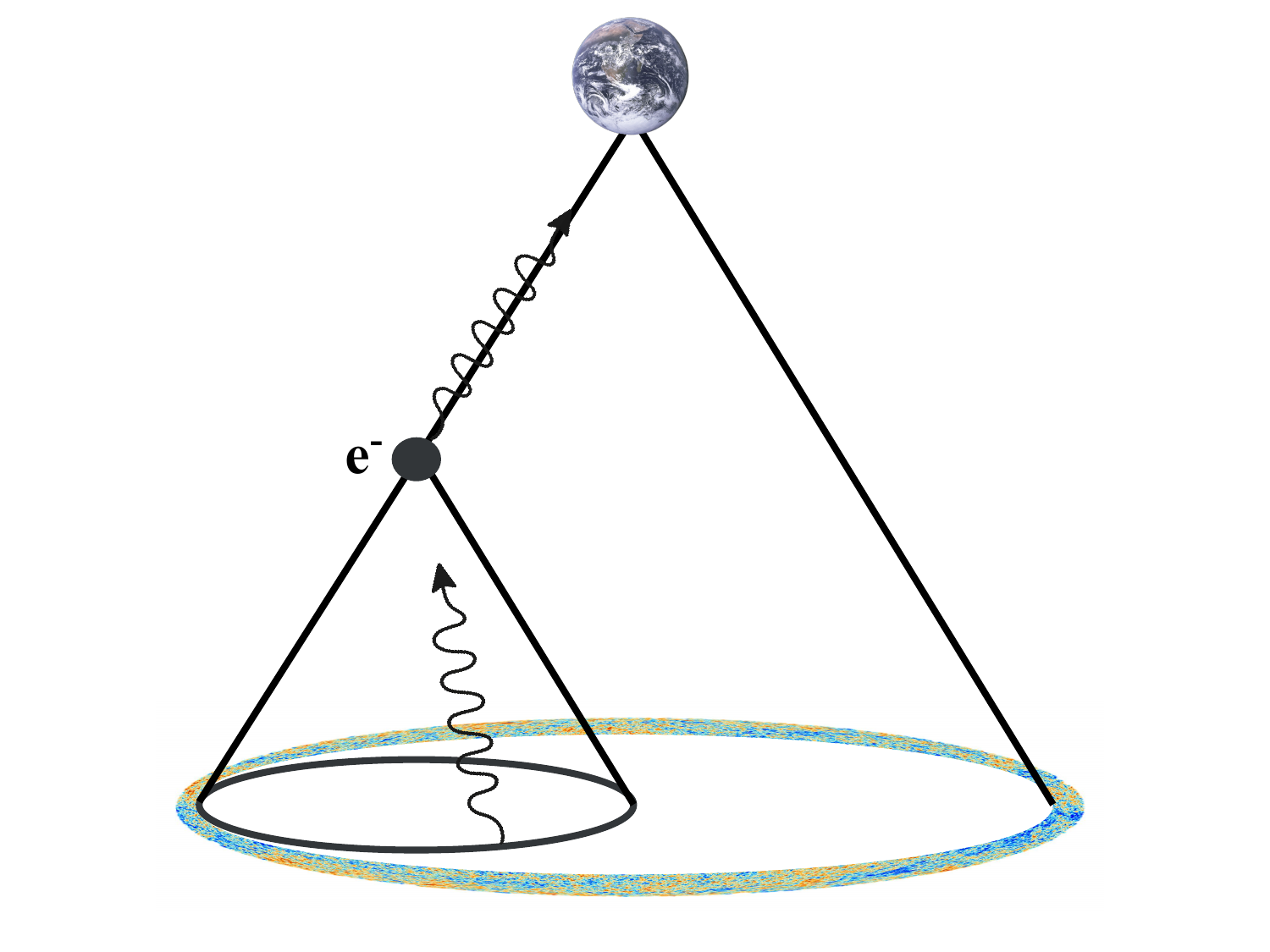}
    \caption{Most of CMB photons we observe travel directly to us along our past light cone, probing a two-dimensional slice through the last scattering surface. In the pSZ effect, CMB photons originating from a different section of the last scattering surface - probed by the electron's past light cone - scatter off a free electron and travel along our line of sight to us. This grants us access to the remote CMB quadrupole as seen from the electron's location. This figure is adapted from Ref.~\cite{Deutsch:2017cja}.}
    \label{fig:pszlightcone}
\end{figure}

Although it will be a challenge to detect the pSZ effect, doing so could have a large impact on our understanding the Universe at the earliest times and on the largest distance scales. This is because the remote quadrupole field probes the Universe {\em inside our past light cone}~\cite{Kamionkowski:1997na}, something that only Thomson scattering can do as illustrated in figure~(\ref{fig:pszlightcone}). This property is shared by the remote dipole field that determines the kSZ effect (see e.g.~\cite{Terrana2017,Deutsch2018a}), and on somewhat smaller distance scales, the reionization $E$ modes. The remote dipole and quadrupole fields therefore have the unique ability to probe the {\em homogeneity} on scales of order the size of the observable Universe, and not simply {\rm isotropy}, because we now have remote observers: the free electrons.  In principle, this can promote our measurement of the largest modes from a two-dimensional cross section to a fully three-dimensional map at last scattering. Indeed, measurements of the remote dipole field have already been used to place the tightest existing constraints on homogeneity and large-scale isocurvature modes~ \cite{Krywonos:2024mpb,Gandhi:2026opo}. 

Because they probe the largest observable scales, measurements of the remote dipole and quadrupole fields are uniquely promising in their ability to help us definitively resolve `CMB anomalies'~\cite{Cayuso:2019hen} including the low amplitude of power on large angular scales, alignment of the quadrupole and octupole, and the dipolar power asymmetry; see e.g. \cite{Planck:2019evm} for a review of these anomalies. The remote quadrupole field has a large coherence length, implying that one must probe it at high redshift to obtain independent cosmological information~\cite{Seto:2000uc,Abramo:2006gp,Bunn:2006mp,Portsmouth:2004mk,Seto:2005de}. 
However, Refs.~\cite{Seto:2005de,Cayuso:2019hen,Adil_2025} have demonstrated that the gain in information with future CMB experiments can be significant enough to determine if the CMB anomalies are due to new physics or simply statistically unlikely manifestations of $\Lambda$CDM. Further, the remote quadrupole is a complementary probe of primordial gravitational waves~\citep{2012PhRvD..85l3540A, Deutsch:2018umo,CMB-HD:2022bsz} as well as a redshift-dependent probe of birefringence~\citep{Lee:2022udm, Hotinli:2022wbk, Gon:2023nbg, Namikawa:2023zux}, dark energy~\cite{Baumann:2003xb}, large-scale homogeneity~\cite{Philcox:2022dht}, and  modified gravity~\cite{Pan:2019dax}. The pSZ effect can also constrain the optical depth beyond the cosmic variance limit imposed by the primary CMB~\citep{Meyers:2017rtf}. Although we do not focus on it here, we note that if the kpSZ effect is detected in future CMB experiments~\citep{Hotinli:2022wbk,Yasini:2016pby} it could be used to constrain primordial non-Gaussianity and cosmic birefringence~\citep{Hotinli:2022wbk}.

The goal of this paper is to outline a novel strategy for detection of the pSZ effect, and apply it to current data. Given the smallness of the pSZ signal, how can we rise to the challenge of making a detection? The key lies in the fact that the pSZ effect is strongly non-Gaussian. Returning to figure~(\ref{fig:qusim}), the pSZ signal is a modulation of the amplitude of small-scale optical depth fluctuations by the remote quadrupole field. This couples large angular scales -- the remote quadrupole field -- to small angular scales -- the optical depth. The pSZ signal is correlated both with the primary CMB temperature and polarization anisotropies through its dependence on the remote quadrupole field and with tracers of LSS through its dependence on the optical depth. Therefore, the signal can be captured by measuring bispectra involving CMB polarization, CMB temperature or polarization, and a tracer of LSS. This is in close analogy with CMB-galaxy-galaxy bispectra associated with the kSZ effect~\cite{Smith:2018bpn}. Previous work on the pSZ effect has focused on observations of polarization at the location of clusters~\cite{Sunyaev1980,Sazonov:1999zp,Seto:2000uc,Seto:2005de,Baumann:2003xb,Abramo:2006gp,Hall:2014wna,Liu:2016fqc,Adil_2025}, quadratic estimators for the remote quadrupole field~\cite{2012PhRvD..85l3540A,Deutsch:2017cja,Deutsch:2017ybc}, and quadratic estimators for the anisotropic optical depth Ref.~\cite{Dvorkin_2009}~\footnote{This was in the context of patchy reionization; see e.g.~\cite{Roy:2022muv} for an application at late times}. The pSZ bispectrum formalism introduced here unifies these various approaches.

The bispectrum amplitude estimator is proportional to the optical depth bias, a multiplicative bias that arises from a mis-match between the model for the anisotropic optical depth used in the estimator weights and the truth. Therefore, the bispectrum amplitude estimator can be re-interpreted as an estimator for the optical depth bias. This is also the case for the kSZ bispectrum estimator, which has been applied to existing datasets to constrain the kSZ optical depth bias at the $\sim 5 \ \%$ level ~\cite{Bloch_2024,McCarthy:2024nik,McCarthy:2024nik,Hotinli:2025tul,Chaussidon:2026vmn}. For kSZ the optical depth bias is a highly informative probe of baryonic feedback; see e.g. ~\cite{Siegel:2025ivd,Bigwood:2025kur,Chaussidon:2026vmn,McCarthy:2025brx}. Using external information (e.g. from kSZ) to fix the optical depth bias, we use bispectra involving low-$\ell$ CMB polarization to derive estimators for the optical depth to reionization (see also Ref.~\cite{Meyers:2017rtf}) and the tensor-to-scalar ratio (see also  Ref.~\cite{Alizadeh_2012}).  

In this paper, we utilize CMB data from \textit{Planck} and the Atacama Cosmology Telescope (ACT) as well as the unWISE galaxy catalog and \textit{Planck}-based maps of the cosmic infrared background (CIB). A simple forecast indicates that existing data combinations yield a signal-to-noise of order one. We therefore do not expect a detection, but rather present a proof of principle analysis that can be used as a template for future observations. 

The plan of the paper is as follows. In Sec.~\ref{sec:pSZandcross} we describe the pSZ effect and our modeling of cross-correlations with galaxies, the CIB, and the primary CMB temperature and polarization anisotropies. In Sec.~\ref{sec:pszandbispectrum} we outline the pSZ bispectrum estimator and how it can be used to summarize existing pSZ statistics. In Sec.~\ref{sec:bqtaurestimators} we use the pSZ bispectrum amplitude estimator to derive estimators for the optical depth bias, the optical depth to reionization, and the tensor-to-scalar ratio. We describe the datasets we consider and our analysis pipeline in Sec.~\ref{sec:analysis_pipeline}. We also perform a simple forecast of the expected signal-to-noise from existing and near-term data, and validate our pipeline with simulations. We apply our analysis pipeline to individual frequency maps in Sec.~\ref{sec:reconstructions} to investigate the impact of foregrounds. We then apply our pipeline to component-separated maps in Sec.~\ref{sec:component_separated_recons}, which provide the lowest noise measurements of pSZ. In Sec.~\ref{sec:constraints} we present constraints on the optical depth bias $b_q$, optical depth to reionization $\tau_{\rm rei}$ and tensor-to-scalar ratio $r$; we conclude in Sec.~\ref{sec:conclusions}.  
Throughout our analysis we assume \textit{Planck} 2018 cosmological parameters~\cite{PlanckCollaboration2020} $\{A_s = 2.1\mathrm{e-}9, n_s=0.967, \tau_{\rm rei}=0.054,H_0=67.3,\Omega_bh^2=0.022,\Omega_ch^2=0.12,\Omega_K=0,H_0=67.3 \mathrm{\,km\,s^{-1}Mpc^{-1}}\}$.

\section{The polarized SZ effect and cross-correlations}\label{sec:pSZandcross}

The contribution to the $Q$ and $U$ Stokes parameters from the pSZ effect is given by 
\begin{eqnarray}\label{eq:pSZ_definition}
    (Q \pm iU)_{\rm pSZ}(\hat{n}) = - \frac{\sqrt{6}}{10} \int d\chi \ \dot{\tau} (\hat{n},\chi) e^{-\tau(\hat{n},\chi)} \ q^{\pm} (\hat{n},\chi)
\end{eqnarray}
where $\hat{n}$ is the line of sight and $\chi$ is the radial comoving distance. The (inhomogeneous) differential optical depth $\dot{\tau}(\hat{n},\chi)$ is defined by 
\begin{eqnarray}
    \dot{\tau} (\hat{n},\chi) \equiv \sigma_T a(\chi) \bar{n}_e(\chi) \left(1 + \delta_e (\hat{n},\chi) \right)
\end{eqnarray}
which carries information about astrophysics through its dependence on the mean $\bar{n}_e(\chi)$ and fluctuations $\delta_e (\hat{n},\chi)$ in the number density of free electrons; $\sigma_T$ is the Thomson cross section and $a(\chi)$ the scale factor. The pSZ effect carries cosmological information through its dependence on the remote quadrupole field defined by
\begin{eqnarray}\label{eq:remotequad}
    q^{\pm} (\hat{n},\chi) = \sum_{m=-2}^2 \Theta_{2m} (\hat{n},\chi) {}_{\pm 2} Y_{2m} (\hat{n})
\end{eqnarray}
where $\Theta_{2m} (\hat{n},\chi)$ are the $\ell = 2$ multipole moments of the CMB temperature 
observed at the spacetime position $\hat{n}\chi$ along our past light cone and ${}_{\pm 2} Y_{2m} (\hat{n})$ are the spin-2 harmonics. The remote quadrupole field receives contributions from both scalar and tensor modes.

The pSZ effect is correlated with LSS through the inhomogeneous optical depth and with the primary CMB through the remote quadrupole field. Modeling these correlations is a crucial component in the analysis we present below. We outline our modeling assumptions for these quantities in the following subsections.

\subsection{Modeling the inhomogeneous optical depth and correlations with tracers of large scale structure}

In our analysis we require a model for the galaxy density, cosmic infrared background (CIB), and electron density fields. There are a variety of approaches to modeling these fields, including halo models and simulations. An accurate model is crucial in the high signal-to-noise regime to make unbiased inferences about cosmology. We do not expect to be in this regime here, and therefore choose to adopt a simplistic model that is consistent with existing observations. We model the galaxy, CIB, and electron density fields as scale-dependent projected biased tracers of the underlying dark matter field. As we discuss below, modeling inaccuracies associated with our simplistic approach can be subsumed into a single nuisance parameter - the optical depth bias. 

For the galaxy survey, the observed (two-dimensional) overdensity field is
\begin{eqnarray}\label{eq:galdens}
    \delta_g(\nhat) &=& (N^g(\hat{n})-\Bar{N}^g)/\Bar{N}^g \nonumber \\
    &=& \int d \chi\, W_g(\chi) b_g(\chi) \delta(\nhat,\chi)
\end{eqnarray}
where $N^g(\hat{n})$ is the number of galaxies in the pixel $\hat{n}$, $\Bar{N}^g$ is the mean number of galaxies per pixel, $b_g(\chi)$ is the linear galaxy bias, $\delta_g(\nhat,\chi)$ is the three-dimensional density field, and $W_g(\chi)$ is the window function for the photometric galaxy sample defined as 
\begin{eqnarray}
    W_g(\chi) \equiv \frac{1}{N} \frac{dN}{dz}(\chi) H(\chi), \ \ \ \int d\chi W_g(\chi) = 1 
\end{eqnarray}
where $H(\chi)$ is the Hubble parameter. We model the galaxy angular power spectra using the Limber approximation as
 \begin{eqnarray}\label{eq: clgglimber}
     C_\ell^{gg}=\int \frac{d\chi}{\chi^2}\,P_{mm}\left(\chi,k=\frac{\ell+\frac{1}{2}}{\chi}\right)W_g^2(\chi)b_g(\chi)^2+N_{\rm shot}
 \end{eqnarray}
 where $N_{\rm shot} = \big(N_{g,\rm tot}/(4\pi f_{\rm sky})\big)^{-1}$ is the shot noise ($f_{\rm sky}$ is the fraction of the sky covered by the survey) and we calculate the non-linear matter power spectrum $P_{mm}$ using \texttt{CAMB}. 
 Our analysis uses sufficiently small angular scales (the sensitivity peaks around $\ell \sim  200$) of the galaxy survey to neglect additional contributions from redshift space distortions and magnification. 

 For the CIB, we define the window function as 
 \begin{eqnarray}
     W_{\rm CIB} \equiv \left(b_I\frac{dI}{dz}\right)(\chi)H(\chi)
 \end{eqnarray}
 where $I$ is the intensity and $b_I$ is the clustering bias; this window function is computed from data using clustering redshifts~\cite{chiang_2019_tomog,Chiang:2025ujk} in the analysis below and not directly modeled. The CIB power spectrum is given by
 \begin{eqnarray}\label{eq: clCIBauto_limber}
     C_\ell^{\rm CIB}=\int \frac{d\chi}{\chi^2}\,P_{mm}\left(\chi,k=\frac{\ell+\frac{1}{2}}{\chi}\right)W_{\rm CIB}^2(\chi) + N_{\rm shot}
 \end{eqnarray}
 using the Limber approximation, where the shot noise is determined by the number density of resolved sources. 

The differential optical depth field is
\begin{equation}\label{eq:taufield}
    \dot{\tau} (\nhat, \chi) =  W_\tau (\chi)  \delta_e(\nhat,\chi)
\end{equation}
with $W_\tau (\chi)$ defined as \begin{eqnarray}\label{eq: Wtau} 
    W_\tau(\chi) \equiv \sigma_T \bar{n}_{e,0}(1+z(\chi))^2, \ \ \ \int_{0}^{\chi_{\rm max}} d\chi \ W_\tau(\chi) = \tau(\chi_{\rm max}).
\end{eqnarray}
Here, $\sigma_T$ is the Thomson scattering cross section and $\bar{n}_{e,0}$ is the average number density of electrons today, modeled as
\begin{eqnarray}\label{eq: n_e bar} 
    \bar{n}_{e,0} = \frac{f_{\rm gas}\, X\, \Omega_{b,0}\, \rho_{\rm crit,0}}{\mu_e\, m_p}
\end{eqnarray}
where $f_{\rm gas}$ is the mass fraction of baryons that are ionized, $X$ is the fraction of the total electrons that are ionized, $\Omega_{b,0}$ is the baryon density parameter today,  $\rho_{\rm crit,0}$ is the critical density today, and $\mu_em_p$ is the mean baryon mass per electron. Our fiducial modeling choices follow Ref.~\cite{Bloch_2024}, with $X=1$, $\mu_e=1.14$, and $f_{\rm gas} = 0.9$. We model the electron overdensity field $\delta_e(\nhat,\chi)$ in Fourier space as
\begin{eqnarray}
\delta_e(\bm{k},\chi)=b_e(k,\chi)\delta_m(\bm{k},\chi)
\end{eqnarray}
where $b_e$ is the scale-dependent
linear bias from Ref.~\cite{Takahashi2020}:
\begin{equation}
    b_e(z,k) = b_*(z)\left[1+\left(\frac{k}{k_*(z)}\right)^{\gamma(z)} \right]^{-1/2}.
\end{equation}
The parameters are given by
\begin{eqnarray}
    b_*(z) &=& \sqrt{-0.013z+0.971} \nonumber \\
    \gamma(z) &=& 0.10z^2-0.59z+1.91 \\
    k_*(z) &=& (-0.42z^3 + 3.10z^2 -3.24z + 1.86) \,{\rm Mpc^{-1}}. 
    \nonumber
\end{eqnarray}

The differential optical depth auto-spectrum is
\begin{eqnarray}\label{eq: Cltaudot} 
    C^{\dot{\tau}\dot{\tau}}_\ell(\chi)& =& \frac{1}{\chi^2} W_\tau(\chi)^2 b_e\left(\chi,k=\frac{\ell + 1/2}{\chi}\right)^2 P_{mm}\left( \chi, k=\frac{\ell + 1/2}{\chi} \right) \,.
\end{eqnarray}
The most important quantity for the analysis below is the model for the cross-spectra between tracers and the optical depth field. Using the models described above, the galaxy-optical depth cross-spectrum is
\begin{eqnarray}\label{eq: Cltaudotg} 
    C^{\dot{\tau}g}_\ell(\chi)& =& \frac{1}{\chi^2} W_g(\chi) W_\tau(\chi) b_g(\chi) b_e\left(\chi,k=\frac{\ell + 1/2}{\chi}\right) P_{mm}\left( \chi, k=\frac{\ell + 1/2}{\chi} \right),
\end{eqnarray}
and the CIB-optical depth cross-spectrum is
\begin{eqnarray}\label{eq: CltaudotCIB} 
    C^{\dot{\tau}\rm CIB}_\ell(\chi)& =& \frac{1}{\chi^2} W_{\rm CIB}(\chi) W_\tau(\chi)  b_e\left(\chi,k=\frac{\ell + 1/2}{\chi}\right) P_{mm}\left( \chi, k=\frac{\ell + 1/2}{\chi} \right) \ .
\end{eqnarray}
The modeling choices in these spectra are determined by the free parameters in the window functions $W_g(\chi)$, $W_{\rm CIB}(\chi)$ and $W_\tau(\chi)$ as well as the bias terms $b_g(\chi)$ and $b_e\left(\chi,k\right)$; we use the fiducial parameter choices described above in our analysis.

The optical depth bias for kSZ and pSZ arises from a mis-match between the model for the optical depth field and the truth. This could be due to incorrect modeling of the gas-halo connection, galaxy-halo connection, uncertainties in the redshift distribution of the galaxy survey, and environmental/selection effects. Here, we assume that the dominant model uncertainty is in the gas-halo model. As kSZ has been measured at a high significance, we use the kSZ optical depth constraints as a prior on our gas-halo model. The kSZ and pSZ optical depths are not the same quantities, and are survey-dependent, but they both depend on the gas-halo model which is how we use kSZ to inform pSZ. We adjust $k_*(z)$ (as compared to Ref.~\cite{Takahashi2020}) to agree with optical depth bias ($b_v$) constraints from ACT$\times$DESI kSZ velocity reconstruction~\cite{McCarthy:2025brx} following Ref.~\cite{Gandhi:2026opo}.

\subsection{Modeling the remote quadrupole field and correlations with the primary CMB}\label{sec:remotequadfield}

The signal of interest in our analysis is the remote quadrupole field ($q^{\pm} (\hat{n},\chi)$): the projected CMB temperature quadrupole ($\Theta_{2m} (\hat{n},\chi)$) observed from different spacetime locations along our past light cone, as defined in Eq.~(\ref{eq:remotequad}).
This is an explicitly gauge invariant quantity, but in Newtonian gauge it can be decomposed as:
\begin{eqnarray}\label{eq:quadcontributions}
    \Theta_{2m} (\hat{n},\chi) = \Theta_{2m}^{\rm SW} (\hat{n},\chi) + \Theta_{2m}^{\rm D} (\hat{n},\chi) + \Theta_{2m}^{\rm ISW} (\hat{n},\chi) + \Theta_{2m}^{\rm T} (\hat{n},\chi) \ .
\end{eqnarray}
Here, $\Theta_{2m}^{\rm SW}$ is the Sachs-Wolfe contribution from redshifting of intrinsic temperature fluctuations, $\Theta_{2m}^{\rm D}$ is the Doppler contribution from the velocity of the primordial plasma, $\Theta_{2m}^{\rm ISW}$ is the integrated Sachs-Wolfe contribution from the time-evolution of gravitational potentials, and $\Theta_{2m}^{\rm T}$ is the contribution from the decay of tensor modes.
A detailed discussion of these contributions to the remote quadrupole field can be found in Refs.~\cite{Deutsch:2017cja,Deutsch:2017ybc,Deutsch:2018umo,Philcox:2022dht}; we present a summary in Appendix~\ref{app:quadrupolefield}.  

The $q^{\pm} (\hat{n},\chi)$ components of the remote quadrupole are related to the Stokes parameters $Q$ and $U$ by:
\begin{eqnarray}
    q^{Q} (\hat{n},\chi) &=& (q^{+} (\hat{n},\chi) + q^{-} (\hat{n},\chi))/2 \\
    q^{U} (\hat{n},\chi) &=&  (q^{+} (\hat{n},\chi) - q^{-} (\hat{n},\chi))/2i \ .
\end{eqnarray}
In harmonic space, we can also find the remote quadrupole $E$ and $B$ mode components:
\begin{eqnarray}
    q_{\ell m}^{E} (\chi) &=& -(q^{+}_{\ell m}(\chi) + q^{-}_{\ell m}(\chi))/2 \\
    q_{\ell m}^{B} (\chi) &=& -(q^{+}_{\ell m}(\chi) - q^{-}_{\ell m}(\chi))/2i \ .
\end{eqnarray}
Scalar perturbations contribute only to $q^E_{\ell m}$ while primordial tensor perturbations contribute to both $q^E_{\ell m}$ and $q^B_{\ell m}$. Finally, we can go the other direction and write the $Q$ and $U$ components in terms of the $E$ and $B$ modes as
\begin{eqnarray}\label{eq:QandUfromEandB}
    q^{Q} (\hat{n},\chi) &=& - \left[ \sum_{\ell m} q_{\ell m}^{E} (\chi) X_{1; \ell m}(\hat{n}) + i q_{\ell m}^{B} (\chi) X_{2; \ell m}(\hat{n})\right] \\
    q^{U} (\hat{n},\chi) &=& - \left[\sum_{\ell m} q_{\ell m}^{B} (\chi) X_{1; \ell m}(\hat{n}) - i q_{\ell m}^{E} (\chi) X_{2; \ell m}(\hat{n})\right]
\end{eqnarray}
where
\begin{eqnarray}\label{eq:X1X2}
    X_{1; \ell m}(\hat{n}) &=& ({}_{+ 2} Y_{\ell m} (\hat{n}) + {}_{- 2} Y_{\ell m} (\hat{n}))/2 \\
    X_{2; \ell m}(\hat{n}) &=& ({}_{+ 2} Y_{\ell m} (\hat{n}) - {}_{- 2} Y_{\ell m} (\hat{n}))/2 \ .
\end{eqnarray}

The remote quadrupole field and the primary CMB temperature and polarization anisotropies are correlated since both depend on long-wavelength modes of comparable scales. The various auto- and cross-spectra between the CMB and remote quadrupole fields from scalar modes can be computed from 
\begin{equation}\label{eq:scalar_spectra}
    C_\ell^{XY} (\chi, \chi') = 4 \pi \int\frac{d k}{k} \Delta^{X}_\ell(k,\chi)\Delta^{Y}_\ell(k,\chi')\mathcal{P}(k),
\end{equation}
where $XY = \{ TT, EE, TE, Tq_E, E q_E, q_E q_E \}$, $\Delta^{X}_\ell(k,\chi)$ are transfer functions defined in Appendix~\ref{app:quadrupolefield} (note that there is $\chi$-dependence only when $X,Y = q_E$), and $\mathcal{P}(k)$ is the dimensionless primordial scalar power spectrum. For tensor modes we have
\begin{equation}\label{eq:tensor_spectra}
    C_\ell^{XY} (\chi, \chi') = \frac{3}{4 \pi^2} \int\frac{d k}{k} \Delta^{X; h}_\ell(k,\chi)\Delta^{Y; h}_\ell(k,\chi')\mathcal{P}_h(k),
\end{equation}
where $XY = \{TT, EE, TE, BB, Eq_E, Bq_B, q_E q_E, q_B q_B\}$, $\Delta^{X}_\ell(k,\chi)$ are transfer functions defined in Appendix~\ref{app:quadrupolefield} (note that there is $\chi$-dependence only when $X$ or $Y$ is $q_E$ or  $q_B$), and $\mathcal{P}_h(k)$ is the dimensionless primordial tensor power spectrum.

\subsection{The conditional remote quadrupole field}

\begin{figure}
    \centering
    \includegraphics[width=0.5\linewidth]{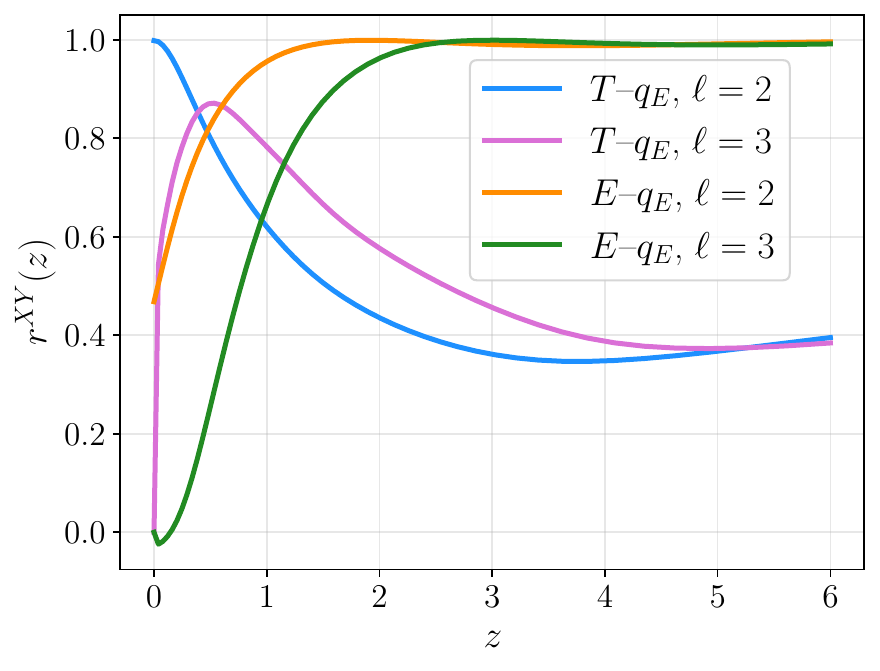}
    \caption{Correlation coefficient between the primary CMB \{$T$, $E$, $B$\} and the scalar remote quadrupole field $q_E$. At low redshift, the $\ell=2$ $E$-mode remote quadrupole ($q_E$) approaches the temperature quadrupole. At high redshift, the $E$-mode remote quadrupole becomes fully correlated with the reionization $E$-modes.}
    \label{fig:rxy}
\end{figure}

Both the remote quadrupole field and the primary CMB temperature and polarization are sourced by large-scale modes at last scattering, and are therefore highly correlated~\cite{Kamionkowski:1997na}. In figure~(\ref{fig:rxy}) we show the correlation coefficient $\big(r_\ell^{XY} (z)= C_{\ell}^{XY}(z)/\sqrt{C_{\ell}^{XX} C_{\ell}^{YY}(z)}\big)$ between the scalar remote quadrupole fields and the primary CMB~\footnote{See Refs.~\cite{Deutsch:2018umo} for a detailed discussion of the correlation between the tensor-induced primary CMB and remote quadrupole.}. In the limit where $z \rightarrow 0$, $r_{\ell=2}^{T q_E} \rightarrow 1$ since the $E$-mode remote quadrupole field is equal to the primary CMB temperature quadrupole. In the limit where $z \rightarrow z_{\rm rei}$, $r_{\ell}^{E q_E} \rightarrow 1$ and $r_{\ell}^{B q_B} \rightarrow 1$ since it is the remote quadrupole field integrated against the visibility function that gives the primary CMB polarization. This is a crucial point, implying that the remote quadrupole field is already significantly constrained at both low and high redshift by existing observations of the CMB temperature and polarization. 

Given a measured temperature and $E$-mode polarization map ($T_{\ell m}^{0},\,E_{\ell m}^{0}$), the maximum likelihood template for the remote quadrupole is 
\begin{eqnarray}\label{eq:qE_template}
    \tilde{q}^E_{\ell m} (\chi) 
    = \frac{1}{C_{\ell}^{TT}C_{\ell}^{EE}-(C_{\ell}^{TE})^2 } && \left[ \left(C_{\ell}^{EE} C_{\ell}^{T q_E} (\chi) - C_{\ell}^{TE} C_{\ell}^{E q_E} (\chi) \right) T_{\ell m}^{0} \right. \\
    &+& \left.  \left(C_{\ell}^{TT} C_{\ell}^{E q_E} (\chi) - C_{\ell}^{TE} C_{\ell}^{T q_E} (\chi) \right) E_{\ell m}^{0} \right]  \nonumber
\end{eqnarray}
where the $C_\ell$'s are $\Lambda$CDM theory spectra (plus expected noise for the auto-correlations); a derivation is given in Appendix~\ref{app:constrained}.
This can be transformed to templates for $Q$ and $U$ using Eq.~\eqref{eq:QandUfromEandB}. 

In addition to this correlated component, a given realization of the remote quadrupole field will have an uncorrelated component, so the full remote quadrupole field is
\begin{eqnarray}\label{eq:qE_conditional}
    q_{\ell m}^{E} (\chi) = \tilde{q}^E_{\ell m} (\chi) + \delta q^{E}_{\ell m} (\chi)
\end{eqnarray}
where $\delta q^{E}_{\ell m}$ is drawn from a multi-variate Gaussian with zero mean and a covariance given by
\begin{eqnarray}
    C_{\ell}^{\delta q_E \delta q_E} (\chi, \chi') = C_{\ell}^{q_E q_E} (\chi, \chi') - C_{\ell}^{\tilde{q}_E \tilde{q}_E} (\chi, \chi') \ .
\end{eqnarray}
There is a significant redshift-redshift correlation along the light cone which must be included in any model. In the presence of primordial tensor modes, there is an analogous split into components that are correlated with the reionization $B$-modes and components that are uncorrelated. 

\begin{figure}[htbp]
    \centering
    
    \begin{subfigure}{0.48\textwidth}
        \centering
        \includegraphics[width=\linewidth]{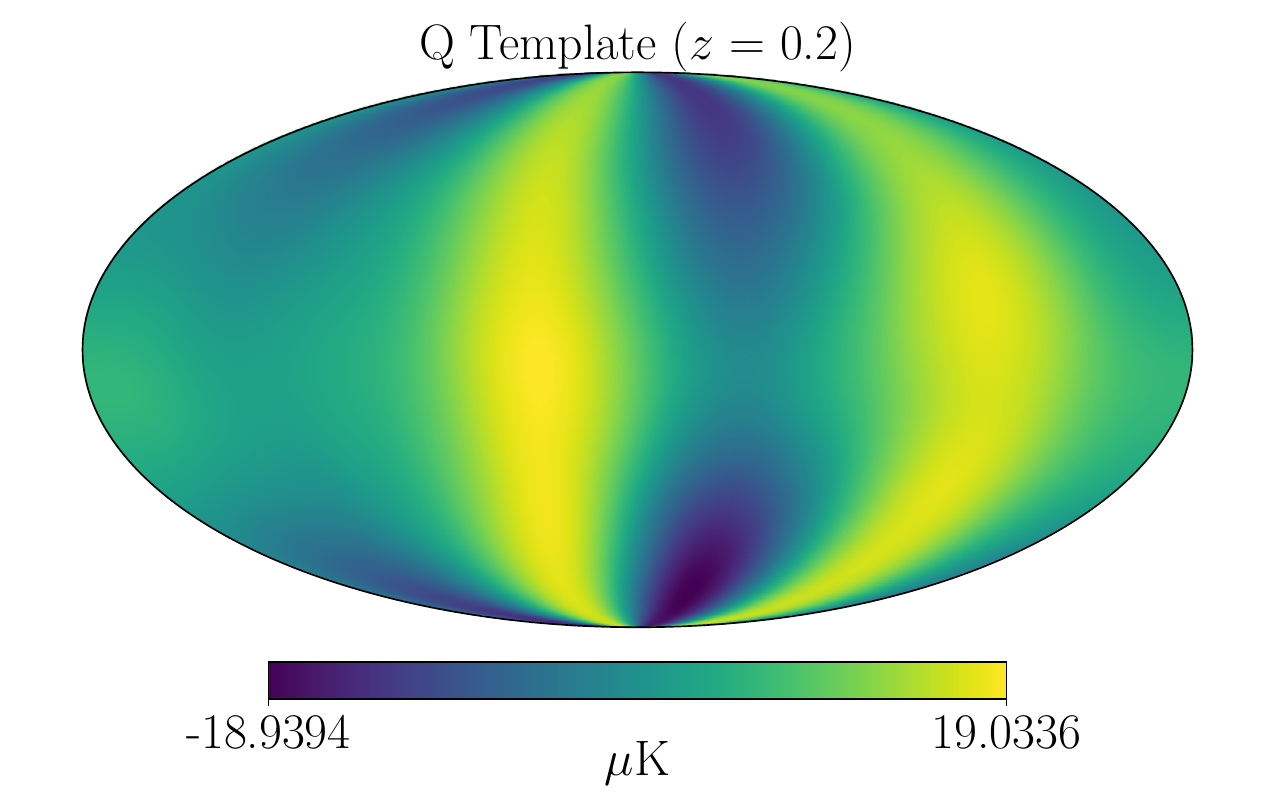}
    \end{subfigure}
    \hfill 
    \begin{subfigure}{0.48\textwidth}
        \centering
        \includegraphics[width=\linewidth]{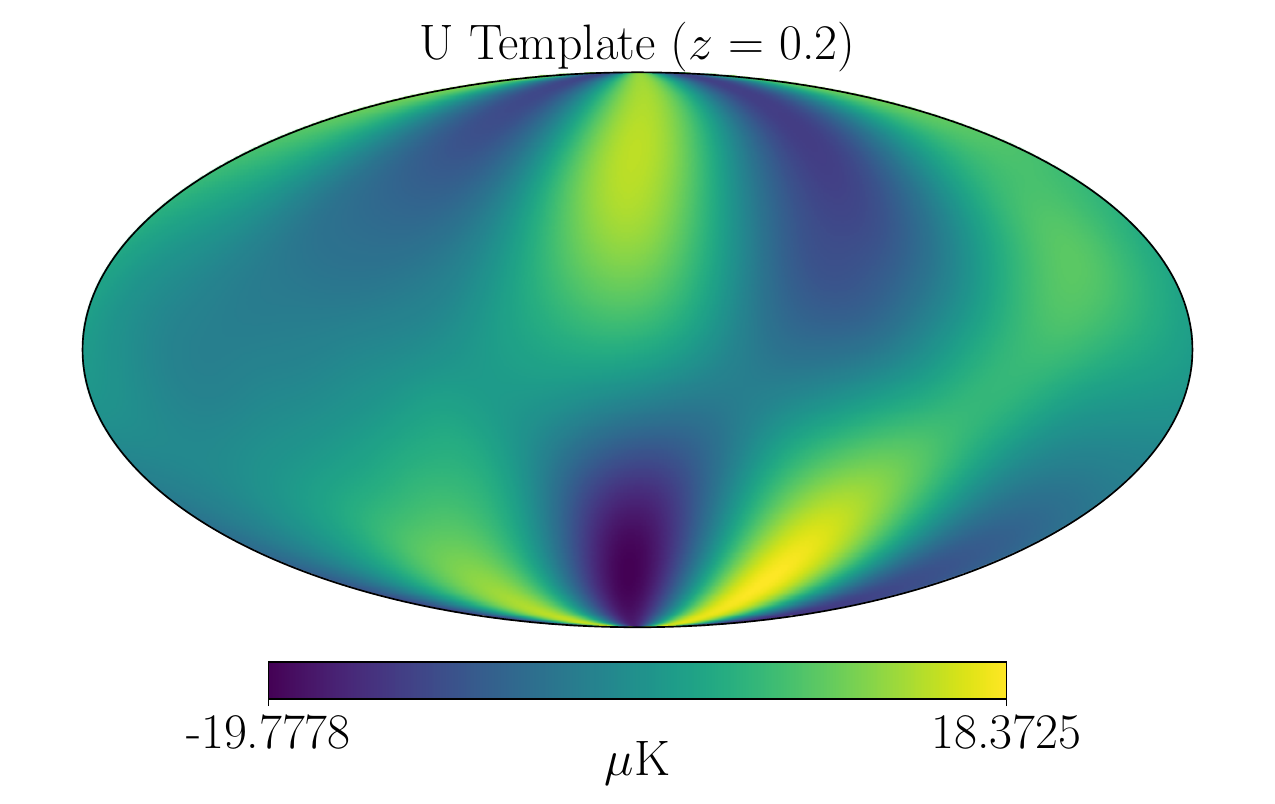}
    \end{subfigure}

    \vspace{0.5cm} 
    \begin{subfigure}{0.48\textwidth}
        \centering
        \includegraphics[width=\linewidth]{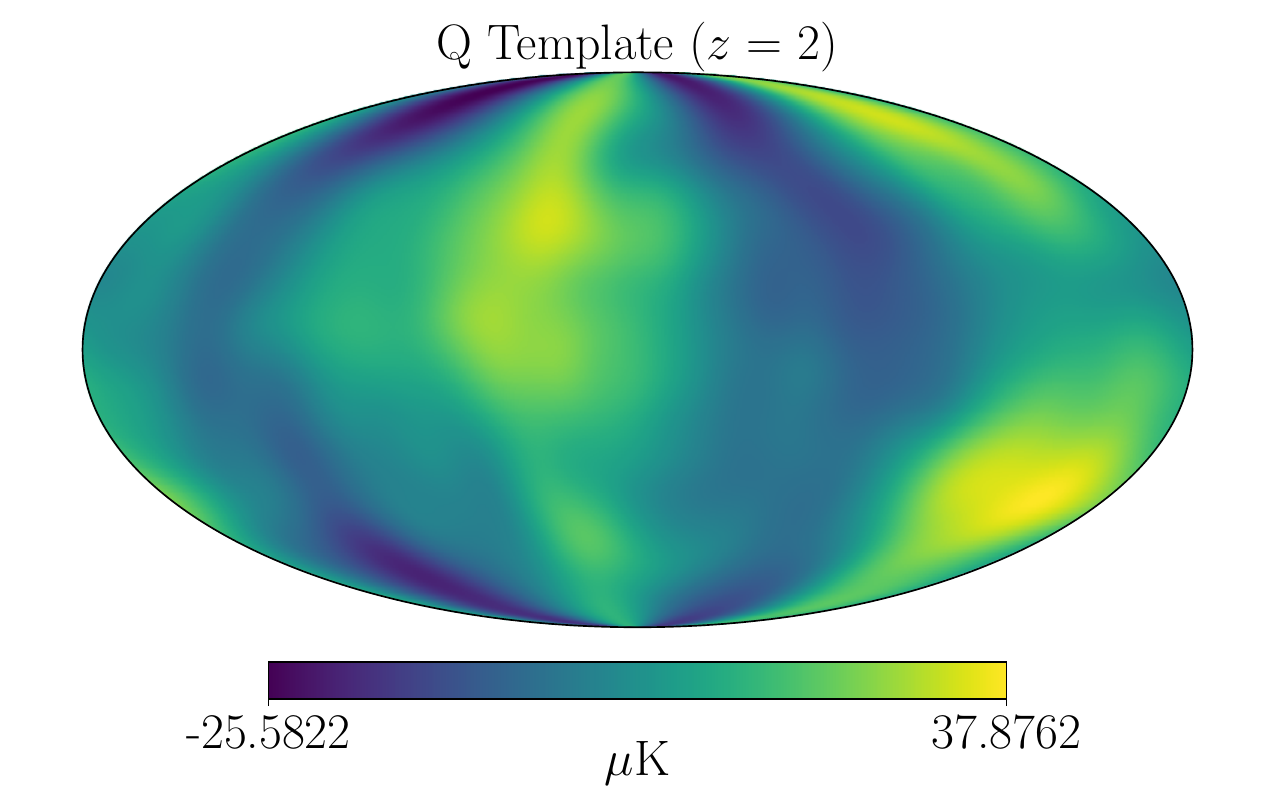}
    \end{subfigure}
    \hfill
    \begin{subfigure}{0.48\textwidth}
        \centering
        \includegraphics[width=\linewidth]{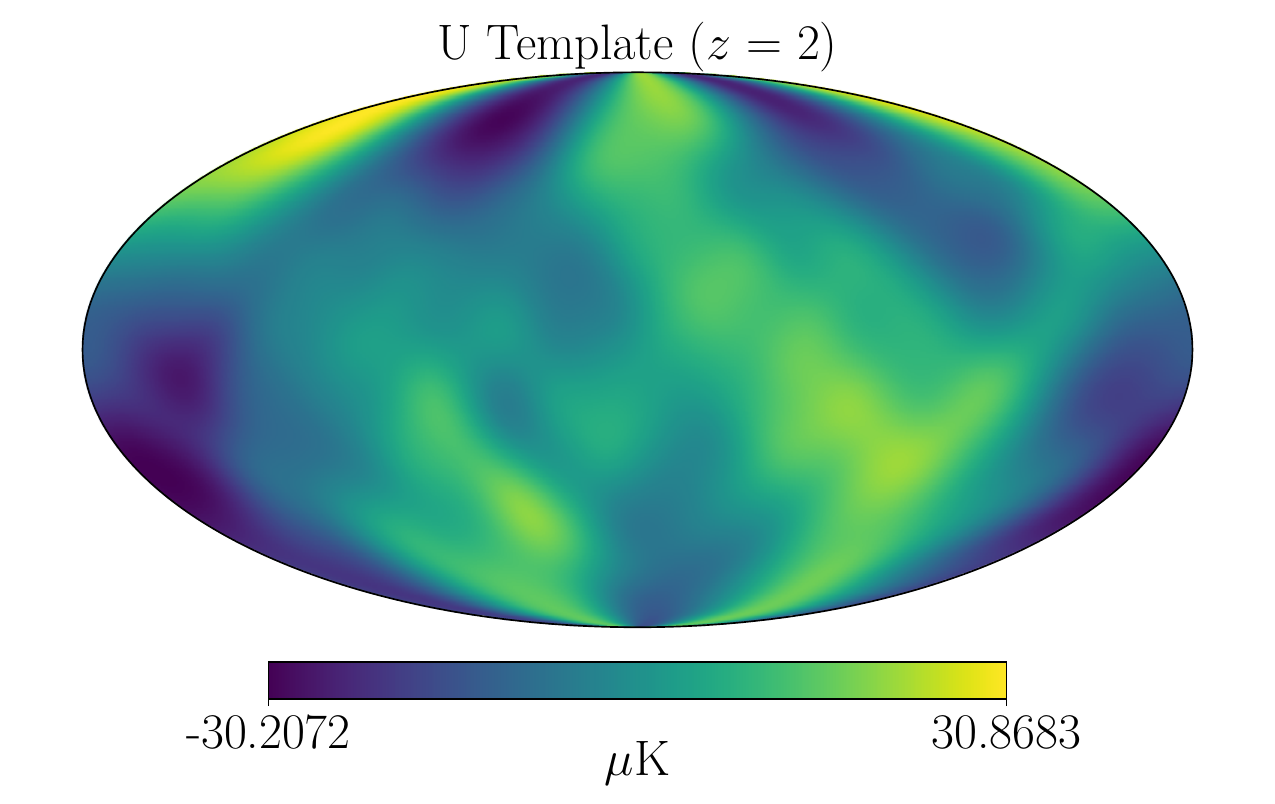}
    \end{subfigure}
    \vspace{0.5cm}

    \begin{subfigure}{0.48\textwidth}
        \centering
        \includegraphics[width=\linewidth]{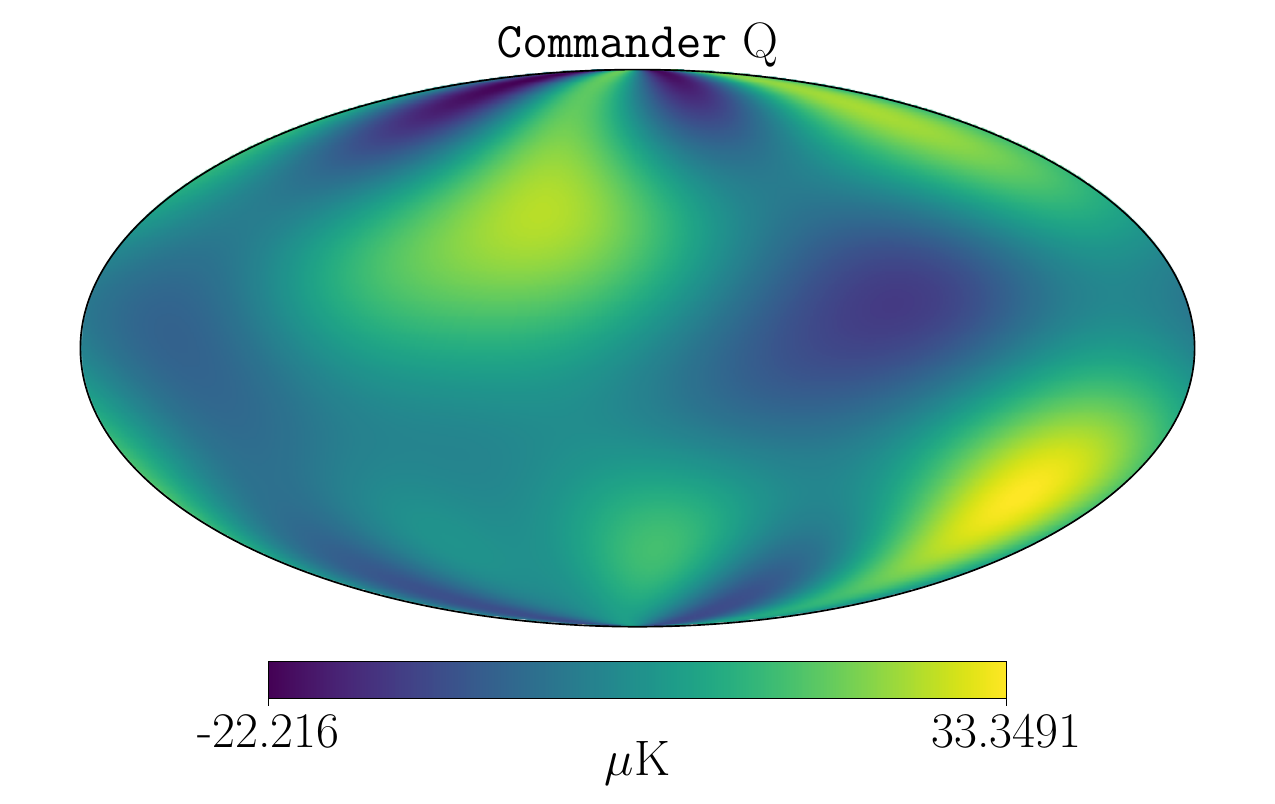}
    \end{subfigure}
    \hfill 
    \begin{subfigure}{0.48\textwidth}
        \centering
        \includegraphics[width=\linewidth]{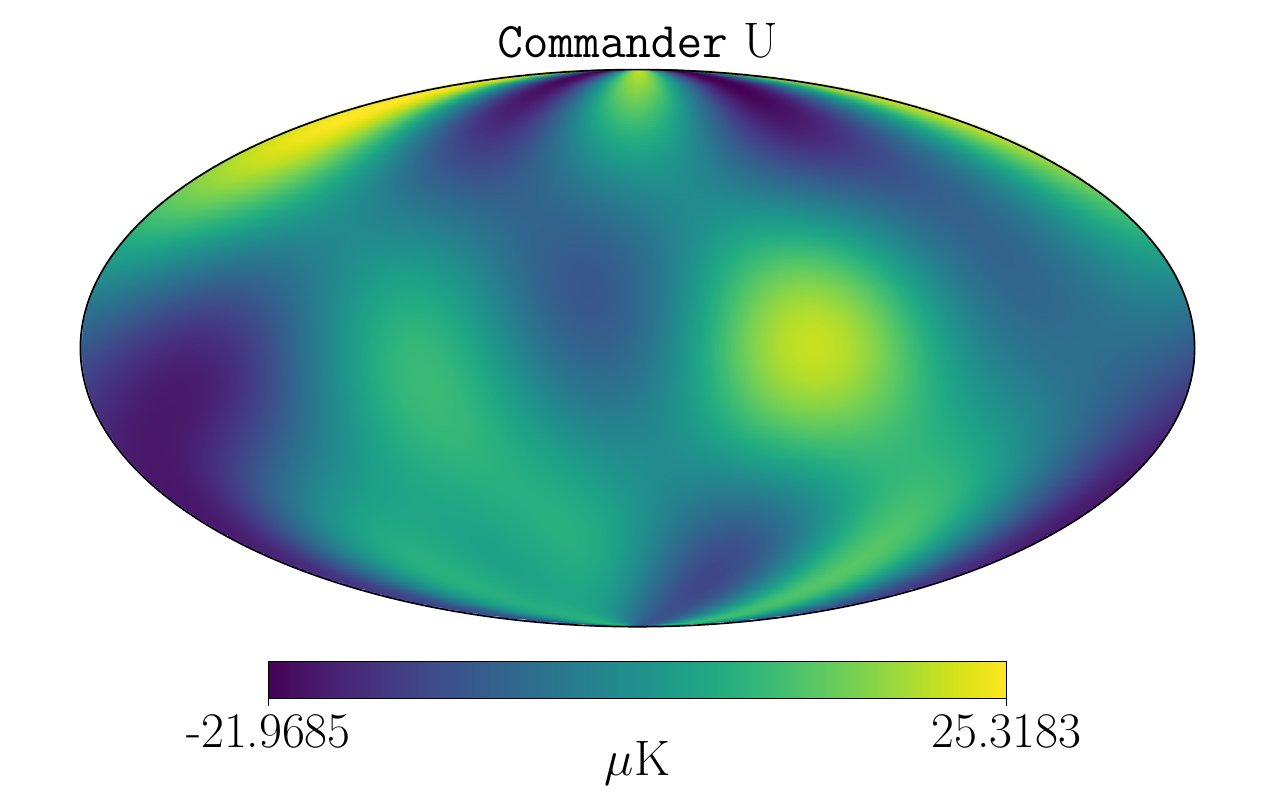}
    \end{subfigure}

    \caption{\textit{Top two rows}: The $Q$ (left) and $U$ (right) remote quadrupole template maps  for $z=0.2$ (upper) and $z=2$ (lower) from Eq.~\eqref{eq:qE_template} using the \texttt{Commander} temperature and $E$-mode polarization. \textit{Bottom row:} The \texttt{Commander} $Q$ (left) and $U$ (right) maps, filtered to $\ell<5$ and rescaled by $-(\sqrt{6}\tau_{\rm rei}/10)^{-1}$ to approximate the remote quadrupole.}
    \label{fig:QUtemplates}
\end{figure}

As described in greater detail in Sec.~\ref{sec:analysis_pipeline}, we use the \textit{Planck} PR3 inpainted \texttt{Commander} CMB map~\cite{PlanckCollaboration2020,Eriksen_2008} as a proxy for the primary CMB temperature and polarization on large angular scales. We use these maps to build a template for the remote quadrupole using Eq.~\eqref{eq:qE_template}. In figure~(\ref{fig:QUtemplates}) we show the template for $q^{Q,U}$ at $z=0.2$ and $z=2.0$, using multipoles up to $\ell_{\rm max}=20$. At $z=0.2$ the remote quadrupole is dominated by the $\ell=2$ mode. Note that the Stokes parameters are composed of spin-2 harmonics (see Eq.~\eqref{eq:QandUfromEandB}), so while $q_{\ell m}^E \rightarrow a_{\ell m}^T$ at low redshift, the map of the Stokes parameters does not directly resemble the temperature quadrupole. At a redshift $z=2$, there is more angular structure in the $q^{Q,U}$ maps. This is because the remote quadrupole field at this distance is de-correlated with the local quadrupole along different lines of sight. At $z=2$, the remote quadrupole field is significantly correlated with the reionization $E$ modes that dominate the primary CMB at low-$\ell$. In the bottom row of figure~(\ref{fig:QUtemplates}), we show the \texttt{Commander} polarization maps that are low-pass filtered ($\ell<5$) and rescaled (by $-(\sqrt{6}\tau_{\rm rei}/10)^{-1}$ to approximate the remote quadrupole at reionization). Comparing the middle and bottom rows of this figure, the remote quadrupole field at $z=2$ is visibly correlated with the \texttt{Commander} map on large angular scales. We pause here to note that the discussion so far has assumed that the template for the remote quadrupole field is signal dominated. This is certainly true for the temperature maps, but not necessarily for polarization. We relax this assumption in our main analysis.

A striking property of the CMB in our Universe is the low temperature quadrupole, $C_2^{\rm obs} = 219 \ {\mu K}^2$ compared with $C_2^{\Lambda \rm CDM} = 1060 \ {\mu K}^2$. Within $\Lambda$CDM, this implies that we do not inhabit a `typical' realization, or equivalently, a typical location. It is also possible that the observed low CMB quadrupole implies new physics, for example a suppression of power on large scales (see e.g.~\cite{Planck:2019evm} for an assessment). Different theories yield different predictions for the redshift dependence of the remote quadrupole field, including the reionization $E$-modes. Indeed, it has been noted e.g. in Ref.~\cite{Givans:2023kbg} that while the temperature quadrupole is low, the $E$-mode quadrupole is high compared to the $\Lambda$CDM expectation. Observations of the full redshift dependence of the remote quadrupole can distinguish different scenarios and improve the constraining power of the CMB temperature and $E$-modes alone~\cite{Cayuso:2019hen,Pan:2019dax}. 

\begin{figure}
    \centering
    \includegraphics[width=\linewidth]{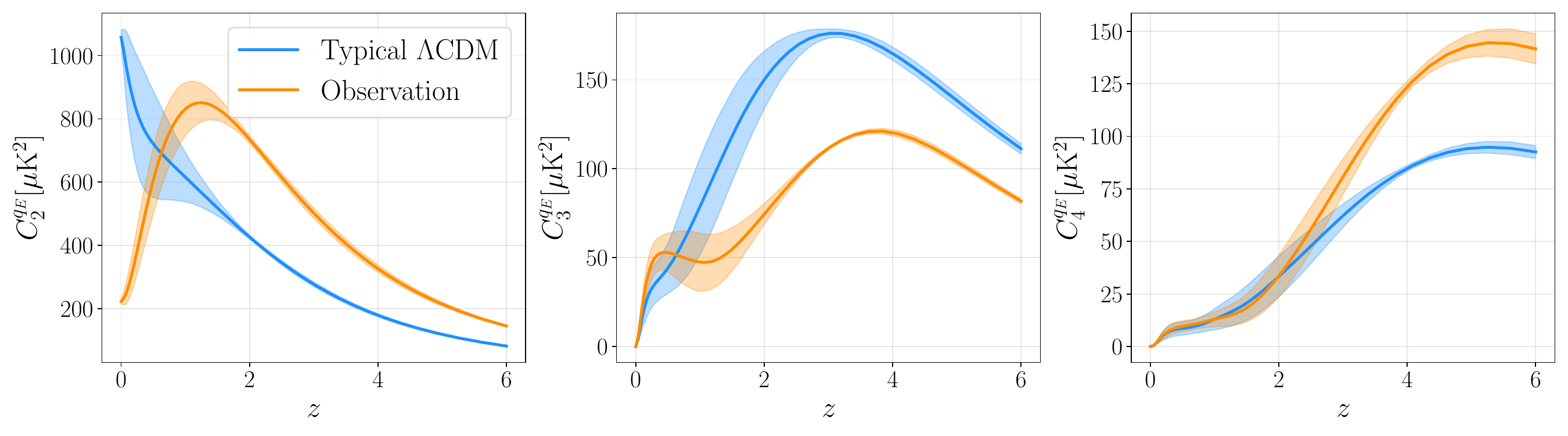}
    \caption{The expected redshift contribution to the $\ell=2-4$ moments of the remote quadrupole signal for realizations conditioned on a typical $\Lambda$CDM draw (blue) or on local observations of $T$ and $E$ (orange). The solid lines represent the mean value averaged over $10^4$ constrained realizations, while the shaded regions indicate the $68\%$ confidence interval. }
    \label{fig:consrealsl2to4}
\end{figure}

To explore the possible variations in the conditional remote quadrupole field, and develop intuition about optimal choices for the tracer field, we create $10^4$ realizations of the remote quadrupole field as defined in Eq.~\eqref{eq:qE_conditional} on a grid of redshifts, conditioned on the observed \texttt{Commander} low-$\ell$ CMB. We create another set of  realizations not constrained to match our observed Universe but rather based on a `typical' $\Lambda$CDM draw for the low-$\ell$ CMB temperature and polarization. In figure~(\ref{fig:consrealsl2to4}), we show the ensemble average of the conditional realizations as well as the variance as a function of redshift for $\ell=2-4$. 

Focusing first on $\ell =2$ (figure~(\ref{fig:consrealsl2to4}), left panel), as expected, the low-redshift ensemble average spectra are equal to the observed (orange curve) or `typical' (blue curve) temperature quadrupole. At high redshift, both approach the remote quadrupole expected from either the observed or `typical' $E$-mode quadrupole. Note at high redshift the slightly larger than typical remote quadrupole for the case conditioned on observations~\footnote{As mentioned previously, we assume here that the template is signal-dominated. The measured $E$ modes are a combination of signal and noise/foreground residuals.}. 

The `typical' conditional remote quadrupole falls monotonically with redshift, while the remote quadrupole conditioned on observations grows at low redshift, reaching a peak around $z \sim 2$, and then falling with redshift. This can be seen as a `reversion to the mean': the local quadrupole is low by chance within $\Lambda$CDM, so as we move down the light cone in redshift, we should expect to see remote quadrupoles increasingly consistent with the rather typical $E$-mode quadrupole.
This is seen at the map-level in figure~(\ref{fig:QUtemplates}). In both cases, the variance in the conditional remote quadrupole (the shaded region around the curves in figure~(\ref{fig:consrealsl2to4})) is largest at low redshift, becoming negligible for $z> 2$. This spread can be thought of as the extra information beyond the CMB temperature and $E$-mode polarization that one might obtain by measuring the remote quadrupole field within $\Lambda$CDM. The mean and variance need not be the same in theories beyond $\Lambda$CDM (e.g. for modified gravity~\cite{Pan:2019dax}).
Moving to $\ell =3$ (figure~(\ref{fig:consrealsl2to4}), center panel) and $\ell =4$ (figure~(\ref{fig:consrealsl2to4}), right panel), the remote quadrupole is zero at low redshift, and increases with redshift to values consistent with the reionization $E$-modes. Note that the magnitude is substantially smaller than the $\ell = 2$ remote quadrupole at low redshift, but comparable at high redshift. This is seen at the map level as the additional structure at $z=2.0$ compared with $z=0.2$ in figure~(\ref{fig:QUtemplates}). 

This analysis demonstrates that \textit{the low CMB quadrupole does not impact the detectability of the remote quadrupole at moderate redshift.} We therefore focus on targeting the remote quadrupole field using tracers at $z \gtrsim 1$, as we now describe in detail.

\section{PSZ and the bispectrum}\label{sec:pszandbispectrum}

The pSZ effect is the product of the optical depth field and remote quadrupole (see figure~(\ref{fig:qusim})), and is therefore an intrinsically non-Gaussian field (e.g. even if the optical depth and remote quadrupole fields were Gaussian, their product would not be). As discussed in the previous section, the optical depth field is correlated with tracers of LSS, whereas the remote quadrupole field is correlated with the low-$\ell$ primary CMB temperature and polarization. In analogy with kSZ ~\cite{Smith:2018bpn}, we can use the optimal bispectrum amplitude estimator to explore the detectability of pSZ using a galaxy survey and the CMB, and to derive a set of quadratic estimators for the remote quadrupole field, optical depth, or tracer density field. For more details on the standard bispectrum estimator formalism we use, refer to Ref.~\cite{Komatsu:2003iq}. 

We first present the general expression for the bispectrum between the low-$\ell$ primary CMB $T$, $E$, $B$, the high-$\ell$ CMB $E$, $B$ and galaxy density. Then, we use this bispectrum as a template to develop an estimator for the bispectrum amplitude. Finally, we demonstrate how this bispectrum amplitude estimator can be decomposed into correlations between various intermediate data products that correspond to estimators previously developed in the literature. 

Taking the spin-2 transform of the pSZ signal in Eq.~\eqref{eq:pSZ_definition} we have
\begin{eqnarray}
    E_{\ell m}^{\rm pSZ} &=& - \frac{\sqrt{6}}{10} \sum_{\ell_1 m_1; \ell_2 m_2} (-1)^m
\sqrt{\frac{(2\ell+1)(2\ell_1+1)(2\ell_2+1)}{4\pi}}
\begin{pmatrix}
\ell & \ell_1 & \ell_2 \\
2 & 0 & -2
\end{pmatrix} 
\begin{pmatrix}
\ell & \ell_1 & \ell_2 \\
-m & m_1 & m_2
\end{pmatrix} \nonumber \\ 
 &\times& \int d\chi \; \dot{\tau}_{\ell_1 m_1} (\chi) e^{-2\tau(\chi)} \ \left( \alpha_{\ell \ell_1 \ell_2}^{E q_E} q^E_{\ell_2 m_2}(\chi) - \alpha_{\ell \ell_1 \ell_2}^{E q_B} q^E_{\ell_2 m_2}(\chi) \right)
\end{eqnarray}
and 
\begin{eqnarray}
    B_{\ell m}^{\rm pSZ} &=& - \frac{\sqrt{6}}{10} \sum_{\ell_1 m_1; \ell_2 m_2} (-1)^m
\sqrt{\frac{(2\ell+1)(2\ell_1+1)(2\ell_2+1)}{4\pi}}
\begin{pmatrix}
\ell & \ell_1 & \ell_2 \\
2 & 0 & -2
\end{pmatrix} 
\begin{pmatrix}
\ell & \ell_1 & \ell_2 \\
-m & m_1 & m_2
\end{pmatrix} \nonumber \\ 
 &\times& \int d\chi \; \dot{\tau}_{\ell_1 m_1} (\chi) e^{-2\tau(\chi)} \ \left( \alpha_{\ell \ell_1 \ell_2}^{B q_E} q^E_{\ell_2 m_2}(\chi) + \alpha_{\ell \ell_1 \ell_2}^{B q_B} q^E_{\ell_2 m_2}(\chi) \right)
\end{eqnarray}
where
\begin{eqnarray}
\alpha^{E q_E}_{\ell \ell' \ell''} = \alpha^{B q_B}_{\ell \ell' \ell''} = \frac{1}{2}\left(1+(-1)^{\ell+\ell'+\ell''}\right), \ \ \ 
\alpha^{B q_E}_{\ell \ell' \ell''} = \alpha^{E q_B}_{\ell \ell' \ell''} = \frac{1}{2}\left(1-(-1)^{\ell+\ell'+\ell''}\right) \ .
\end{eqnarray}
Note that there is always an $E$-mode and $B$-mode contribution from pSZ even when the remote quadrupole is pure $E$-mode. This is a general property of secondaries; see also discussions of lensing~\cite{Lewis:2006fu} and screening~\cite{Dvorkin:2009ah}.

Correlating the pSZ signal with a tracer denoted by $g$ (e.g. galaxy density) and the CMB (denoted by $Z$ and $V$), the bispectra associated with pSZ are:
\begin{eqnarray}\label{eq:pSZbispectra}
B^{ZgV}_{\ell \ell' \ell''}
&=&
-\frac{\sqrt{6}}{10}
\sqrt{\frac{(2\ell+1)(2\ell'+1)(2\ell''+1)}{4\pi}}
\begin{pmatrix}
\ell & \ell' & \ell'' \\
2 & 0 & -2
\end{pmatrix} \nonumber \\
&\times&
 \int d\chi \;
\left[ \alpha^{V q_E}_{\ell \ell' \ell''} C^{Z q_E}_{\ell}(\chi) + \alpha^{V q_B}_{\ell \ell' \ell''} C^{Z q_B}_{\ell}(\chi)\right]
C^{\dot{\tau} g}_{\ell'}(\chi)\,
e^{-2\tau(x)}
\end{eqnarray}
where
\begin{equation}
Z = T, E, B, 
\quad
V = E, B.
\end{equation}
The $V$ terms are the high-$\ell$ CMB polarization which contain the pSZ signal and the $Z$ are the low-$\ell$ CMB temperature and polarization which are correlated with the remote quadrupole field. The non-zero cross-correlations for $Z=T, E, B$ are
\begin{eqnarray}
C^{T q_E}_{\ell}(\chi), \ C^{E q_E}_{\ell}(\chi), \ 
C^{B q_B}_{\ell}(\chi) \ .
\end{eqnarray}

The integral in Eq.~\eqref{eq:pSZbispectra} can be evaluated directly using the model spectra. To make the connection with previous results in the literature more clear, here we approximate the integral following Ref.~\cite{Bloch_2024}. As shown in Ref.~\cite{Bloch_2024} this is an excellent approximation for the large photometric redshift bins we will use in our analysis. We factorize the optical-depth tracer cross power spectrum into the product of a scale-dependent and a redshift-dependent function and approximate $e^{-\tau(z)} \simeq 1$:
\begin{equation}\label{eq:Wqdef}
    C_{\ell}^{\dot{\tau} g} (\chi) e^{-\tau(\chi)} \simeq \bar{C}_{\ell}^{\tau g} \frac{C_{\ell=\bar{\ell}}^{\dot{\tau} g} (\chi)}{\bar{C}_{\ell=\bar{\ell}}^{\tau g}} 
    = \bar{C}_{\ell}^{\tau g} W_q(\chi)
\end{equation}
where $\bar{\chi}$ is a reference redshift, usually the median of the redshift bin, and $\bar{\ell}$ is a reference scale (we set it to 200, discussed in Sec.~\ref{sec:recpipeline}). Then 
\begin{eqnarray}
    \bar{C}_{\ell}^{\tau g} = C_{\ell}^{\dot{\tau} g} (\chi = \bar{\chi}) \Delta \chi
\end{eqnarray} 
and $\Delta \chi$ is chosen so that $\int d\chi W_q(\chi) = 1$. We refer to $W_q(\chi)$ as the quadrupole window function, it traces the redshift-dependence of the galaxy-optical depth cross-power spectrum. We note that $W_q$ is the same as one would obtain in kSZ velocity reconstruction with the same tracer sample; see Ref.~\cite{Bloch_2024}. In this approximation, we simplify the bispectra as 
\begin{eqnarray}\label{eq:pSZbispectra_integrated}
B^{ZgV}_{\ell \ell' \ell''}
&=&
-\frac{\sqrt{6}}{10}
\sqrt{\frac{(2\ell+1)(2\ell'+1)(2\ell''+1)}{4\pi}}
\begin{pmatrix}
\ell & \ell' & \ell'' \\
2 & 0 & -2
\end{pmatrix} \bar{C}_{\ell'}^{\tau g} 
\left[ \alpha^{V q_E}_{\ell \ell' \ell''} C^{Z \bar{q}_E}_{\ell} + \alpha^{V q_B}_{\ell \ell' \ell''} C^{Z \bar{q}_B}_{\ell}\right]
\end{eqnarray}
where we define the fields
\begin{eqnarray}\label{eq:binavgremoteq}
    \bar{q}^E_{\ell m} = \int d\chi \ W_{q}(\chi) q^E_{\ell m}(\chi), \ \ \  \bar{q}^B_{\ell m} = \int d\chi \ W_{q}(\chi) q^B_{\ell m}(\chi) \ .
\end{eqnarray}
This bispectrum is the basic building block we use below. We use the theory spectra $\bar{C}_{\ell}^{\tau g}$ and $C^{Z \bar{q}_E}_{\ell}$, $C^{Z \bar{q}_B}_{\ell}$ as defined in Sec.~\ref{sec:pSZandcross} to compute the expected value of the bispectrum.

The theory bispectrum serves as a template that we use to estimate the amplitude of the bispectrum from data. For a choice of $Z$ and $V$ the bispectrum amplitude estimator on the full sky is given by
\begin{eqnarray}\label{eq:AXgY}
\hat{A}_{ZgV}
= \sigma_{ZgV}^2
\sum 
\begin{pmatrix}
\ell & \ell' & \ell'' \\
m & m' & m''
\end{pmatrix}
\frac{ B^{ZgV}_{\ell \ell' \ell''} }{C_\ell^{ZZ} C_{\ell'}^{gg} C_{\ell''}^{VV}}
 Z_{\ell m} g_{\ell' m'} V_{\ell'' m''}
\end{eqnarray}
where
\begin{eqnarray}\label{eq:oneoversigmafish}
    \frac{1}{\sigma_{ZgV}^2}
=
\sum_{\ell \ell' \ell''}
\frac{\left(B^{ZgV}_{\ell \ell' \ell''}\right)^2}{C_\ell^{ZZ} C_{\ell'}^{gg} C_{\ell''}^{VV}} \ .
\end{eqnarray}
The power spectra and bispectra appearing in the estimator are computed from the theoretical models described in Sec.~\ref{sec:pSZandcross}. 

The bispectrum estimator weights in Eq.~\eqref{eq:AXgY} are appropriate for an ensemble in which the fields $Z,g,V$ are nearly Gaussian and the pSZ signal to noise is small. Varying over realizations of the fields $Z,g,V$, the estimator mean and variance are
\begin{eqnarray}
    \langle A_{ZgV} \rangle = 1, \ \ \ \langle A_{ZgV}^2 \rangle = \sigma_{ZgV}^2 \ .
\end{eqnarray}
The estimator is therefore unbiased and, since the variance is given by the Fisher information as in Eq.~\eqref{eq:oneoversigmafish}, minimum variance.

The computational cost of the sums in the estimator Eq.~\eqref{eq:AXgY} can be greatly simplified by re-casting it in pixel space as an integral over filtered fields. This is possible because the pSZ bispectra Eq.~\eqref{eq:pSZbispectra} are factorizable. We begin by finding the minimum variance combination of the estimators for $V=E,B$:
\begin{eqnarray}\label{eq:AZg}
\hat{A}_{Zg}
= \frac{\sigma_{ZgE}^{-2} \hat{A}_{ZgE} + \sigma_{ZgB}^{-2} \hat{A}_{ZgB}}{\sigma_{ZgE}^{-2}  + \sigma_{ZgB}^{-2}} \ .
\end{eqnarray}
By combining the $V=E,B$ estimators, we make use of all the high-$\ell$ information in CMB polarization. 

Next, define the following set of filtered fields. First,
\begin{eqnarray}\label{eq:tautildedef}
    \tilde{\tau}(\hat{n}) = \sum_{\ell' m'} \frac{\bar{C}^{\tau g}_{\ell'}}{C_{\ell'}^{gg} } g_{\ell' m'} Y_{\ell' m'} (\hat{n}),
\end{eqnarray}
this is equivalent to the maximum-likelihood template for $\tau$ given $g$. Next, we define the inverse-variance filtered CMB polarization maps:
\begin{eqnarray}\label{eq:QtildeEB}
        \tilde{Q}_{EB}(\hat{n}) = -\sum_{\ell'' m''} \left[\frac{E_{\ell'' m'' }}{C_{\ell''}^{EE}} X_{1;\ell'' m''} (\hat{n}) + \frac{i B_{\ell'' m'' }}{C_{\ell''}^{BB}} X_{2;\ell'' m''} (\hat{n}) \right],
\end{eqnarray}
\begin{eqnarray}\label{eq:UtildeEB}
        \tilde{U}_{EB}(\hat{n}) = - \sum_{\ell'' m''} \left[\frac{B_{\ell'' m'' }}{C_{\ell''}^{BB}} X_{1;\ell'' m''} (\hat{n})- \frac{iE_{\ell'' m'' }}{C_{\ell''}^{EE}} X_{2; \ell'' m''} (\hat{n}) \right],
\end{eqnarray}
where $X_{1;\ell m}$ and $X_{2;\ell m}$ are defined in Eq.~(\ref{eq:X1X2}). Finally, 
\begin{eqnarray}\label{eq:UfromZ}
        \hat{Q}_{Z}(\hat{n}) = -\sum_{\ell m} \left[\frac{C_{\ell}^{Z\bar{q}_E}}{C_{\ell}^{ZZ}} Z_{\ell m} X_{1;\ell m} (\hat{n}) + i \frac{C_{\ell}^{Z\bar{q}_B}}{C_{\ell}^{ZZ}} Z_{\ell m} X_{2;\ell m} (\hat{n}) \right],
\end{eqnarray}
\begin{eqnarray}\label{eq:QfromZ}
        \hat{U}_{Z}(\hat{n}) = - \sum_{\ell m} \left[\frac{C_{\ell}^{Z\bar{q}_B}}{C_{\ell}^{ZZ}} Z_{\ell m} X_{1;\ell m} (\hat{n}) - \frac{iC_{\ell}^{Z\bar{q}_E}}{C_{\ell}^{ZZ}} Z_{\ell m} X_{2; \ell m} (\hat{n}) \right],
\end{eqnarray}
this is the maximum-likelihood template for the remote quadrupole field in the tracer's redshift bin given a measurement of $Z$.
Each of these fields can be efficiently computed using fast spherical harmonic transforms implemented in e.g. \texttt{healpix}. With these definitions, we re-write Eq.~\eqref{eq:AZg} in the more computationally efficient form:
\begin{equation}\label{eq:A_estimator}
\hat{A}_{Zg} =  - \frac{\sqrt{6}}{10}
\sigma_{Zg}^2 \int d^2 \hat{n} \ \tilde{\tau}(\hat{n}) \left[\tilde{Q}_{EB}(\hat{n}) \hat{Q}_{Z}(\hat{n}) + \tilde{U}_{EB}(\hat{n}) \hat{U}_{Z}(\hat{n}) \right]
\end{equation}
where
\begin{eqnarray}\label{eq:A_estimator_variance}
\sigma_{Zg}^{-2} &\equiv& \sigma_{ZgE}^{-2}+\sigma_{ZgB}^{-2} \nonumber \\
&=& \frac{6}{100} \sum_{\ell,\ell', \ell''} \frac{(2\ell+1)(2\ell'+1)(2\ell''+1)}{4 \pi} \begin{pmatrix}
\ell & \ell' & \ell'' \\
2 & 0 & -2
\end{pmatrix}^2  
\frac{(\bar{C}_{\ell'}^{\tau g})^2 }{C_{\ell'}^{gg} C_{\ell''}^{EE} C_{\ell''}^{BB} } \\
&\times &
\left[ \left(|\alpha^{E\bar{q}_E}_{\ell \ell' \ell''}|^2 C_{\ell''}^{BB}+|\alpha^{B\bar{q}_E}_{\ell \ell' \ell''}|^2 C_{\ell''}^{EE}\right) \frac{(C_{\ell}^{Z\bar{q}_E})^2}{C_{\ell}^{ZZ}} +  \left(|\alpha^{E\bar{q}_B}_{\ell \ell' \ell''}|^2 C_{\ell''}^{EE}+|\alpha^{B\bar{q}_B}_{\ell \ell' \ell''}|^2 C_{\ell''}^{BB}\right) \frac{(C_{\ell}^{Z\bar{q}_B})^2}{C_{\ell}^{ZZ}} \right]. \nonumber
\end{eqnarray}
Note that from now on we will drop the bar over $\bar{q}_E/\bar{q}_B$, but will always be referring to the remote quadrupole averaged over the window function $W_q$.

The bispectrum amplitude estimator Eq.~\eqref{eq:A_estimator} can be used to detect the presence of the pSZ signal directly from data. It is also possible to decompose Eq.~\eqref{eq:A_estimator} to describe a set of pSZ statistics that can be viewed as interesting data products in their own right. These intermediate statistics are useful for studying the remote-quadrupole, the pSZ signal, and the optical depth field. The different ways of decomposing the bispectrum estimator are as follows:
\begin{enumerate}
    \item {\bf Remote quadrupole reconstruction correlated with a remote quadrupole template}: The bispectrum amplitude estimator can be split up as a quadratic estimator for the remote quadrupole field and the maximum-likelihood template for the remote quadrupole field (Eq.~(\ref{eq:QfromZ})). We define quadratic estimators for the $Q$ and $U$ components of the remote quadrupole fields defined in Eq.~\eqref{eq:binavgremoteq} as
    \begin{eqnarray}\label{eq:QandUrealspaceQEs}
    \hat{Q}_q (\hat{n}) = -\frac{\sqrt{6}}{10} N^q \tilde{\tau}(\hat{n}) \tilde{Q}_{EB}(\hat{n}), \ \ \ \hat{U}_q (\hat{n}) = -\frac{\sqrt{6}}{10} N^q \tilde{\tau}(\hat{n}) \tilde{U}_{EB}(\hat{n}) \ .
    \end{eqnarray}
    The bispectrum amplitude estimator Eq.~\eqref{eq:A_estimator} becomes
    \begin{equation}
\hat{A}_{Zg} =  
\sigma_{Zg}^2 \int d^2 \hat{n} \ \left[\hat{Q}_{q}(\hat{n}) \hat{Q}_{Z}(\hat{n}) + \hat{U}_{q}(\hat{n}) \hat{U}_{Z}(\hat{n}) \right]
\end{equation}
    and the variance decomposes as
    \begin{eqnarray}\label{eq:sigmaZg_QE}
        \sigma_{Zg}^{-2} &=& \sum_{\ell} (2\ell+1) \left[\frac{(C_{\ell}^{Zq_E})^2}{C_{\ell}^{ZZ} N_\ell^{q_E}} + \frac{(C_{\ell}^{Zq_B})^2}{C_{\ell}^{ZZ} N_\ell^{q_B}} \right]
    \end{eqnarray}
    where the quadrupole reconstruction noise is
    \begin{eqnarray}
    (N_\ell^{q_E})^{-1} &=& \frac{6}{100} \sum_{\ell', \ell''} \frac{(2\ell'+1)(2\ell''+1)}{4 \pi} \begin{pmatrix}
    \ell & \ell' & \ell'' \\
    2 & 0 & -2
    \end{pmatrix}^2  
    \frac{\left(|\alpha^{Eq_E}_{\ell \ell' \ell''}|^2 C_{\ell''}^{BB}+|\alpha^{Bq_E}_{\ell \ell' \ell''}|^2 C_{\ell''}^{EE}\right)(\bar{C}_{\ell'}^{\tau g})^2 }{C_{\ell'}^{gg} C_{\ell''}^{EE} C_{\ell''}^{BB} } \nonumber \\
    &\simeq& (N^q)^{-1}
    \end{eqnarray}
    \begin{eqnarray}
    (N_\ell^{q_B})^{-1} &=& \frac{6}{100} \sum_{\ell', \ell''} \frac{(2\ell'+1)(2\ell''+1)}{4 \pi} \begin{pmatrix}
    \ell & \ell' & \ell'' \\
    2 & 0 & -2
    \end{pmatrix}^2  
    \frac{\left(|\alpha^{Eq_B}_{\ell \ell' \ell''}|^2 C_{\ell''}^{BB}+|\alpha^{Bq_B}_{\ell \ell' \ell''}|^2 C_{\ell''}^{EE}\right)(\bar{C}_{\ell'}^{\tau g})^2 }{C_{\ell'}^{gg} C_{\ell''}^{EE} C_{\ell''}^{BB} } \nonumber \\
    &\simeq& (N^q)^{-1}
    \end{eqnarray}
    and 
    \begin{eqnarray}\label{eq:Nq}
        (N^q)^{-1} = \frac{3}{100} \sum_{\ell'} \frac{(2\ell'+1)}{4 \pi}  \frac{\left( C_{\ell'}^{BB}+C_{\ell'}^{EE}\right)(\bar{C}_{\ell'}^{\tau g})^2 }{C_{\ell'}^{gg} C_{\ell'}^{EE} C_{\ell'}^{BB} } \ .
    \end{eqnarray}
 The variance of the bispectrum estimator Eq.~\eqref{eq:sigmaZg_QE} is simply the signal-to-noise squared of the cross-correlation between the maximum-likelihood template and the quadratic estimator for the remote quadrupole field (in the limit where the pSZ signal is small compared to the reconstruction noise). Taking the harmonic transform of Eq.~\eqref{eq:QandUrealspaceQEs}, we obtain the harmonic-space quadratic estimators $\hat{q}_{\ell m}^{E}$ and $\hat{q}_{\ell m}^{B}$ first derived in Ref.~\cite{Alizadeh_2012} and used in a wide variety of forecasts, e.g.~\cite{Deutsch:2017cja,Deutsch2018a,Pan:2019dax,Cayuso:2019hen,Deutsch:2018umo,Philcox:2022dht}. 
    \item {\bf Polarized SZ template correlated with CMB polarization}: We can instead decompose the bispectrum estimator into a pSZ template correlated with CMB polarization. We define a template for the pSZ contribution sourced by the optical depth correlated with the galaxy sample:
    \begin{eqnarray}
        \tilde{Q}_{\rm pSZ} (\hat{n}) = -\frac{\sqrt{6}}{10} \tilde{\tau}(\hat{n}) \hat{Q}_Z(\hat{n}), \ \ \  \tilde{U}_{\rm pSZ} (\hat{n}) = -\frac{\sqrt{6}}{10} \tilde{\tau}(\hat{n}) \hat{U}_Z(\hat{n}) \ .
    \end{eqnarray}
    In this case, Eq.~\eqref{eq:A_estimator} decomposes as
\begin{equation}\label{eq:pSZtemplateA}
\hat{A}_{Zg} = 
\sigma_{Zg}^2 \int d^2 \hat{n} \ \left[\tilde{Q}_{EB}(\hat{n}) \tilde{Q}_{\rm pSZ}(\hat{n}) + \tilde{U}_{EB}(\hat{n}) \tilde{U}_{\rm pSZ}(\hat{n}) \right]
\end{equation}
    with variance
    \begin{eqnarray}\label{eq:sigmaZg_QE_psz}
        \sigma_{Zg}^{-2} &=& \sum_{\ell'} \frac{(2\ell'+1)}{2} \left[ \frac{C_{\ell'}^{EE; \rm pSZ}}{C_{\ell'}^{EE}} + \frac{C_{\ell'}^{BB; \rm pSZ}}{C_{\ell'}^{BB}} \right]
    \end{eqnarray}
    where 
    \begin{eqnarray}
        C_{\ell'}^{EE;\rm pSZ} = C_{\ell'}^{BB;\rm pSZ} \simeq \frac{6}{100} \frac{(\bar{C}_{\ell'}^{\tau g})^2 }{C_{\ell'}^{gg}} \sum_\ell \frac{(2\ell+1)}{4\pi} \left[\frac{(C_{\ell}^{Zq_E})^2}{C_{\ell}^{ZZ}} + \frac{(C_{\ell}^{Zq_B})^2}{C_{\ell}^{ZZ}} \right]
    \end{eqnarray}
    is the power spectrum of the pSZ template. The bispectrum variance therefore describes the total signal-to-noise squared of the cross-correlation of the CMB polarization and the pSZ template. For a set of delta function sources, Eq.~\eqref{eq:pSZtemplateA} reduces to a weighted sum of the product of the pSZ template and CMB polarization. The resulting estimator has been used in previous literature to forecast the ability of measurements in the direction of clusters to constrain pSZ, e.g. Ref.~\cite{Hall:2014wna}. 
    \item {\bf Optical depth reconstruction correlated with an optical depth template}: Finally, we can instead split up the estimator into an optical depth template and an optical depth reconstruction. We define a quadratic estimator for the optical depth associated with the tracer as:
    \begin{eqnarray}\label{eq:tauestimator}
        \hat{\tau}_Z (\hat{n}) = - \frac{\sqrt{6}}{10} \left[N_{\ell'}^\tau \left( \tilde{Q}_{EB} \hat{Q}_{Z} + \tilde{U}_{EB} \hat{U}_{Z}\right)_{\rm HT} \right]_{\rm HT^{-1}} .
    \end{eqnarray}
    Here, $( \ )_{\rm HT}$ indicates a (spin-2) spherical harmonic transform while, $[ \ ]_{\rm HT^{-1}}$ indicates an inverse spherical harmonic transform, and $N_{\ell'}^\tau$ is defined as 
    \begin{eqnarray}
        (N_{\ell'}^\tau)^{-1} = \frac{6}{100} \frac{C_{\ell'}^{EE} + C_{\ell'}^{BB}}{C_{\ell'}^{EE} C_{\ell'}^{BB}} \sum_\ell \frac{(2\ell+1)}{4\pi} \left[\frac{(C_{\ell}^{Zq_E})^2}{C_{\ell}^{ZZ}} + \frac{(C_{\ell}^{Zq_B})^2}{C_{\ell}^{ZZ}} \right] .
    \end{eqnarray}
    In this case, Eq.~\eqref{eq:A_estimator} decomposes as
    \begin{equation}
\hat{A}_{Zg} =  
\sigma_{Zg}^2 \int d^2 \hat{n} \ \tilde{\tau} (\hat{n}) \hat{\tau}_Z (\hat{n})
\end{equation}
    with variance
    \begin{eqnarray}\label{eq:sigmaZg_tau}
        \sigma_{Zg}^{-2} &=& \sum_{\ell'} (2\ell'+1) \frac{(C_{\ell'}^{\tau g})^2}{C_{\ell'}^{gg} N_{\ell'}^\tau } 
    \end{eqnarray} 
    which is the signal-to-noise squared of the cross-correlation between the template and reconstruction of the optical depth. The estimator Eq.~\eqref{eq:tauestimator} is equivalent to the quadratic estimator introduced in Ref.~\cite{Dvorkin_2009} for patchy reionization if we simply extended the definition of the templates $\hat{Q}_Z, \ \hat{U}_Z$ to include the un-screened CMB polarization. A similar estimator has also been considered for screening in non-linear structure  e.g.~\cite{Roy:2022muv,Pirvu:2023lch,Mondino:2024rif}. 
\end{enumerate}

For the purposes of evaluating the bispectrum amplitude estimator Eq.~\eqref{eq:A_estimator}, a discussion of how to factorize terms in the integral is irrelevant. However, the three interpretations of the bispectrum estimator described above single out a number of intermediate data products. These include: the optical depth template $\tilde{\tau}$, the remote-quadrupole template $\{ \hat{Q}_Z, \hat{U}_Z \}$, the template for the pSZ signal $\{\tilde{Q}_{\rm pSZ}, \tilde{U}_{\rm pSZ} \}$, the quadratic estimator for the optical depth field $\hat{\tau}_Z$, and the quadratic estimator for the remote quadrupole field $\{\hat{Q}_q , \hat{U}_q \}$. One can include information in both the cross-spectra and the auto-spectra of these data products, or cross-correlate with other cosmological fields, to obtain more information about the optical depth, galaxy sample, or remote quadrupole field. These correlations summarize a hierarchy of pSZ n-point correlation functions. In the following, we focus on the reconstructed remote quadrupole field $\{\hat{Q}_q , \hat{U}_q \}$ as our intermediate data product of choice, computing the auto-spectra for various data combinations and cross-correlating with templates $\{ \hat{Q}_Z, \hat{U}_Z \}$ for the remote quadrupole field. We leave it to future work to study the properties of the pSZ templates and reconstructed optical depth field $\hat{\tau}_Z$.

\section{Estimators for the optical depth bias, optical depth to reionization, and tensor-to-scalar ratio}\label{sec:bqtaurestimators}

In Eq.~\eqref{eq:A_estimator}, we wrote a generic estimator for the bispectrum amplitude. This was derived assuming data on the full sky. It is unbiased and minimum variance over an ensemble that includes variations over both the signal and noise components of the large angular scale CMB ($Z$). To apply to data, we must generalize the basic bispectrum estimator to data on the masked sky in an ensemble where the signal component of $Z$ is held fixed. For scalar modes, the basic bispectrum amplitude estimator Eq.~\eqref{eq:A_estimator} is proportional to the product of $b_q$ and $\tau$; for tensor modes it is proportional to $r$. We therefore specialize the amplitude estimator to target several parameters of interest: the optical depth bias $b_q$, the optical depth to reionization $\tau_{\rm rei}$, the tensor-to-scalar ratio $r$. These cases provide a representative sample of the range of physics that can be probed with measurements of the pSZ effect.

\subsection{Optical depth bias}

The quadratic estimator for the remote quadrupole field Eq.~\eqref{eq:QandUrealspaceQEs} is unbiased only when the model relating the galaxy sample to the optical depth is accurate. This model enters the bispectrum Eq.~\eqref{eq:pSZbispectra_integrated} through $\bar{C}_\ell^{\tau g}$. If the true spectrum $[ \bar{C}_{\ell}^{\tau g} ]^{\rm t}$ differs from the model spectrum $\bar{C}_{\ell}^{\tau g}$, the mean of the quadratic estimator Eq.~\eqref{eq:QandUrealspaceQEs} is 
\begin{eqnarray}
    \langle \hat{Q}_q (\hat{n}) \rangle = b_q \bar{Q}_q (\hat{n}), \ \ \ \langle \hat{U}_q (\hat{n}) \rangle = b_q \bar{U}_q (\hat{n})
\end{eqnarray}
where 
\begin{eqnarray}\label{eq:bqdefinition}
b_q \equiv \frac{\sum_{\ell'} \frac{(2\ell'+1)}{4 \pi}  \frac{\left( C_{\ell'}^{BB}+C_{\ell'}^{EE}\right)\bar{C}_{\ell'}^{\tau g} [\bar{C}_{\ell'}^{\tau g}]^{\rm t} }{C_{\ell'}^{gg} C_{\ell'}^{EE} C_{\ell'}^{BB} }}{\sum_{\ell'} \frac{(2\ell'+1)}{4 \pi}  \frac{\left( C_{\ell'}^{BB}+C_{\ell'}^{EE}\right)(\bar{C}_{\ell'}^{\tau g})^2 }{C_{\ell'}^{gg} C_{\ell'}^{EE} C_{\ell'}^{BB} }} \ .
\end{eqnarray}
Propagating this to the bispectrum amplitude estimator Eq.~\eqref{eq:A_estimator} we obtain
\begin{equation}
\langle \hat{A}_{Zg} \rangle = b_q  \ .
\end{equation}
The bispectrum amplitude estimator is therefore an estimator for $b_q$. If our gas-halo model is correct, $b_q = 1$ and a significant measurement of $\hat{A}_{Zg}$ constitutes a detection of the pSZ effect. If the model does not match reality, a departure of $b_q$ from unity informs us about astrophysics and helps us to correct the model.

We extend the basic amplitude estimator by forming the unbiased minimum variance joint estimator for $b_q$ using the bispectrum for both $Z=T$ and $Z=E$. We assume that the galaxy survey and small-scale CMB polarization are measured in the presence of a mask $M(\hat{n})$. We also assume that we have access to a full-sky measurement of $T$ and $E$ on large angular scales, in an ensemble where the signal component of $T$ and $E$ are held fixed. Under these assumptions, the estimator for $b_q$ is:
\begin{eqnarray} \label{eq:bqest}
    \hat{b}_q = - \frac{\sqrt{6}}{10}
\sigma_{b_q}^2 \int d^2 \hat{n} \ M(\hat{n}) \tilde{\tau}(\hat{n}) \left[\tilde{Q}_{EB}(\hat{n}) \hat{Q}(\hat{n}) + \tilde{U}_{EB}(\hat{n}) \hat{U}(\hat{n}) \right].
\end{eqnarray}
Now the $\hat{Q}(\hat{n})$ and $\hat{U}(\hat{n})$ are defined as
\begin{eqnarray}\label{eq:QUtemps}
    \hat{Q}(\hat{n}) &\equiv& - \sum_{\ell m} \tilde{q}_{\ell m}^E X_{1;\ell m} \\
    \hat{U}(\hat{n}) &\equiv& - \sum_{\ell m} i \tilde{q}_{\ell m}^E X_{2;\ell m}
\end{eqnarray}
where $\tilde{q}_{\ell m}^E$, given in Eq.~\eqref{eq:qE_template}, is the maximum-likelihood template for the $E$-mode remote quadrupole given observations of both $T$ and $E$ (denoted $T_0$, $E_0$). Note that since we assume the template is measured on the full sky, $\hat{Q}(\hat{n})$ and $\hat{U}(\hat{n})$ remain pure $E$-mode.

In an ensemble where only the noise component of $T$ and $E$ are varied, the estimator variance on the cut sky is
\begin{eqnarray}\label{eq:bvvar}
    \sigma_{b_q}^{-2} \equiv \sum_\ell (2\ell+1) \frac{\mathbf{C}^{Zq_E}_\ell \cdot (\mathbf{C}^{ZZ}_\ell)^{-1} (\mathbf{C}^{ZZ}_\ell)_{0, \rm cut} (\mathbf{C}^{ZZ}_\ell)^{-1} \cdot \mathbf{C}^{Zq_E}_\ell}{N_\ell^{Eq_E}}
\end{eqnarray}
where $\mathbf{C}^{Zq_E}_\ell = \{ C^{Tq_E}_\ell, C^{Eq_E}_\ell\}$, $\mathbf{C}^{ZZ}_\ell = \{ \{ C^{TT}_\ell, C^{TE}_\ell \}, \{ C^{TE}_\ell, C^{EE}_\ell \} \}$, and $(\mathbf{C}^{ZZ}_\ell)_{0, \rm cut}$ is filled with spectra on the cut sky where the signal is fixed, and the noise is given by an ensemble average. This accounts for the fact that the $T$, $E$ signal is fixed to our observed realization. In practice, we cannot distinguish the signal and noise components of $T$ and $E$, and therefore approximate $(\mathbf{C}^{ZZ}_\ell)_{0, \rm cut}$ by the measured cut-sky spectra. Under these assumptions, the ensemble-average of the estimator is $b_q$:
\begin{eqnarray}
    \langle \hat{b}_q \rangle = b_q \ .
\end{eqnarray}
For an accurate model of the galaxy-optical depth cross-correlation, this will be nearly equal to unity. 

The quantity $b_q$ is closely related to the optical depth bias introduced in the context of the kSZ effect; we will denote this as $b_v$, and for a single photometric galaxy bin it is defined as 
\begin{eqnarray}\label{eq:bvdefinition}
    b_v \equiv \frac{\sum_{\ell'} \frac{(2\ell'+1)}{4 \pi}  \frac{\bar{C}_{\ell'}^{\tau g} [\bar{C}_{\ell'}^{\tau g}]^{\rm t} }{C_{\ell'}^{gg} C_{\ell'}^{TT} }}{\sum_{\ell'} \frac{(2\ell'+1)}{4 \pi}  \frac{(\bar{C}_{\ell'}^{\tau g})^2 }{C_{\ell'}^{gg} C_{\ell'}^{TT} }} \ .
\end{eqnarray}
The kSZ-galaxy-galaxy bispectrum can be used to write an estimator for $b_v$: 
\begin{eqnarray}
    \hat{b}_v = - 
\sigma_{b_v}^2 \int d^2 \hat{n} \ M(\hat{n})\tilde{\tau}(\hat{n}) \tilde{v}_g (\hat{n}) \tilde{T}(\hat{n})
\end{eqnarray}
where $\tilde{\tau}$ is defined in Eq.~\eqref{eq:tautildedef}, $\tilde{v}_g$ is a template for the remote dipole field constructed from e.g. galaxy density, and
\begin{eqnarray}
    \tilde{T}(\hat{n}) \equiv \sum_{\ell'' m''} \frac{T_{\ell'' m''}}{C_{\ell''}^{TT}} Y_{\ell'' m''}(\hat{n}) \ .
\end{eqnarray}
The kSZ optical depth bias was first constrained with \textit{Planck} and unWISE data~\cite{Bloch_2024}, and subsequently measured at high significance using ACT temperature maps in cross-correlation with SDSS~\cite{Lague:2024czc} and DESI galaxies~\cite{McCarthy:2024nik,McCarthy:2025brx,Lai:2025qdw,Hotinli:2025tul,Chaussidon:2026vmn}. For instance, Ref.~\cite{Hotinli:2025tul} used DESI Legacy Imaging Surveys with ACT DR5 and found $b_v=0.45_{-0.05}^{+0.06}$ (11.6$\sigma$ velocity-galaxy cross-correlation detection). Whereas Ref.~\cite{Chaussidon:2026vmn} used DESI DR2 with ACT DR6 and found for the luminous red galaxies $b_v=0.221_{-0.0217}^{+0.0216}$, emission line galaxies $b_v=0.115_{-0.0254}^{+0.0255}$, and quasars $b_v=0.227_{-0.0529}^{+0.0528}$  (21$\sigma$ cumulative detection). The optical depth bias values depend on the sample, but overall these analyses indicate that the gas is more spread out than expected from simulations.  

Comparing Eq~\eqref{eq:bqdefinition} to Eq.~\eqref{eq:bvdefinition}, note that although both $b_v$ and $b_q$ depend on the same cross-spectra $\bar{C}_{\ell}^{\tau g}$, they are not equal. In addition, for the same fiducial and true $\bar{C}_{\ell}^{\tau g}$, different galaxy surveys and CMB experiments will yield different numerical values for $b_v$ and $b_q$ since the noise properties will be different. However, it is possible to use $b_v$ and $b_q$ for a variety of data combinations to jointly constrain a set of model parameters underlying $\bar{C}_{\ell}^{\tau g}$. As discussed in greater detail in Sec.~\ref{sec:snr}, kSZ is more sensitive to the gas distribution at halo-scales than pSZ.

\subsection{Optical depth to reionization}

The pSZ bispectra $B_{\ell \ell' \ell''}^{EgE}$ and $B_{\ell \ell' \ell''}^{EgB}$ are proportional to the optical depth to reionization $\tau_{\rm rei}$ at low-$\ell$. This is because the bispectrum contains a factor of $C_{\ell}^{E q_E}$, which isolates the contribution from reionization $E$-modes. If the estimator weights are produced assuming a fiducial value $\tau_{\rm rei}$ and the true value is $[\tau_{\rm rei}]^t$, then an estimator for the optical depth to reionization is obtained from Eq.~\eqref{eq:bqest} by restricting the template to be based on the $E$-modes and re-scaling:
\begin{eqnarray}\label{eq:tauest}
    \hat{\tau}_{\rm rei} = - \frac{\sqrt{6}}{10} 
\sigma_{\tau_{\rm rei}}^2 \int d^2 \hat{n} \ M(\hat{n}) \tilde{\tau}(\hat{n}) \left[\tilde{Q}_{EB}(\hat{n}) \frac{\hat{Q}_E(\hat{n})}{\tau_{\rm rei}} + \tilde{U}_{EB}(\hat{n}) \frac{\hat{U}_E(\hat{n})}{\tau_{\rm rei}} \right]
\end{eqnarray}
where 
\begin{eqnarray}
    \hat{Q}_E(\hat{n}) &\equiv& - \sum_{\ell m} \tilde{q}_{\ell m}^{E} X_{1;\ell m} \\
    \hat{U}_E(\hat{n}) &\equiv& - \sum_{\ell m} i \tilde{q}_{\ell m}^{E} X_{2;\ell m} \\
    \tilde{q}^{E}_{\ell m} (\chi) 
    &=& \frac{C_{\ell}^{E q_E}  E_{\ell m} }{C_{\ell}^{EE}} \ .\label{eq:tautemp}
\end{eqnarray}
Here, $\tilde{q}_{\ell m}^{E}$ is the maximum-likelihood template for the $E$-mode remote quadrupole using only the observation of $E$. We have also defined
\begin{eqnarray}
    \sigma_{\tau_{\rm rei}}^2 &=& \tau_{\rm rei}^2 \sigma^2_{Eg} \\
    \sigma^{-2}_{Eg} &\equiv&  \sum_\ell (2\ell+1) \frac{(C_\ell^{Eq_E})^2 (C_\ell^{EE})_{0, \rm cut}}{(C_\ell^{EE})^2 N_\ell^{Eq_E}} \ .
\end{eqnarray}
In general, the estimator mean is
\begin{eqnarray}
    \langle \hat{\tau}_{\rm rei} \rangle = b_q [\tau_{\rm rei}]^t \ .
\end{eqnarray}
The measured optical depth is unbiased only when $b_q=1$. This can in principle be accomplished by using the kSZ effect to measure $b_v$, and translating the associated constraints on $\bar{C}_{\ell}^{\tau g}$ to fix $b_q$; here we will assume an accurate model and set $b_q=1$. 

\subsection{Tensor-to-scalar ratio}

The pSZ bispectra $B_{\ell \ell' \ell''}^{BEq}$ and $B_{\ell \ell' \ell''}^{BBq}$ are proportional to the amplitude of primordial tensor modes, which we parameterize through the tensor-to-scalar ratio $r$. An estimator for $r$ is:
\begin{eqnarray}\label{eq:rest}
    \hat{r} = - \frac{\sqrt{6}}{10}
\sigma_{r}^2 \int d^2 \hat{n} \ M(\hat{n}) \tilde{\tau}(\hat{n}) \left[\tilde{Q}_{EB}(\hat{n}) \hat{Q}_B(\hat{n}) + \tilde{U}_{EB}(\hat{n}) \hat{U}_B(\hat{n}) \right]
\end{eqnarray}
where 
\begin{eqnarray}
    \sigma_{r}^2 &=& \sigma_{Bg}^{2}(r=1) \\
    \sigma^{-2}_{Bg} &\equiv&  \sum_\ell (2\ell+1) \frac{(C_\ell^{Bq_B}(r=1))^2 (C_\ell^{BB})_{0, \rm cut}}{(C_\ell^{BB})^2 N_\ell^{Bq_B}} 
\end{eqnarray}
and $\{ \hat{Q}_B (\hat{n}),\hat{U}_B (\hat{n}) \}$ are signal templates. If there was a measurement of the low-$\ell$ $B$-mode signal from primordial tensors, we could construct
\begin{eqnarray}\label{eq:QUBmodetemplate}
    \hat{Q}_B (\hat{n}) = - i \sum_{\ell m}  \frac{C_{\ell}^{Bq_B}(r=1) }{C_{\ell}^{BB}} B_{\ell m} X_{2;\ell m} (\hat{n}), \ \ \ \hat{U}_{B}(\hat{n}) = - \sum_{\ell m} \frac{C_{\ell}^{Bq_B}(r=1)}{C_{\ell}^{BB}} B_{\ell m} X_{1;\ell m} (\hat{n}) \ .
\end{eqnarray}
from observed $B_{\ell m}$. However, this is not the case. In the absence of an empirical template we instead evaluate Eq.~\eqref{eq:QUBmodetemplate} for an ensemble of fixed templates $B_{\ell m}$ spanning a basis of possible $B$-mode quadrupoles and compute the estimator Eq.~\eqref{eq:rest} for each. We define an orthonormal basis for quadrupole orientations by expanding $\sum_m a_{2m}Y_{2m}(\hat{n}) = Q_{ij}n_in_j$ in Cartesian components and matching 
coefficients to obtain the following:             
  \begin{eqnarray}
      Q_{zz} \propto a_{2,0},  \quad Q_{xz} \propto \mathrm{Re}(a_{2,1}), \quad Q_{yz} \propto \mathrm{Im}(a_{2,1}), \quad 
  Q_{xx}-Q_{yy} \propto \mathrm{Re}(a_{2,2}), \quad Q_{xy} \propto \mathrm{Im}(a_{2,2}) \ .   
  \end{eqnarray}
    We can then define five orthonormal basis states $e^{(\alpha)}$, each with unit norm $\sum_m|a_{2m}^{B,(\alpha)}|^2=1$:                   
  \begin{eqnarray}\label{eq:tensorbasis}
      e^{(0)}&:& a^B_{2,0} = 1,\,      
      e^{(1)}: a^B_{2,\pm 1} = \pm\frac{1}{\sqrt{2}},\, e^{(2)}: a^B_{2,\pm 1} = \frac{i}{\sqrt{2}} \nonumber \\   
      e^{(3)}&:& a^B_{2,\pm 2} = \frac{1}{\sqrt{2}},\, e^{(4)}: a^B_{2,2} = \frac{i}{\sqrt{2}},\ a^B_{2,-2} = -\frac{i}{\sqrt{2}}
  \end{eqnarray}                 
  corresponding to the $Q_{zz}$, $Q_{xz}$, $Q_{yz}$, $Q_{xx}-Q_{yy}$, and $Q_{xy}$ orientations of the quadrupole tensor respectively.  
  
  For each basis state $e^{(\alpha)}$, we substitute $B_{\ell m}=\sqrt{5C_\ell^{BB}} a_{2m}^{B,(\alpha)}\delta_{\ell,2}$ into       Eq.~\eqref{eq:QUBmodetemplate} to obtain templates $\hat{Q}_B^{(\alpha)}(\hat{n})$ and $\hat{U}_B^{(\alpha)}(\hat{n})$, applying   
  the Wiener filter $C_\ell^{Bq_B}(r=1)/C_\ell^{BB}$ to the basis $a_{\ell m}$s before creating the polarization maps. We evaluate the estimator Eq.~\eqref{eq:rest} for each template direction, obtaining five constraints on $r$.

\section{Analysis pipeline}\label{sec:analysis_pipeline}

In this section, we describe the datasets we use and the details of the analysis pipeline. For pSZ quadrupole reconstruction, we need CMB polarization maps cross-correlated with a tracer field of the optical depth, using only the small scales of both observables. Primarily we use ACT, because it is high-resolution and low-noise. We also use \textit{Planck} because it has a larger number of frequency channels, allowing us to better assess foreground contamination. For both surveys, we use component-separated maps for foreground mitigation. For the tracer field, we consider both the unWISE galaxy survey and the CIB. Both of these tracers have a large number density of objects, large sky overlap with our CMB survey, and coverage at $z\sim1-2$ where the remote quadrupole peaks. Finally, for the bispectrum we need CMB temperature and polarization maps on large-scales to make a template of the remote quadrupole field. We use \textit{Planck} for this because it has the largest sky coverage.

\subsection{\textit{Planck} and ACT data}

In this work, we use \textit{Planck} Data Release 3 (PR3) temperature and polarization maps~\footnote{\textit{Planck} data products can be found \href{here}{https://pla.esac.esa.int/\#home}.} \cite{PlanckCollaboration2020}, including both individual frequency maps and component-separated maps. The individual frequency maps we consider are 100, 143, 217, and 353 GHz, which are debeamed with a Gaussian beam of FWHM 9.68, 7.30, 5.02, and 4.94 arcminutes, respectively. \textit{Planck} polarization maps are dominated by galactic foregrounds (synchrotron at low frequencies and thermal dust at high frequencies) on scales $\ell \lesssim 10^2$, primary CMB $E$-modes on intermediate scales $10^2 \lesssim \ell \lesssim 10^3$ (at 100, 143, and 217 GHz), and instrumental noise on small scales $\ell \gtrsim 10^3$ (the effective polarization white-noise levels are $83.4$, $62.2$, $87.7$, $129$ $\mu$K-arcmin, respectively). As described in more detail below, we high-pass filter to remove scales strongly dominated by thermal dust. We mask the CMB maps using the HFI polarization point source mask and an apodized galactic cut retaining 50\% of the sky  \cite{Planck:2018yye}.

We also analyze \textit{Planck} component-separated maps. The \textit{Planck} collaboration provides four such products - \texttt{Commander}~\cite{Eriksen_2008}, \texttt{SMICA}~\cite{Planck:2018yye}, \texttt{SEVEM}~\cite{Leach:2008fi,2012MNRAS.420.2162F}, and \texttt{NILC}~\cite{BasakDelabrouille2012} - each applying a different foreground separation methodology. \texttt{SMICA} (Spectral Matching Independent Component Analysis) linearly weights frequency maps in harmonic space to minimize the blackbody spectrum's variance; we use this product for small angular scale polarization. The \texttt{SMICA} maps have less power than the individual frequency maps on large-scales due to the removal of galactic foregrounds, and less power on small-scales due to the reduction in the effective noise (corresponding to a white noise of roughly $52$ $\mu$K-arcmin).  We mask \texttt{SMICA} using the \texttt{SMICA}-based confidence mask which retains 78\% of the sky \cite{Planck:2018yye}. For the large-scale temperature and polarization template for the remote quadrupole field (Eq.~(\ref{eq:qE_template})), our primary choice is \texttt{Commander}, a Bayesian method that fits a parametric model - including instrumental noise, astrophysical foregrounds, and the CMB - to the multi-frequency data. We additionally construct the remote quadrupole template from the \texttt{SMICA}, \texttt{SEVEM}, and \texttt{NILC} maps to validate our results; a comparison of all four methods is presented in Appendix~\ref{app:compsep}. All maps are provided at HEALPix resolution $N_{\rm side}=2048$, smoothed with an effective 5 arcminute FWHM Gaussian beam which we deconvolve. 

When using \texttt{Commander} to create a remote quadrupole field template using Eq.~(\ref{eq:qE_template}) we need to model the noise power spectra. We do this by calculating the power spectra for 300 \texttt{Commander} noise (instrumental noise + residual systematics) simulations and taking the mean as our fiducial model \cite{Planck:2018yye}. In Appendix~\ref{app:compsep}, we do the same for the other three component-separated methods, using their respective noise simulations. The component-separated maps used for the quadrupole template are the inpainted versions provided by \textit{Planck}, in which the full map is Wiener filtered and the galactic masked region is filled using Gaussian constrained realizations~\cite{Planck:2018yye}. The Wiener filtering suppresses power at high $\ell$ where instrumental noise dominates, so the masked inpainted and non-inpainted maps differ on small angular scales. However, at $\ell \leq 20$ the Wiener filter weight is $\approx 1$, so both maps give consistent power spectra on these scales; we verify this explicitly by comparing the masked power spectra of the original and inpainted Commander maps. The maps on these scales are simply constrained realizations; therefore while Ref.~\cite{Planck:2018yye} recommends caution when using the inpainted maps for analysis, they provide a convenient mechanism to deal with $E$/$B$ mixing caused by masking on large angular scales. We further assess the error due to inpainting in Appendix~\ref{app:compsep}.

We use ACT Data Release 6 (DR6) maps, which cover $\sim 45\%$ of the sky~\footnote{ACT data products can be found \href{https://lambda.gsfc.nasa.gov/product/act/act_dr6.02/}{here}.}. Specifically, we use the ACT+\textit{Planck} co-adds in the bands 77-112 GHz (`90 GHz'), 124-172 GHz (`150 GHz'), and 182-277 GHz (`220 GHz') with the point source signal removed (`srcfree') \cite{ACT:2025xdm}. 
We rotate to galactic coordinates and degrade to $N_{\rm side}=2048$ - to match the resolution of our galaxy and CIB tracer maps. The degrading is done in harmonic space, by removing $\ell>3N_{\rm side}-1$. We debeam using a FWHM of 5 arcminutes instead of the actual ACT beams. This is done to mitigate the impact of small scale systematics on the estimator by effectively filtering out high $\ell$ modes. Additionally, we use the ACT DR6+\textit{Planck} needlet internal linear combination (NILC) blackbody temperature and polarization maps\footnote{The $T$ and $E$ maps are available at \url{https://lambda.gsfc.nasa.gov/product/act/act_dr6.02/act_dr6.02_nilc_info.html}, we thank W. Coulton for providing the corresponding $B$ map.} made using the method of Ref.~\cite{ACT:2023wcq}. We similarly rotate and degrade them, and debeam using FWHM of 5 arcminutes.
 To create an ACT footprint mask, we take the full resolution ACT maps, make a binary mask based off the non-zero pixels, then rotate and degrade the mask. The mask is then reconverted to binary values by setting values lower (higher) than 0.5 to 0 (1). We finally increase the mask by $2^\circ$ to remove any residual galactic contamination and bright sources near the boundary; the resulting $f_{\rm sky}$ is 49\%.

\subsection{unWISE data}

The unWISE galaxy catalogue \cite{Schlafly_2019} has a redshift range $0\leq z \leq 2$ and contains more than 500 million galaxies which come from the NEOWISE dataset \cite{2010AJ....140.1868W,Mainzer2014}. The specific galaxy catalogue we use is described in Ref.~\cite{Krolewski:2019yrv}, which has three samples termed `blue', `green', `red' in order of increasing median redshift. For this work, we use the blue and green bins, which have the largest number density and the best photometric redshift characterization. We exclude the red sample because it has the smallest number density and has been shown to have larger systematics \cite{Krolewski:2021yqy}.

We form the galaxy density map from the number counts $N^g(\hat{n})$ per pixel at $N_{\rm side}=2048$ following Eq.~(\ref{eq:galdens}). The unWISE blue sample has $\Bar{N}^g=2.8$ or a number density of $0.95$ $\mathrm{arcmin}^{-2}$, and the green sample has $\Bar{N}^g=1.5$ or a number density of $0.51$ $\mathrm{arcmin}^{-2}$. We use the mask of Ref.~\cite{Krolewski2020} which uses a galactic plane cut that keeps 70\% of the sky, as well as masking stars, planetary nebulae, and bright sources, for an overall sky coverage of 59\%. 

\subsection{CIB data}

The \textit{Planck} 2015 data release included CIB maps \cite{Planck:2016frx} created from the generalized needlet internal linear combination (GNILC) component separation method \cite{2011MNRAS.418..467R}. GNILC decomposes multi-frequency maps into a needlet basis and applies frequency-dependent weights at each needlet scale to disentangle the CIB anisotropies from Galactic dust emission. We use the 353, 545, and 857 GHz CIB maps which cover 63\% of the sky. 

We use the bias-weighted CIB redshift distributions from Ref.~\cite{Chiang:2025ujk} obtained using clustering redshifts as developed in Refs.~\cite{menard_2013_tomog, chiang_2019_tomog}. This technique uses $\sim 3\times 10^6$ spectroscopic objects from the Sloan Digital Sky Survey I-IV to calculate the tracer bias-scaled, intensity-weighted redshift distribution of the sample ($b_I\, dI/dz$). They consider a variety of CIB maps from \textit{Planck} PR3, \textit{Herschel}, and \textit{IRAS} data. We use the best fit $b_I\, dI/dz$ distributions of Ref.~\cite{Chiang:2025ujk} for the CIB in the \textit{Planck} 353, 545, and 857 GHz bands.

\begin{figure}
    \centering
    \includegraphics[width=0.8\linewidth]{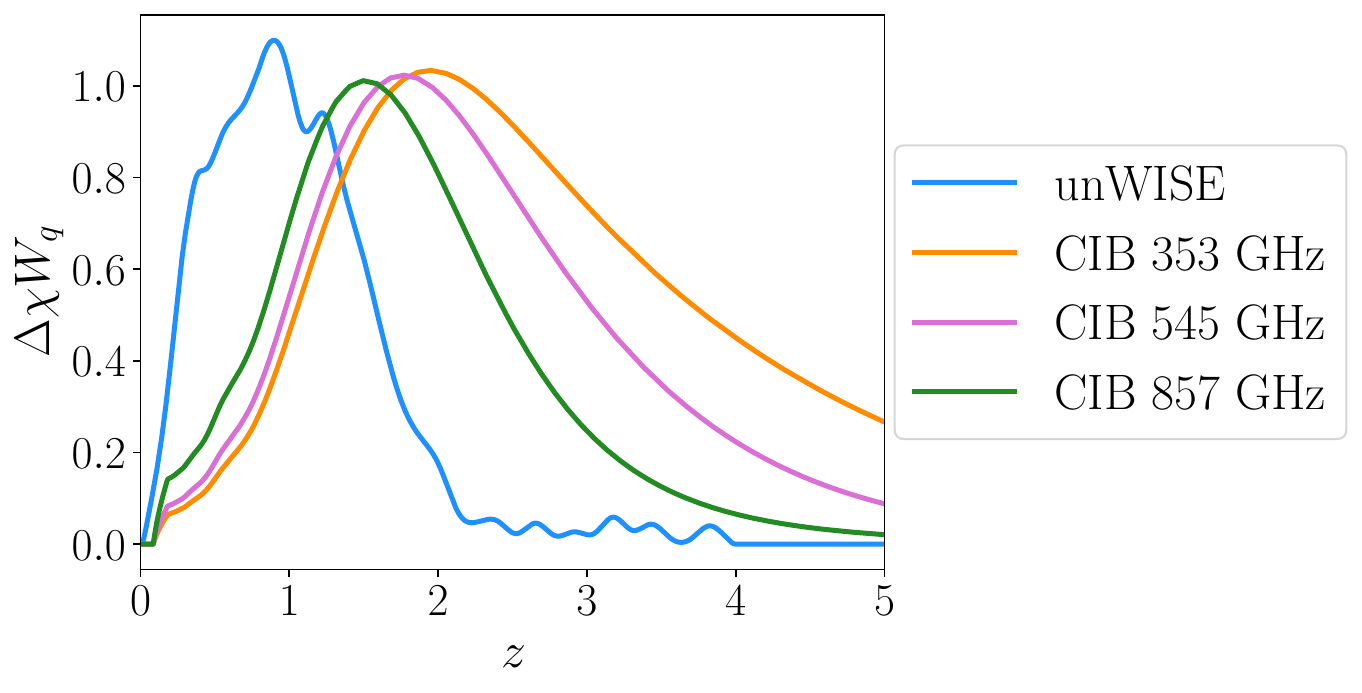}
    \caption{The unWISE (`blue' + `green' bins) and CIB quadrupole window functions. }
    \label{fig:wq}
\end{figure}

In figure~(\ref{fig:wq}) are the window functions $W_q$ (see Eq.~\eqref{eq:Wqdef}) for unWISE and the three CIB channels. This shows that unWISE peaks at $z\sim 1$ compared to the CIB which peaks at $z\sim 1.5-2$ and extends to higher redshifts. Comparing to Fig.~\ref{fig:consrealsl2to4}, these tracers, and in particular the CIB, captures most of the peak in the remote quadrupole signal.

\subsection{Expected signal to noise}\label{sec:snr}

Before describing our analysis pipeline in detail, we forecast the detectability of the remote quadrupole signal for various data combinations. We keep the same tracer fields, unWISE and \textit{Planck} CIB, but forecast for \textit{Planck}, ACT, Simons Observatory (SO) \cite{Ade2019} and LiteBIRD \cite{LiteBIRD:2022cnt}. SO is a ground-based CMB survey, the large-aperture telescope (LAT) will map roughly 40\% of the sky with $\sim11$ $\mu$K-arcmin polarization sensitivity, whereas the small aperture telescope (SAT) will cover 10\% with $\sim4$ $\mu$K-arcmin sensitivity - we forecast separately for both configurations. LiteBIRD is a satellite mission designed to achieve 2 $\mu$K-arcmin polarization sensitivity. As an overall measure of sensitivity for these future surveys, we forecast constraints on the optical depth bias $b_q$ and the optical depth to reionization $\tau_{\rm rei}$. It will take a more futuristic survey to place competitive constraints on the tensor-to-scalar ratio $r$, see Refs.~\cite{Alizadeh_2012,Deutsch:2017cja} for forecasts.

For the pSZ reconstruction auto-spectrum, the signal is $C_\ell^{\hat{q}_E\hat{q}_E} = p^2 C_\ell^{q_Eq_E,0} + N^q$ where $p$ denotes either $b_q$ or $\tau_{\rm rei}$ (forecasted separately), $N^q$ is defined in Eq.~\eqref{eq:Nq}, and $C_{\ell}^{q_Eq_E,0}$ is the true realization signal spectra. We model $C_\ell^{q_Eq_E,0}$ by conditioning on \texttt{Commander}: for $p=b_q$ this is the power spectrum of Eq.~(\ref{eq:qE_template}) and for $p=\tau_{\rm rei}$ it is the power spectrum of Eq.~(\ref{eq:tautemp}). The $p^2$ scaling of the auto-spectrum gives $\partial C_\ell^{\hat{q}\hat{q}}/\partial p = 2p C_\ell^{qq,0}$ and the Fisher matrix is
\begin{equation}
      F^{\rm auto}_{p} = \sum_\ell (2\ell+1)\,
      \left[2\left(\frac{C_\ell^{q_Eq_E,0,\rm masked}}{N^q}\right)^2
      + 2\left(\frac{C_\ell^{q_Bq_B,0,\rm masked}}{N^q}\right)^2\right],
      \label{eq:Fisher_auto}
\end{equation}
where $C_\ell^{q_Eq_E,0,\rm masked}$ and $C_\ell^{q_Bq_B,0,\rm masked}$ are the masked unit-amplitude ($p{=}1$) signal spectra. There is no intrinsic $C_\ell^{q_Bq_B,0}$ signal, the non-zero $C_\ell^{q_Bq_B,0,\rm masked}$ term arises entirely from mask-induced $E$-to-$B$ leakage. 
 The resulting constraint is
\begin{equation}
      \sigma_{p}^{\rm auto} = \frac{1}{\sqrt{F^{\rm auto}_{p}}}.
      \label{eq:sigma_bq_auto}
\end{equation}

We also use the local CMB as a noiseless template of $C_\ell^{qq,0}$ - as in the bispectrum approach of Eq.~(\ref{eq:bqest})). The cross-spectrum between the reconstruction $\hat{q}$ and this noiseless template scales linearly in $p$ (rather than as $p^2$). Combining this cross-spectrum term with the auto-spectrum gives the total
\begin{equation}
  F^{\rm all}_{p} = \sum_\ell (2\ell+1)\, \left[\frac{C_\ell^{q_Eq_E,0,\rm masked}}{N^q} + \frac{C_\ell^{q_Bq_B,0,\rm masked}}{N^q}\right]
  \label{eq:Fisher_all}
\end{equation}
and the error on $p$ is
\begin{equation}
      \sigma_{p}^{\rm all} = \frac{1}{\sqrt{F^{\rm all}_{p}}}\,.
      \label{eq:sigma_bq_all}
\end{equation}

To create $C_\ell^{q_Eq_E,0,\rm masked}$, $C_\ell^{q_Bq_B,0,\rm masked}$, we use the same masks for \textit{Planck} and ACT as we use in the data reconstruction (described in Sec.~\ref{sec:recpipeline}). For SO we use the same mask as for ACT, and for LiteBIRD the same mask as \textit{Planck}. We apply masks because simply rescaling by $f_{\rm sky}$ is insufficient at these angular scales: unlike at high $\ell$ where many modes contribute per multipole, at $\ell \sim 2$-10 the mask geometry directly determines which orientations of the quadrupole are accessible, and different mask shapes at fixed $f_{\rm sky}$ can yield different constraining power. We calculate $N^q$ using CMB theory spectra modeled as $C_\ell^{EE,\rm tot} = C_\ell^{EE} + N_\ell^{EE}$ (and likewise for $BB$), where $C_\ell^{EE,BB}$ are the $\Lambda$CDM signal spectra. The noise power spectrum models are described in Appendix~\ref{app:noisemodel}.

In table~(\ref{tab:SNR_forecast}) are the expected levels of reconstruction noise for different data combinations. The reconstruction noise decreases substantially from current data to next-generation experiments, ranging from $N^q = 1.82 \times 10^{-8}$ for \textit{Planck} 100 GHz $\times$ unWISE down to $N^q = 6.34 \times 10^{-11}$ for SO LAT 93 GHz $\times$ CIB 353 GHz, a reduction of more than two orders of magnitude. The CIB 353 GHz tracer consistently outperforms unWISE for reconstruction noise across all CMB maps. Lower reconstruction noise could potentially be achieved by co-adding data sets, for example SO LAT and LiteBIRD. We defer a detailed assessment of this scenario to future work that incorporates detailed modeling of foreground mitigation. 

\begin{figure}
    \centering
    \includegraphics[width=0.9\linewidth]{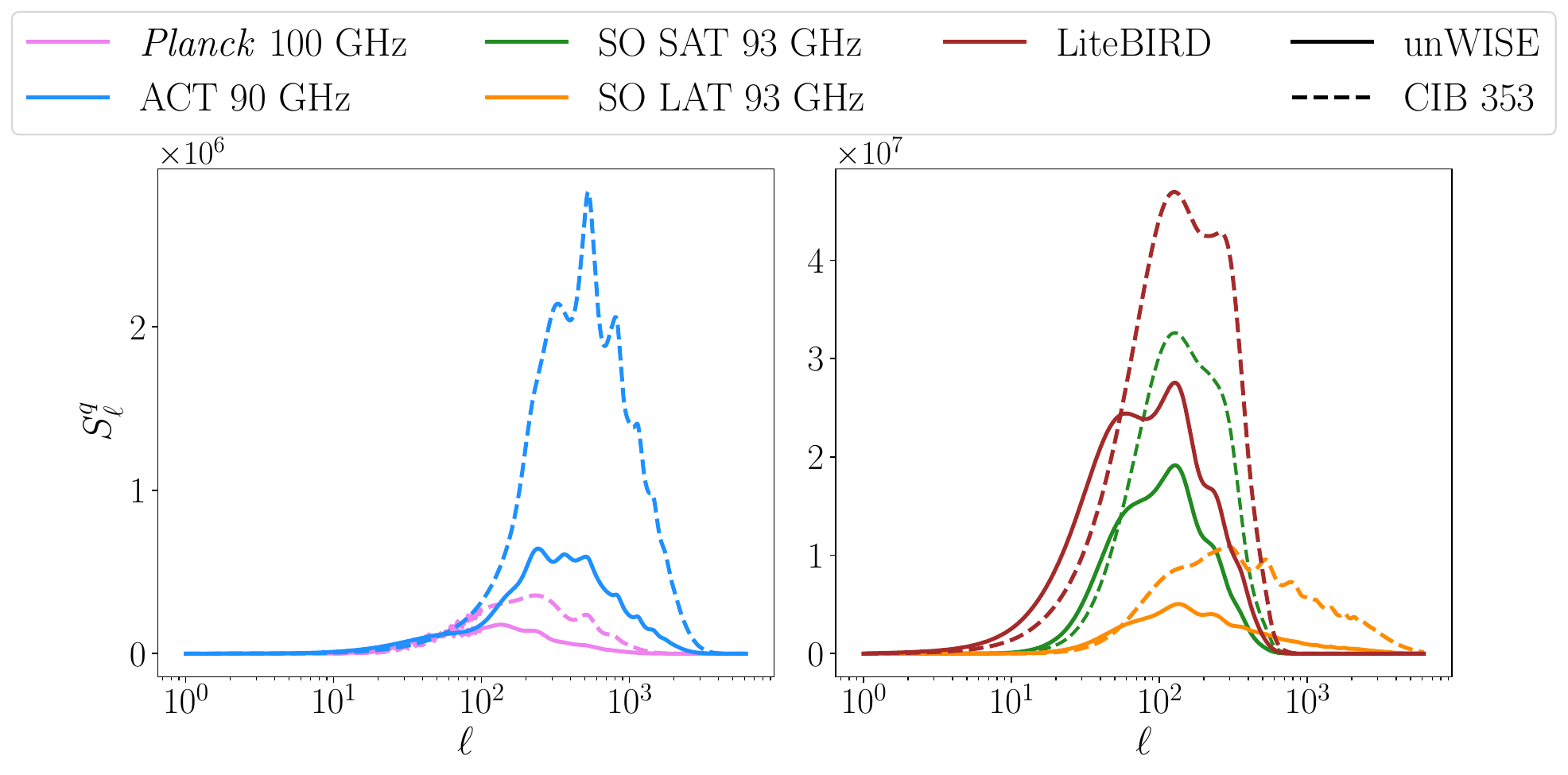}
    \caption[The reconstruction noise summand.]{The reconstruction noise summand, $S^q_\ell$, for current CMB datasets (left): \textit{Planck} 100 GHz (light purple) and ACT 90 GHz (blue). On the right are future datasets: SO SAT 93 GHz (green), SO LAT 93 GHz (orange), LiteBIRD (brown). The solid lines use unWISE as the tracer and the dashed lines use CIB 353 GHz. The summand is inversely related to the reconstruction noise, so the higher the curve the lower the noise. Additionally, this figure illustrates that the peak sensitivity is around $\ell \sim \mathcal{O}(100)$. }
    \label{fig:summandq}
\end{figure}

In figure~(\ref{fig:summandq}) we also plot the summand ($S^q_\ell$) of the forecasted reconstruction noise ($N^q=(\sum S^q_\ell)^{-1}$):
\begin{eqnarray}
    S^q_\ell = \frac{3}{100} \frac{(2\ell+1)}{4 \pi}  \frac{\left( C_{\ell}^{BB}+C_{\ell}^{EE}\right)(\bar{C}_{\ell}^{\tau g})^2 }{C_{\ell}^{gg} C_{\ell}^{EE} C_{\ell}^{BB} } \ .
\end{eqnarray} 
The shape of the summand illustrates which CMB multipoles contribute most to the estimator variance. On the left, comparing \textit{Planck} and ACT shows that ACT achieves lower reconstruction noise, and that CIB 353 GHz provides lower noise than unWISE across both instruments. The peak of the summand shifts from $\ell \sim 100$ for \textit{Planck} $\times$ unWISE to $\ell \sim 300$ for ACT $\times$ unWISE, reflecting ACT's higher angular resolution. Using CIB 353 GHz instead of unWISE shifts the peak to even smaller angular scales for both instruments, with ACT benefiting more strongly since its beam does not suppress these modes. In the right panel, SO LAT shows a similar sensitivity to ACT, whereas SO SAT and LiteBIRD peak at larger angular scales around $\ell \sim 100$. 

These results show that the constraining power is concentrated at $\ell \sim \mathcal{O}(100)$, in the two-halo regime of the galaxy–gas cross-correlation. Unlike kSZ velocity reconstruction, which is sensitive to one-halo scales and therefore to the profile of gas within individual halos, this demonstrates that pSZ quadrupole reconstruction with these datasets carries little information about gas profiles. The benefit of this insensitivity to baryonic feedback is that the pSZ signal is not suppressed as much as kSZ.

The forecasted constraining power on $b_q$ and $\tau_{\rm rei}$ for auto- and all-(auto \& cross) spectra is listed in table~(\ref{tab:SNR_forecast}). For \textit{Planck} and ACT, neither tracer yields $\sigma_{b_q} < 1$, reflecting the noise-dominated nature of current data. Looking ahead, LiteBIRD $\times$ CIB 353 GHz achieves $\sigma_{b_q}^{\rm all} = 0.769$, making this the first combination projected to detect a pSZ remote quadrupole signal at greater than $1\sigma$.

The \textit{Planck} fiducial value $\tau_{\rm rei} = 0.054$ provides a natural benchmark: experiments achieving $\sigma_{\tau_{\rm rei}} \lesssim \tau_{\rm rei}$ would be sensitive enough to detect a nonzero $\tau_{\rm rei}$ at $\gtrsim 1\sigma$ and begin to provide meaningful constraints. With current data, neither \textit{Planck} nor ACT approach this threshold. The most promising combination is LiteBIRD $\times$ CIB 353 GHz ($\sigma_{\tau_{\rm rei}} = 0.0405$), which achieves $\sigma_{\tau_{\rm rei}} < \tau_{\rm rei}^{\rm \textit{Planck}}$. 

The optical depth to reionization is the worst constrained parameter in $\Lambda$CDM. Precisely measuring $\tau_{\rm rei}$ is important for cosmology as its value is degenerate with proposed new physics, such as time-evolving dark energy or a neutrino mass that conflicts with current particle physics experimental bounds \cite{Sailer:2025lxj,Sullivan:2026tas}. It is difficult to constrain $\tau_{\rm rei}$ because it is a faint signal on very large scales in CMB polarization, which are cosmic variance limited and subject to galactic foreground contamination. The power of instead using pSZ to constrain $\tau_{\rm rei}$ is that it uses small scale observations, where cosmic variance is negligible. The relevant foreground contamination is also of a different nature: rather than diffuse galactic emission at the largest scales, the dominant concerns are extragalactic polarized sources. Together, these features mean that pSZ avoids the two main obstacles to constraining $\tau_{\rm rei}$ from the primary CMB polarization: cosmic variance on large angular scales and galactic foreground contamination. Therefore, these forecasts are promising that in the near-term pSZ will be able to provide a cross-check on the magnitude of $\tau_{\rm rei}$.

\begin{table}[ht]
\centering
\setlength{\tabcolsep}{8pt}
\begin{tabular}{l|l|c|c|c|c|c}
\toprule
\textbf{CMB map} & \textbf{Tracer} & $\bm{N^q\;\times 10^{10}}$ & $\bm{\sigma_{b_q}^{\rm all}}$ & $\bm{\sigma_{b_q}^{\rm auto}}$ &
$\bm{\sigma_{\tau_{\rm rei}}^{\rm all}}$ & $\bm{\sigma_{\tau_{\rm rei}}^{\rm auto}}$ \\
\midrule
\multirow{2}{*}{\textit{Planck} 100\,GHz} & unWISE  & $182$   & $13.2$  & $388$  & $0.686$ & $1.24$ \\
                                            & CIB 353 & $50.6$  & $6.87$  & $113$  & $0.362$ & $0.595$ \\
\hline
\multirow{2}{*}{ACT 90\,GHz}              & unWISE  & $17.5$  & $5.57$  & $120$  & $0.298$ & $0.547$ \\
                                            & CIB 353 & $3.63$  & $2.67$  & $28.8$ & $0.141$ & $0.243$ \\
\hline
\multirow{2}{*}{SO SAT 93\,GHz}           & unWISE  & $2.40$  & $2.06$  & $16.7$ & $0.110$ & $0.206$ \\
                                            & CIB 353 & $1.06$  & $1.44$  & $8.61$ & $0.0761$ & $0.147$ \\
\hline
\multirow{2}{*}{SO LAT 93\,GHz}           & unWISE  & $2.89$  & $2.27$  & $20.1$ & $0.121$ & $0.224$ \\
                                            & CIB 353 & $0.567$ & $1.05$  & $4.71$ & $0.0555$ & $0.118$ \\
\hline  
\multirow{2}{*}{LiteBIRD}                 & unWISE  & $1.46$  & $1.19$  & $3.42$ & $0.0615$ & $0.195$ \\
                                            & CIB 353 & $0.634$ & $0.769$ & $1.72$ & $0.0405$ & $0.120$ \\
\bottomrule
\end{tabular}
\caption{Forecasted reconstruction noise $N^q$ and constraints on $b_q$ and $\tau_{\rm rei}$ for each CMB map--tracer combination, including noise modeling.}
\label{tab:SNR_forecast}
\end{table}

\subsection{Reconstruction pipeline and bispectrum amplitude estimator}\label{sec:recpipeline}

\textbf{Pre-computed power spectra.} To calculate the quadrupole reconstruction noise $N_q$ in Eq.~(\ref{eq:Nq}), we pre-compute the following power spectra. The galaxy- and CIB-optical depth cross-power spectra $\{\bar{C}_\ell^{\tau g},\,\bar{C}_\ell^{\tau \rm CIB}\}$ are evaluated at a reference redshift $\chi=\bar{\chi}$, which is the median redshift of the $dN/dz$. For $\bar{C}_{\ell=\bar{\ell}}^{\tau g}$ and $\bar{C}_{\ell=\bar{\ell}}^{\tau \rm CIB}$ we use a reference scale $\bar{\ell}=200$ because this is roughly our peak sensitivity, as shown in figure~(\ref{fig:summandq}). We estimate $C_\ell^{gg}$ from the power spectrum of the masked unWISE number count map, rescaled by $f_{\rm sky}^{-1}$; the CIB power spectrum is estimated analogously as $C_\ell^{\rm CIB}=C_\ell^{\rm CIB,\,masked}/f_{\rm sky}$. We apodize the GNILC CIB footprint by $2^\circ$ and mask bright point sources with flux thresholds at $0.035 \ {\rm MJy}/{\rm Sr}$ for 353 GHz, $0.10 \ {\rm MJy}/{\rm Sr}$ for 545 GHz, and $0.18 \ {\rm MJy}/{\rm Sr}$ for 857 GHz. The brightest sources contribute significantly to the CIB auto power spectrum. The statistical anisotropy induced by these bright sources leads to poor agreement between the estimator pre-factor $N_q$ and the reconstruction power spectra; see Sec.~\ref{sec:reconstructions}. We therefore adjust the flux cuts until good empirical agreement is found between the estimator pre-factor $N_q$ and the reconstruction power spectra. The resulting masks are listed in table~(\ref{tab:specmasks}). The CMB power spectra $C_\ell^{EE}$ and $C_\ell^{BB}$ are computed from the masked, debeamed CMB maps. For \textit{Planck} individual frequency maps, we apply a galactic cut retaining $60\%$ of the sky which is apodized by a $2^{\circ}$ kernel as well as the \textit{Planck} point source masks. For \textit{Planck} component separated maps we apodize the \texttt{SMICA} confidence mask with a $2^{\circ}$ kernel. For ACT maps we apodize the union of the ACT footprint and the CIB footprint with a $2^{\circ}$ kernel and apply the union of \textit{Planck} frequency point source masks. The masks and associated sky coverage are listed in table~(\ref{tab:EEBBspectramasks}). With these choices, we find good empirical agreement between the estimator pre-factor $N_q$ and the reconstruction power spectra; see Sec.~\ref{sec:component_separated_recons}.

\textbf{Filtered fields for the quadratic estimator.} We construct the filtered fields entering Eq.~(\ref{eq:QandUrealspaceQEs}) as follows. The optical depth proxy $\tilde{\tau}(\hat{n})$ is built from the unmasked unWISE number density map (or CIB map): we take its spherical harmonic transform and multiply the $a_{\ell m}$'s by the filter $\bar{C}_\ell^{\tau g}/C_\ell^{gg}$ (or $\bar{C}_\ell^{\tau \rm CIB}/C_\ell^{\rm CIB}$), then apply a top-hat pass-band $[\ell_{\min},\ell_{\max}]$. These $\ell_{\min}$, $\ell_{\max}$ choices are listed in table~(\ref{tab:filterlowhigh}). This removes the largest scales to avoid potential systematics and the smallest scales to avoid pixel-scale artifacts. We then inverse spherical harmonic transform to obtain $\tilde{\tau}(\hat{n})$. Without a passband, we find disagreement between the estimator variance and $N_q$ at low-$\ell$. We check that by filtering to sufficiently high $\ell_{\rm min}$ the analytic noise $N_q$ agrees with the empirical variance of the reconstruction at low $\ell$. This has a cost in terms of the reconstruction noise levels. 

The filtered CMB polarization fields $\tilde{Q}_{EB}(\hat{n})$ and $\tilde{U}_{EB}(\hat{n})$ (Eqs.~\ref{eq:QtildeEB}--\ref{eq:UtildeEB}) are built from the unmasked CMB $Q$ and $U$ maps. We decompose these into $a_{\ell m}^E$ and $a_{\ell m}^B$ via a spin-2 harmonic transform, multiply by $1/(B_\ell C_\ell^{EE})$ and $1/(B_\ell C_\ell^{BB})$ respectively, where $B_\ell$ is a 5 arcminute Gaussian beam that suppresses small-scale power, and apply the same passband filter as for the tracer fields. An inverse spin-2 harmonic transform then yields $\tilde{Q}_{EB}(\hat{n})$ and $\tilde{U}_{EB}(\hat{n})$. The reconstructed quadrupole maps $\hat{Q}_q(\hat{n})$ and $\hat{U}_q(\hat{n})$ are then masked. The masks used for various data combinations are listed in table~(\ref{tab:recmasks}).

\begin{figure}[htbp]
    \centering
    
    \begin{subfigure}{0.48\textwidth}
        \centering
        \includegraphics[width=\linewidth]{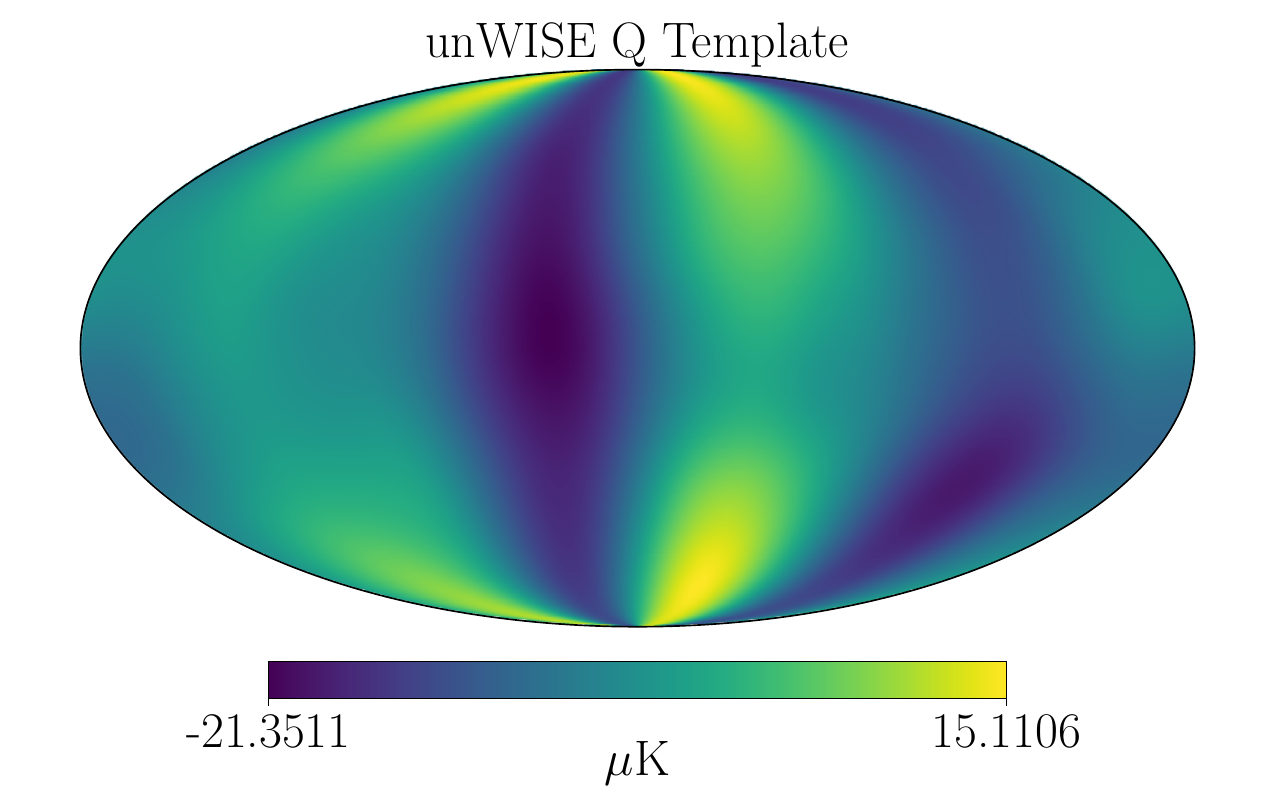}
    \end{subfigure}
    \hfill 
    \begin{subfigure}{0.48\textwidth}
        \centering
        \includegraphics[width=\linewidth]{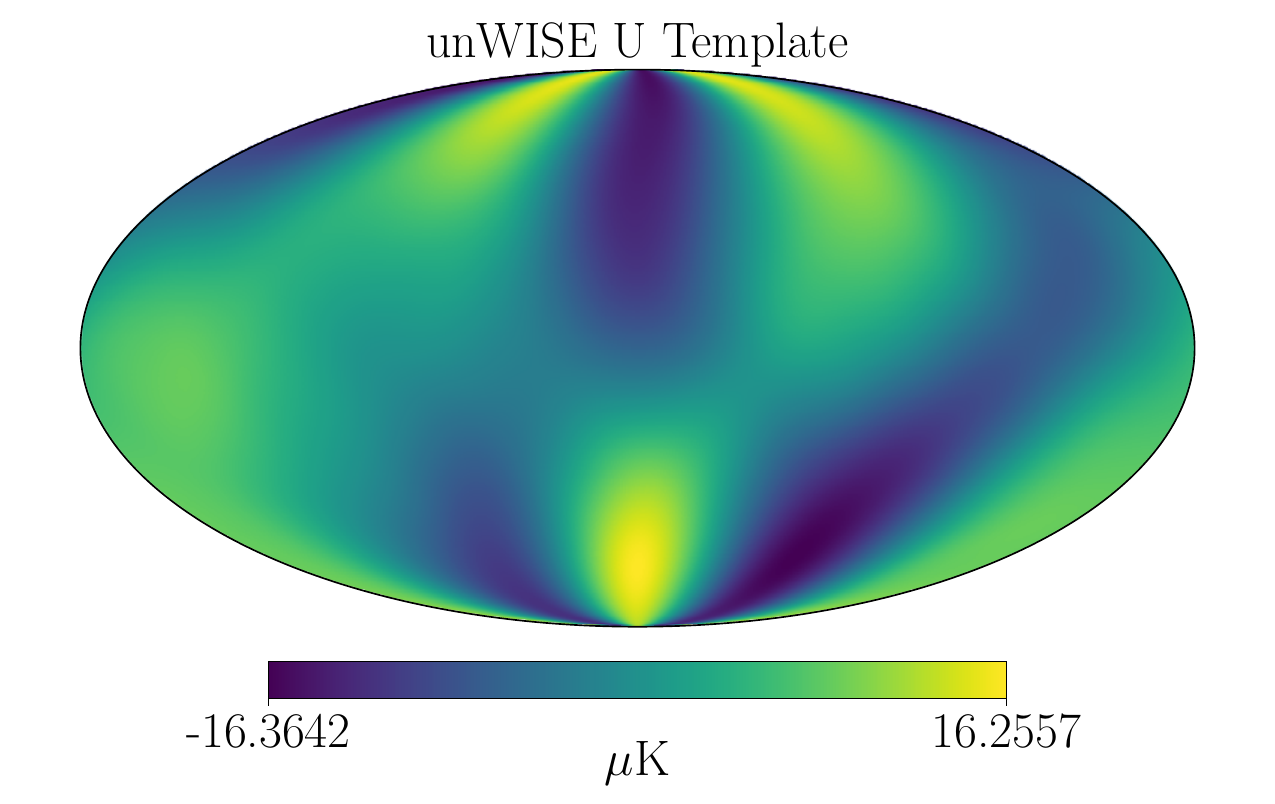}
    \end{subfigure}

    \vspace{0.5cm} 
    \begin{subfigure}{0.48\textwidth}
        \centering
        \includegraphics[width=\linewidth]{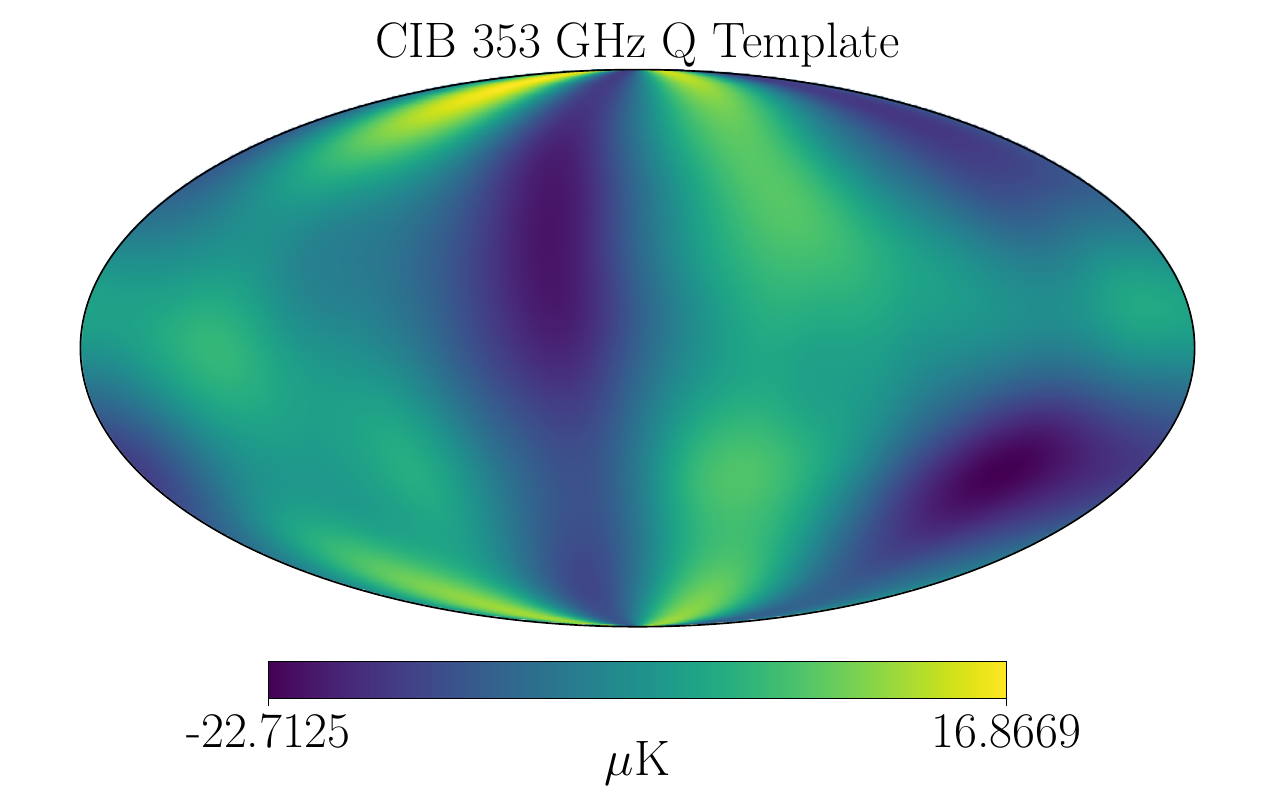}
    \end{subfigure}
    \hfill
    \begin{subfigure}{0.48\textwidth}
        \centering
        \includegraphics[width=\linewidth]{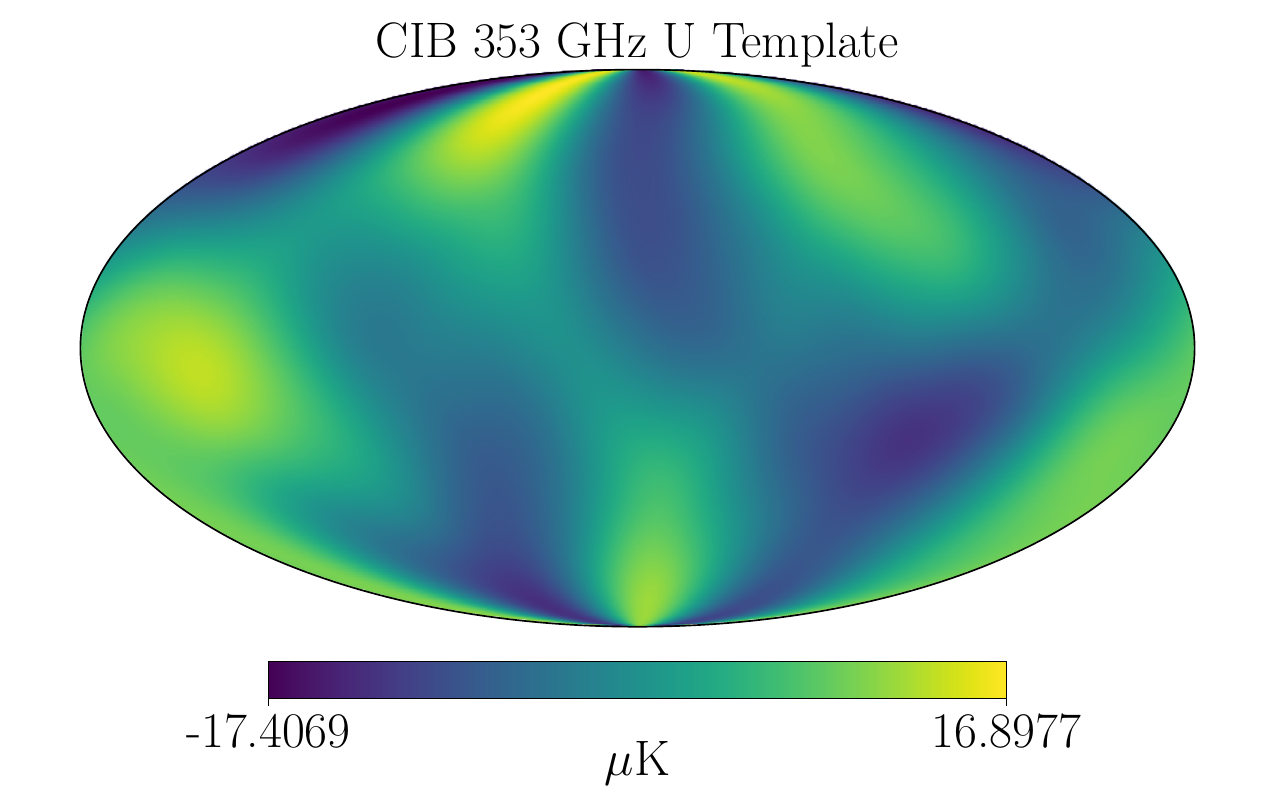}
    \end{subfigure}

    \caption[Q and U remote quadrupole template maps for unWISE and CIB 353 GHz.]{The Q (left) and U (right) remote quadrupole template maps  for unWISE (upper) and CIB 353 GHz (lower) from Eq.~\eqref{eq:qE_template} using the \texttt{Commander} temperature and $E$-mode polarization.}
    \label{fig:QUtemplates_unwise_CIB}
\end{figure}

\textbf{Estimators for $b_q$, $\tau_{\rm rei}$, and $r$.} The estimators for $b_q$ (Eq.~\eqref{eq:bqest}), $\tau_{\rm rei}$ (Eq.~\eqref{eq:tauest}), and $r$ (Eq.~\eqref{eq:rest}) are discretized by taking 
\begin{equation}
    \int d^2\hat{n} M(\hat{n}) \rightarrow \sum_{\hat{n} \in M \neq 0} \Omega_{\rm pix}
\end{equation}
where $M(\hat{n})$ is the reconstruction mask and $\Omega_{\rm pix}$ is the pixel area. The estimators are based on the same filtered fields $\tilde{\tau}$, $\tilde{Q}_{EB}$, $\tilde{U}_{EB}$ in the quadratic estimator. The additional inputs are the remote quadrupole templates $\{ \hat{Q}(\hat{n}), \hat{U}(\hat{n})\}$, $\{ \hat{Q}_E(\hat{n}), \hat{U}_E(\hat{n})\}$, and $\{ \hat{Q}_B(\hat{n}), \hat{U}_B(\hat{n})\}$ introduced in the previous section as well as the estimator variances. The \texttt{Commander}-based templates are built from the inpainted full-sky \texttt{Commander} maps. We spherical harmonic transform the \texttt{Commander} $T$/$Q$/$U$ maps to obtain $a_{\ell m}^T$, $a_{\ell m}^E$, debeam them, and apply the filter of Eq.~(\ref{eq:qE_template}) using theory spectra $\{C_\ell^{TT},\,C_\ell^{TE},\,C_\ell^{EE},\,C_\ell^{Tq_E},\,C_\ell^{Eq_E}\}$. These are calculated from $\Lambda$CDM following Eq.~(\ref{eq:scalar_spectra}) with $\Delta_\ell^{q_E}(k,\chi)$ integrated over the $W_q$ window function. The auto-correlation spectra ($C_\ell^{TT},\,\,C_\ell^{EE}$) also include noise, which we model by calculating the noise power spectrum from the public \texttt{Commander} noise simulations and taking the mean. We mask these noise simulations with a galactic cut retaining $60\%$ of the sky and rescale the power spectrum by $f_{\rm sky}$ to estimate it on the full sky. For the templates, we include multipoles $\ell=2$--$20$ and null higher $\ell$s; the correlation between the local CMB and the remote quadrupole falls off rapidly with multipole, so there is minimal gain from extending beyond $\ell_{\rm max}=20$. Summing the filtered $T$ and $E$ contributions and performing an inverse spin-2 harmonic transform yields the template maps $\hat{Q}(\hat{n})$ and $\hat{U}(\hat{n})$, which are shown in figure~(\ref{fig:QUtemplates_unwise_CIB}) for unWISE and CIB 353 GHz. 

For the estimator variance $\sigma_{b_q}^{-2}$ (Eq.~(\ref{eq:bvvar})), we use the same theory spectra as described above. We factorize the numerator as:
\begin{eqnarray}
\mathbf{C}^{Zq_E}_\ell \cdot (\mathbf{C}^{ZZ}_\ell)^{-1} (\mathbf{C}^{ZZ}_\ell)_{0,\rm cut} (\mathbf{C}^{ZZ}_\ell)^{-1} \cdot \mathbf{C}^{Zq_E}_\ell = \mathbf{C}^{Zq_E}_\ell \cdot (\mathbf{C}^{ZZ}_\ell)^{-1}\cdot (\mathbf{C}^{Z\tilde{q}_E}_\ell)_{0,\rm cut}\,.
\end{eqnarray} 
The $(\mathbf{C}^{Z\tilde{q}_E}_\ell)_{0,\rm cut}$ term is the cross-spectrum between the \texttt{Commander} $Z_0\in{T_0,E_0}$ fields and the Wiener-filtered quadrupole template $\tilde{q}_E$ (Eq.~\ref{eq:qE_template}) after applying the reconstruction mask to the quadrupole template in pixel space.

\textbf{Validation with simulations.} We validate both the quadratic estimator and the optical depth bias estimator using simulations of ACT 90\,GHz and unWISE, and with CIB 353\,GHz as a representative CIB case. We generated 1500 noise-only, signal-plus-noise, and signal-only simulations. We perform the simulations at $N_{\rm side}=1024$ ($\ell_{\rm max}=3071$), which from figure~(\ref{fig:summandq}), captures the majority of the signal-to-noise in the estimator.

For the galaxy tracer, synthetic signal $a_{\ell m}$'s are drawn from the theory clustering power spectra ($C_\ell^{gg}$ without shot noise) and synthetic noise $a_{\ell m}$'s from $N_{\rm shot}$. These are combined to form a mock galaxy realization. The signal optical depth map is formed by filtering the signal-only $a_{\ell m}$'s by $\bar{C}_\ell^{\tau g}/C_\ell^{gg}$, while the noise optical depth map uses the full signal-plus-noise mock. This produces a $\tilde{\tau}$ that is correlated with the tracer realization. 
  
For CIB 353\,GHz, the signal $a_{\ell m}$'s are drawn from the theory $C_\ell^{\rm CIB}$ and the noise from $C_\ell^{\rm noise} = |C_\ell^{\rm data} - C_\ell^{\rm CIB}|$. Here we have taken the difference of the data and theory power spectra to make mock noise power spectra. The mock $\tilde{\tau}$ maps follow the same filtering procedure as implemented for galaxies.

For the CMB realizations, we create mock CMB (signal and noise) maps and inject a mock pSZ signal. First, the mock CMB signal and noise map (without the injected pSZ signal) is generated from the inverse-variance-filtered power spectra $1/C_\ell^{EE}$ and $1/C_\ell^{BB}$.
Then, the remote quadrupole field signal is drawn as a constrained realization, conditioned on a simulated \texttt{Commander} CMB signal-only $T$ and $E$ following Eq.~(\ref{eq:qE_conditional}). We use a simulated \texttt{Commander} map here so that we can separate the signal and noise. A pSZ signal map is then formed as
\begin{equation}
  (\hat{Q} \pm i\hat{U})_{\rm pSZ}(\hat{n})
  = -\,b_q\,\frac{\sqrt{6}}{10}\,\hat{q}_E(\hat{n})\,\tilde{\tau}^{\rm sig}(\hat{n})
\end{equation}
with injected amplitude $b_q = 1$. This is inverse-variance filtered in the same way as the CMB maps and added to the CMB realizations to create our mock signal+noise CMB maps. For the signal only case, we only use the pSZ signal map with no mock CMB. The same reconstruction masks are applied as in the analysis of the real data.

The $\hat{Q}(\hat{n})$ and $\hat{U}(\hat{n})$ remote quadrupole template maps are calculated following Eq.~(\ref{eq:QUtemps}). However, instead of using the observed \texttt{Commander} map, we again use the simulated \texttt{Commander} CMB signal map and add noise realizations - so the signal is fixed and the noise varies. The noise realizations are Gaussian draws from the \texttt{Commander} noise simulation power spectra. For the prefactor, $\sigma_{b_q}^2$, the noise contribution to the $(\mathbf{C}_\ell^{ZZ})_{0,\rm cut}$ term is calculated using the \texttt{Commander} noise simulation power spectra. We apply the same prefactor for each realization. 

\begin{figure}
    \centering
    \includegraphics[width=\linewidth]{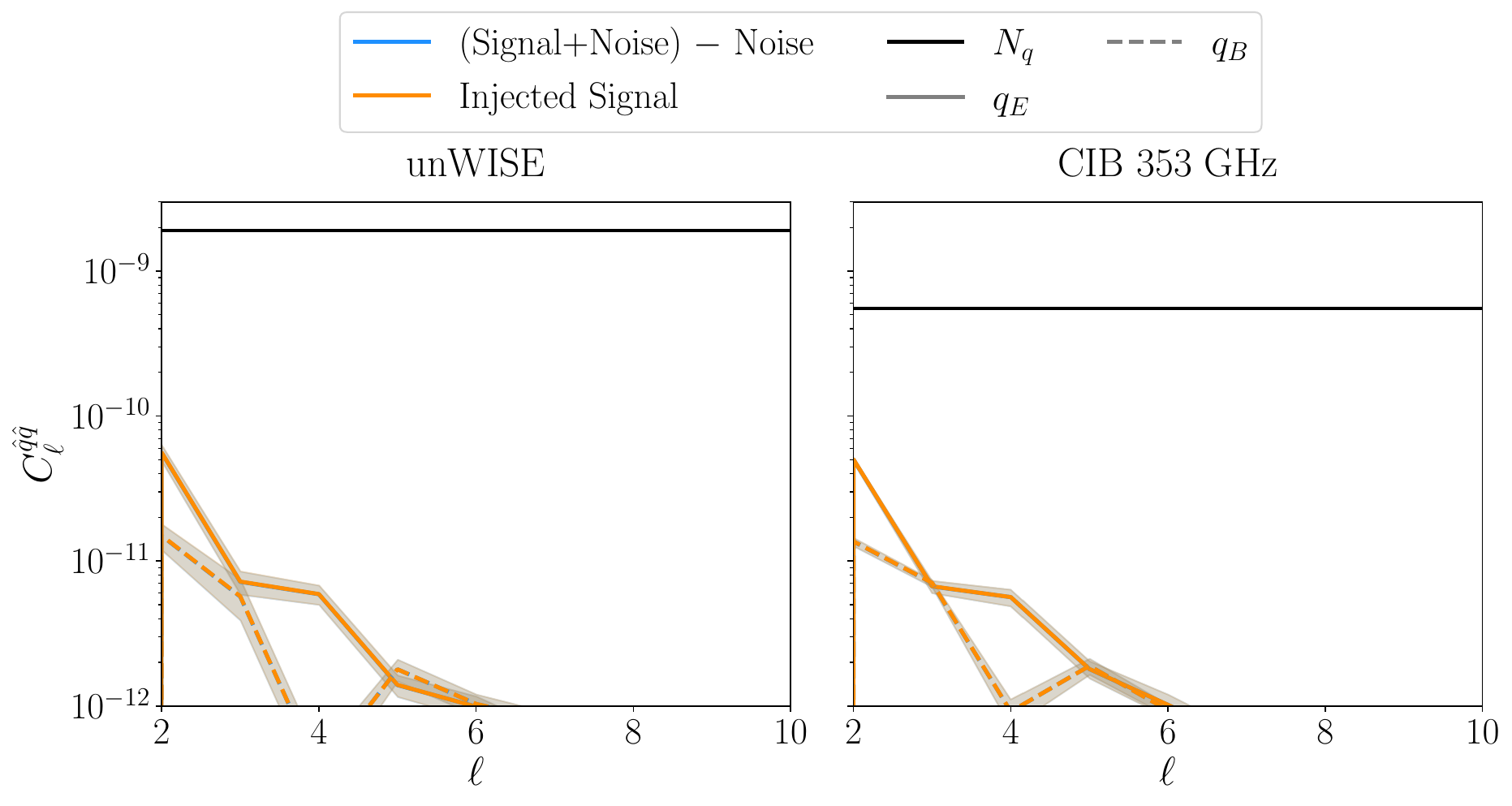}
    \caption{The reconstructed remote quadrupole spectra from simulations of ACT 90 GHz $\times$ unWISE (left) and ACT 90 GHz $\times$ CIB 353 GHz (right). The colored lines are the mean over all simulations and the shaded regions are $\pm 1\sigma$ scatter across simulations. Solid lines are used for the $q_E$ spectra and dashed lines for the $q_B$ spectra. The (signal+noise)$-$noise curve (blue) is the difference between the signal+noise and noise-only reconstructions, and the injected signal curve (orange) is the injected signal power spectrum (with $b_q=1$). The blue and orange curves completely overlap. The horizontal black line is the analytic reconstruction noise $N_q$.
    }
    \label{fig:simspec_unwiseCIB}
\end{figure}

Figure~(\ref{fig:simspec_unwiseCIB}) shows the mean quadratic estimator power spectra averaged over all realizations. Note that although the injected signal is pure $E$ mode, masking induces a $B$ mode component. We plot the difference between the signal+noise and the noise-only reconstructions to show that they exactly reproduce the injected signal spectra. This confirms that the quadratic estimator is unbiased. The noise-only reconstruction is also consistent with the analytic noise level at all $\ell$. 

\begin{figure}
    \centering
    \includegraphics[width=\linewidth]{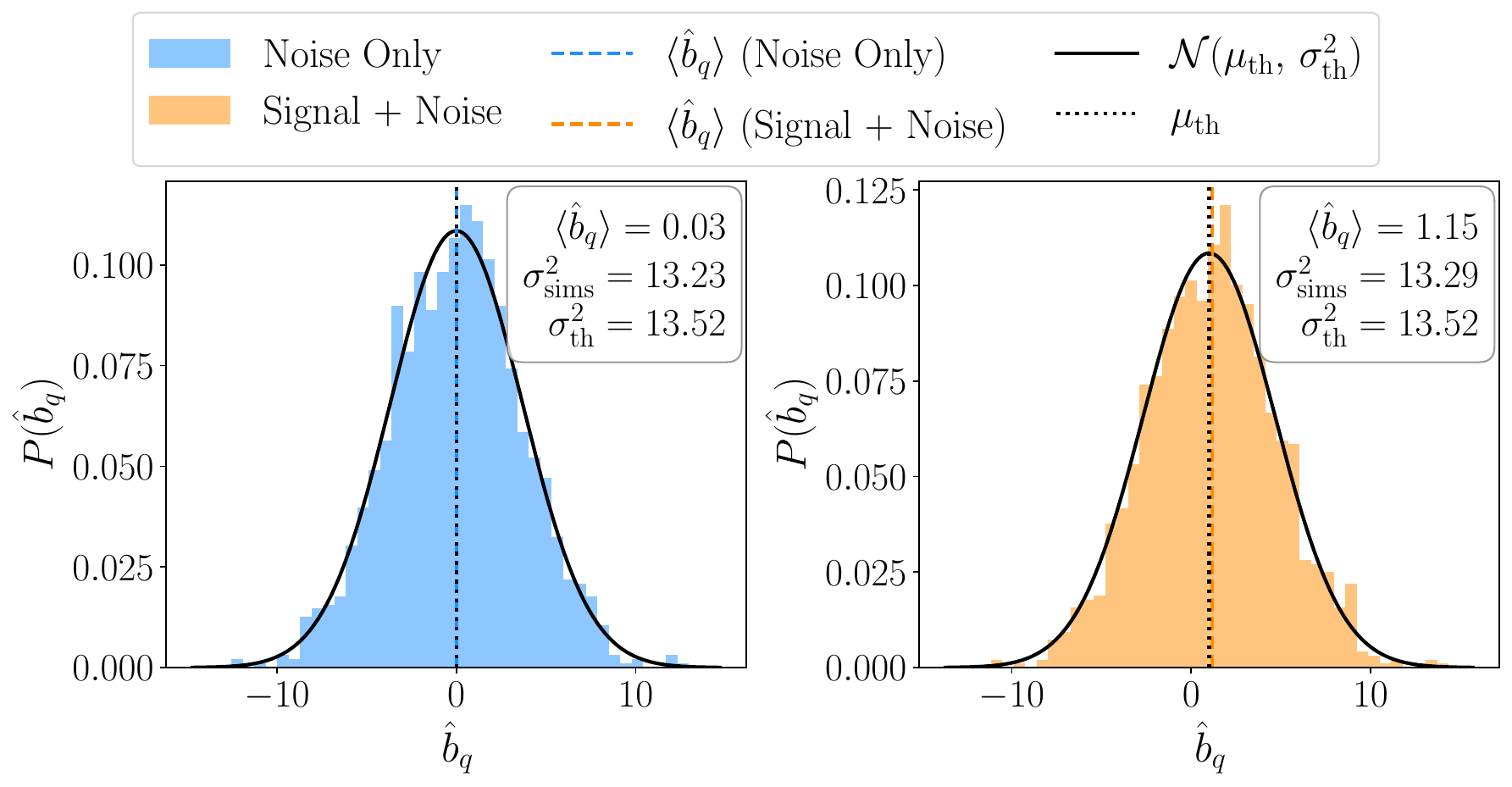}
    \caption{Distribution of the bispectrum amplitude estimator $\hat{b}_q$ over 1500 simulations of ACT 90 GHz $\times$ unWISE, for noise-only realizations (left, blue) and signal+noise realizations with injected amplitude $b_q = 1$ (right, orange). The black curve is the predicted Gaussian $\mathcal{N}(\mu_{\rm th},\,\sigma^2_{\rm th})$ with mean $\mu_{\rm th} = 0$ (noise-only) or $\mu_{\rm th} = 1$ (signal-plus-noise) and variance $\sigma^2_{\rm th}$ from Eq.~(\ref{eq:bvvar}). The dashed and dotted vertical lines mark the simulation mean $\langle\hat{b}_q\rangle$ and the theoretical expectation $\mu_{\rm th}$ respectively. }
    \label{fig:simhist_unwise}
\end{figure}

\begin{figure}
    \centering
    \includegraphics[width=\linewidth]{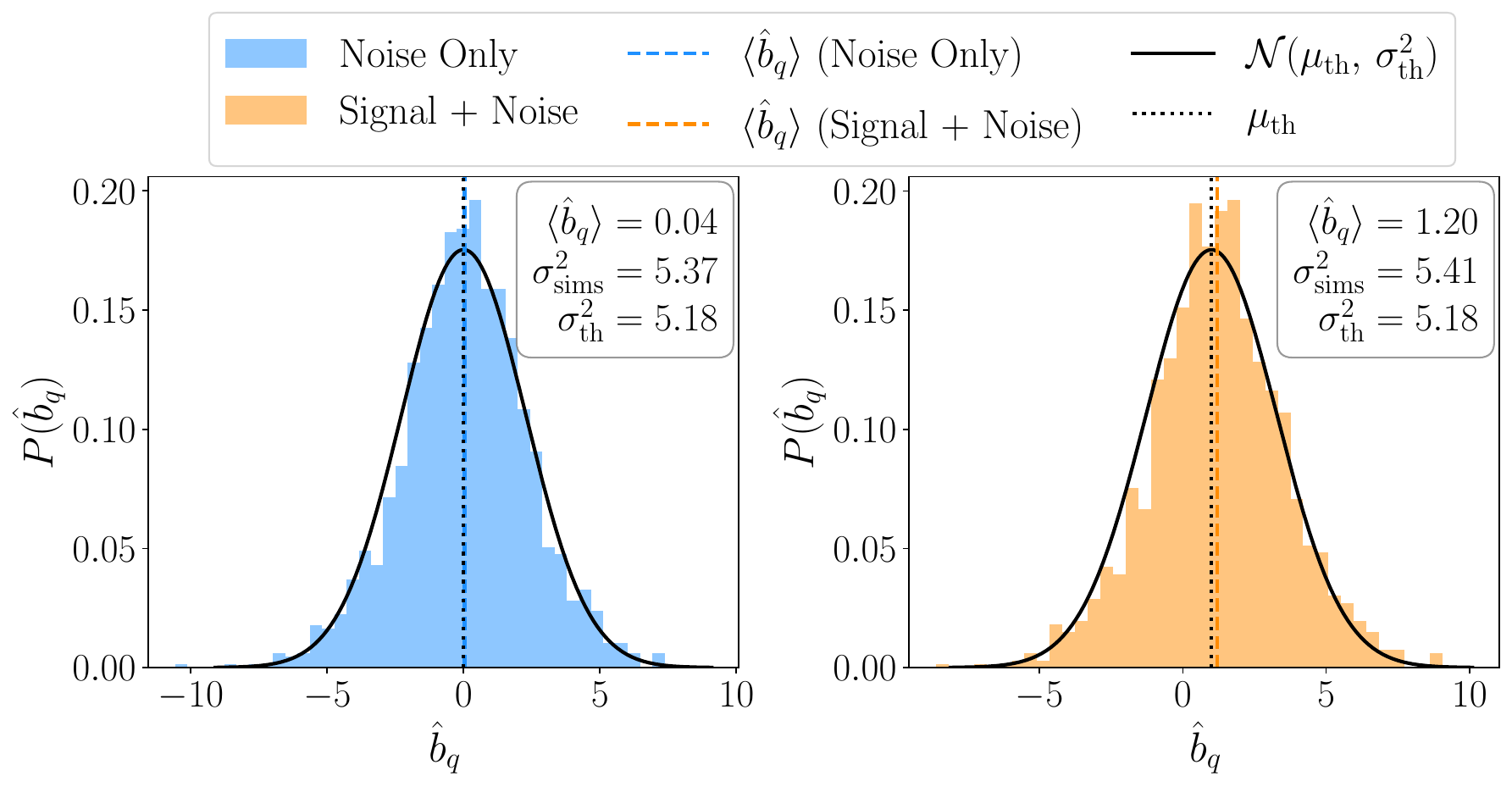}
    \caption{Same as figure~(\ref{fig:simhist_unwise}) except ACT 90 GHz $\times$ CIB 353 GHz.}
    \label{fig:simhist_CIB353}
\end{figure}

Figures~(\ref{fig:simhist_unwise}-\ref{fig:simhist_CIB353}) show the distribution of $\hat{b}_q$ over the 1500 simulations for the noise-only and signal-plus-noise cases using the maps described above on the masked sky. We overplot the predicted Gaussian with mean $0$ (noise-only) or $b_q=1$ (signal+noise) and variance $\sigma_{\rm th}^2$ which is simply $\sigma_{b_q}^2$ from Eq.~(\ref{eq:bvvar}) evaluated on the simulated maps. The simulated variance agrees with the expected value from the estimator pre-factor to within a few percent: for unWISE, noise only is $\sigma_{\rm sims}^2 = 13.23$ (signal+noise $\sigma_{\rm sims}^2 = 13.29$) versus $\sigma_{\rm th}^2 = 13.52$ ($2\%$ underestimate); for CIB 353 GHz, noise only is $\sigma_{\rm sims}^2 = 5.37$ (signal+noise $\sigma_{\rm sims}^2 = 5.41$) versus $\sigma_{\rm th}^2 = 5.18$ ($4\%$ overestimate). For both tracers, the recovered noise-only means are consistent with zero. The signal+noise means are somewhat biased, with $\langle \hat{b}_q \rangle=1.15$ for unWISE and $\langle \hat{b}_q \rangle=1.20$ for CIB 353 GHz. To confirm the presence of a bias even without reconstruction noise, we evaluate the estimator on the pSZ signal component alone. We find $\langle \hat{b}_q \rangle=1.12$ for unWISE and $\langle \hat{b}_q \rangle=1.17$ for CIB 353 GHz, confirming a $\sim 10-20 \%$ bias on measurements of the optical depth bias. This estimator bias can be corrected by dividing $\sigma_{b_q}^2$ by the signal-only amplitude recovered from simulations. This has the effect of reducing the variance of $\hat{b}_q$. Given that we must simulate each data combination used below to find the appropriate correction factor, and that an accurate estimate of the estimator bias appropriate for these cases would require more detailed simulations, we accept that the variance predicted by the estimator pre-factor $\sigma_{b_q}^2$ from Eq.~(\ref{eq:bvvar}) is a slight over-estimate ($\sim 10-20\%$) and do not make numerical corrections to the estimator pre-factor in our results below.

Extensions of this work would use more realistic simulations. Currently, the CMB noise is drawn from Gaussian realizations. More realistic noise realizations could account for the spatially varying noise due to the atmosphere and telescope scan strategy. The tracer maps are also generated from Gaussian realizations of the theory power spectra. Additionally, we do not model any foreground contamination, assuming that ACT 90 GHz is clean. Despite these idealizations, the close agreement between the theory and simulation results demonstrates that the estimator is nearly unbiased and Eq.~(\ref{eq:bvvar}) provides a reliable prediction of the variance.

\section{Individual-frequency reconstructions and foregrounds}\label{sec:reconstructions}

\begin{figure}
    \centering
    \includegraphics[width=\linewidth]{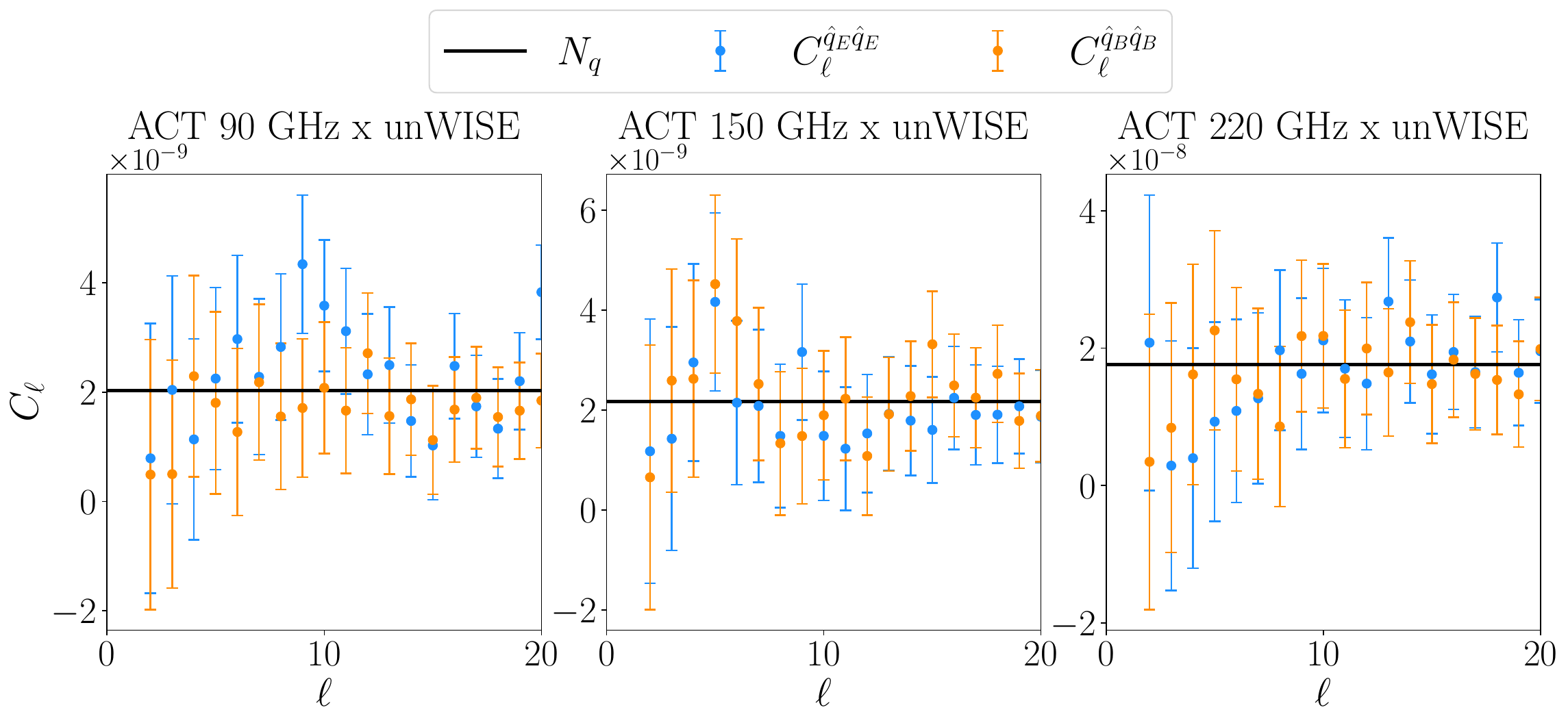}
    \caption{The remote quadrupole reconstruction angular power spectra, $C_\ell^{q_Eq_E}$ (blue) and $C_\ell^{q_Bq_B}$ (orange), for ACT individual frequency maps cross-correlated with unWISE. The black line is the reconstruction noise and the errors bars are the 1$\sigma$ cosmic variance error on the noise $\sigma_\ell=\sqrt{2/((2\ell+1)f_{\rm sky})}N_q$. }
    \label{fig:ACTinvidunwise_specs}
\end{figure}

We perform pSZ quadrupole reconstructions using individual frequency maps from \textit{Planck} and ACT. We present the ACT results here and \textit{Planck} results in Appendix~\ref{app:QEPlanck}. The ACT $\times$ unWISE reconstruction spectra are shown in figure~(\ref{fig:ACTinvidunwise_specs}). They are all consistent with reconstruction noise, demonstrating that there are no significant systematics biasing the reconstruction. 

\begin{figure}
    \centering
    \includegraphics[width=\linewidth]{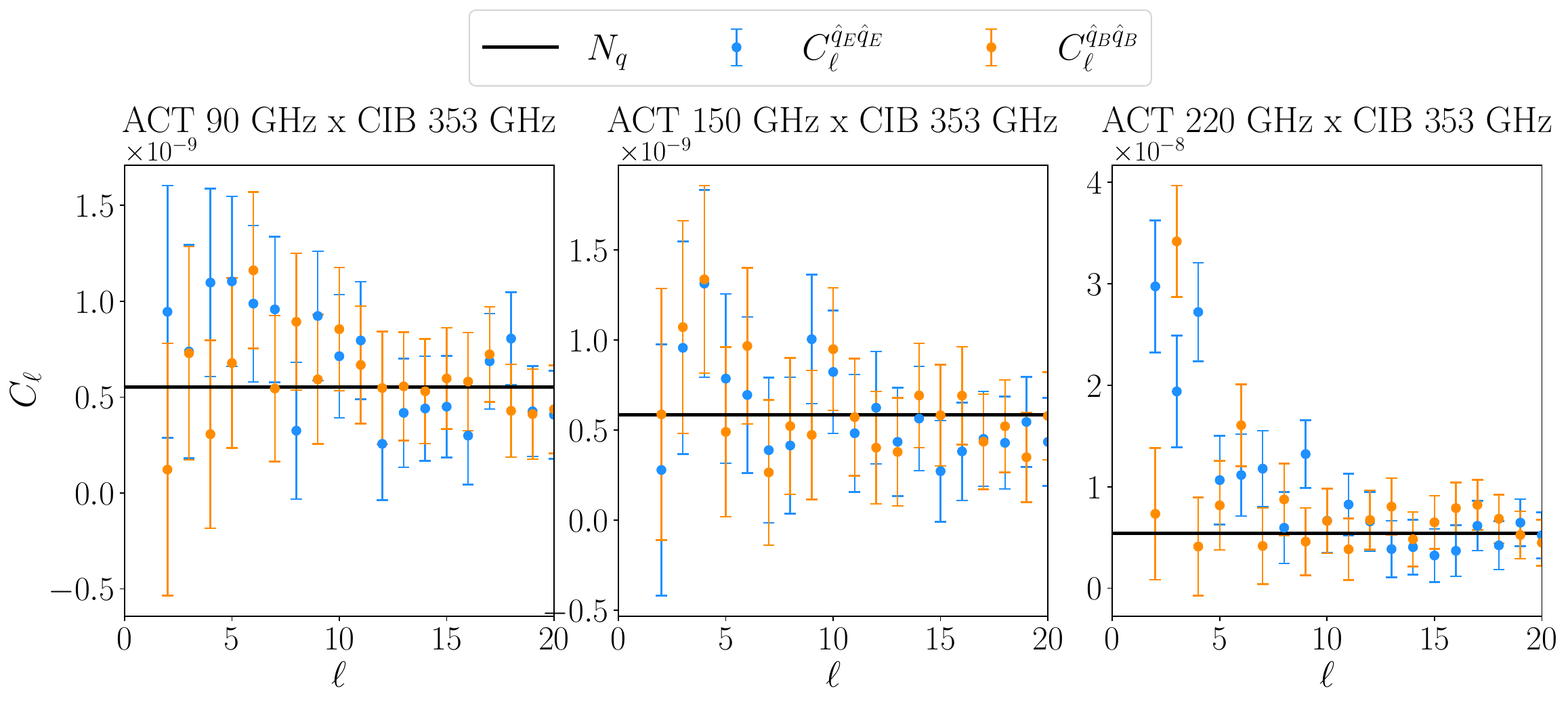}
    \caption{The same as figure~(\ref{fig:ACTinvidunwise_specs}) except for ACT $\times$ CIB 353 GHz.}
    \label{fig:ACTinvidCIB353_specs}
\end{figure}

The ACT $\times$ CIB 353 GHz reconstruction spectra are shown in figure~(\ref{fig:ACTinvidCIB353_specs}). There is more power than expected from reconstruction noise at low-$\ell$; this excess increases with frequency with the largest upturn for ACT 220 GHz. Excess power is similarly seen in ACT $\times$ CIB 545 GHz (figure~(\ref{fig:ACT_CIB545_recspec})) and CIB 857 GHz (figure~(\ref{fig:ACT_CIB857_recspec})), which are presented in Appendix~\ref{app:QEACT}. This behavior is also seen in the \textit{Planck} 217 and 353 GHz $\times$ CIB reconstructions described in Appendix~\ref{app:QEPlanck}. At higher CMB polarization frequencies there is a larger dust contribution to the maps. The observed frequency dependence suggests that residual galactic dust in the CIB maps is correlated with dust emission in the \textit{Planck} maps. To show up as excess power in the estimator, the cross-correlation must be anisotropic, for example due to residual galactic foregrounds or anisotropic depth. The ACT and \textit{Planck} results indicate that when using CIB as the tracer, we should choose CMB frequency channels in the 90-150 GHz range as they have the least dust contamination.

\begin{figure}[htbp]
    \centering
    \begin{subfigure}{0.48\textwidth}
        \centering
        \includegraphics[width=\linewidth]{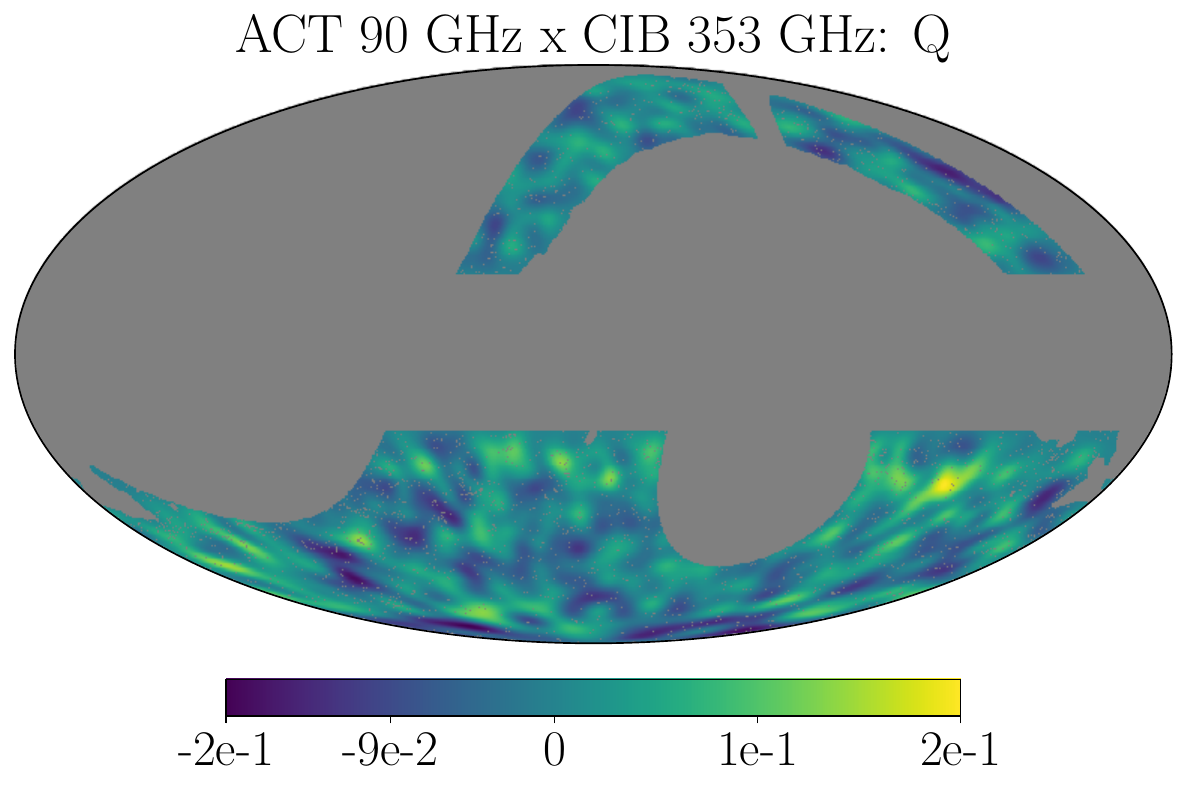}
    \end{subfigure}
    \hfill 
    \begin{subfigure}{0.48\textwidth}
        \centering
        \includegraphics[width=\linewidth]{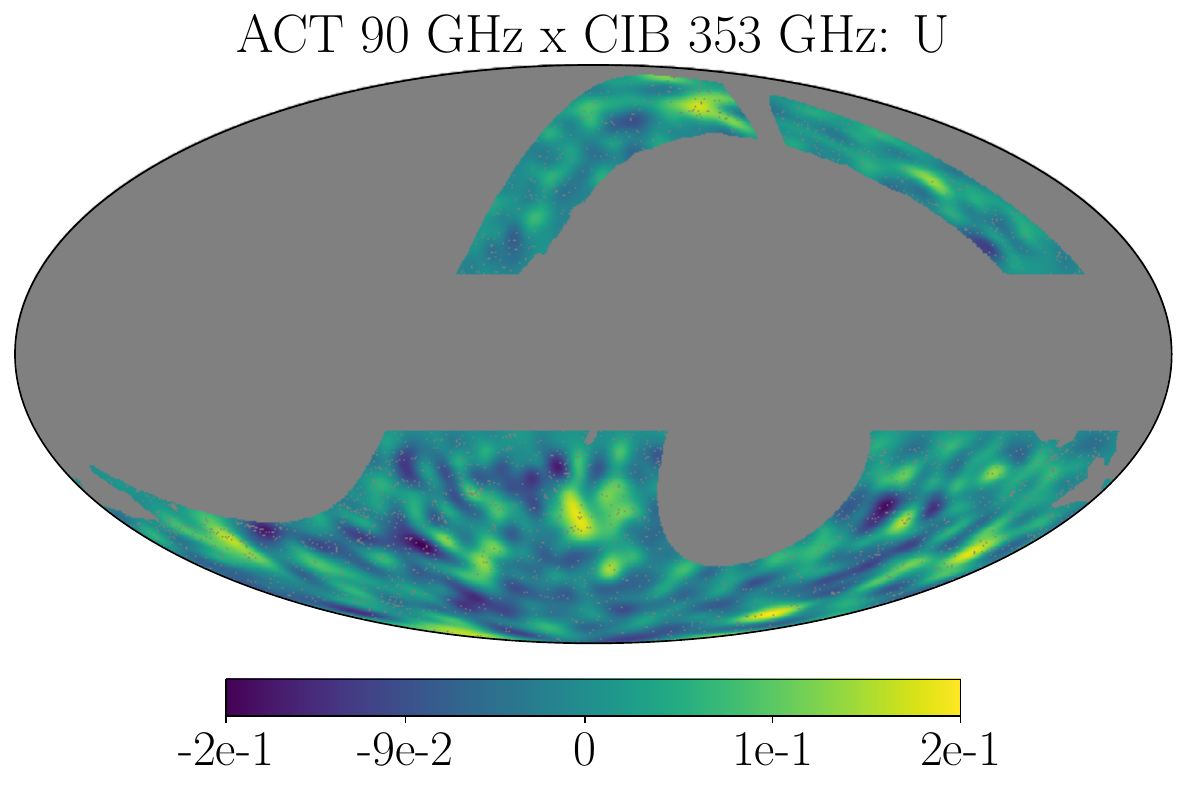}
    \end{subfigure}
    \caption{The $Q$ (left) and $U$ (right) remote quadrupole reconstruction maps for ACT 90 GHz $\times$ CIB 353 GHz. }
    \label{fig:QUreconACT90CIB353}
\end{figure}

Figure~(\ref{fig:QUreconACT90CIB353}) shows the Q and U maps for ACT 90 GHz $\times$ CIB 353 GHz. This further illustrates that we do not observe any clear structure in our map (eg. in comparison to the remote quadrupole template maps shown in figure~(\ref{fig:QUtemplates})); the reconstruction is consistent with noise.

Table~(\ref{tab:Nrec}) lists the reconstruction noise values for all data combinations. This clearly illustrates that the CIB maps yield significantly lower noise than unWISE. The best tracer is CIB 353 GHz (with $\sim4\times$ lower noise than unWISE). Additionally, ACT 220 GHz has substantially higher noise across all tracers ($\sim10\times$) due to the higher noise in the CMB maps; in the context of remote quadrupole reconstruction this channel therefore primarily serves as a systematics check. 

\begin{table}[h]                       
  \centering
  \renewcommand{\arraystretch}{1.3} 
  \begin{tabular}{l|c|c|c|c}
      \toprule 
      \multicolumn{5}{c}{$\bm{N_q \times 10^{9}}$} \\
      \toprule
      \textbf{CMB Map} & \textbf{unWISE} & \textbf{CIB 353 GHz} & \textbf{CIB 545 GHz} & \textbf{CIB 857 GHz} \\
      \midrule
      ACT 90 GHz  & $2.03$ & $0.551$ & $0.732$ & $0.888$ \\
      \hline
      ACT 150 GHz & $2.17$ & $0.585$ & $0.778$ & $0.951$ \\
      \hline  
      ACT 220 GHz & $17.6$ & $5.44$  & $7.19$  & $8.50$  \\
      \hline 
      ACT NILC    & $1.54$ & $0.436$ & $0.577$ & $0.733$ \\
      \hline 
      \textit{Planck} \texttt{SMICA} & $6.71$ & $2.16$ & $2.76$ & $4.45$ \\ 
      \bottomrule 
  \end{tabular}
  \caption{Reconstruction noise $N_q$ for each large-scale structure tracer combination with ACT CMB and \textit{Planck} \texttt{SMICA} polarization maps. }
  \label{tab:Nrec}                         
  \end{table}

\section{Reconstruction with \textit{Planck} and ACT component-separated maps}\label{sec:component_separated_recons}

\begin{figure}
    \centering
    \includegraphics[width=0.7\linewidth]{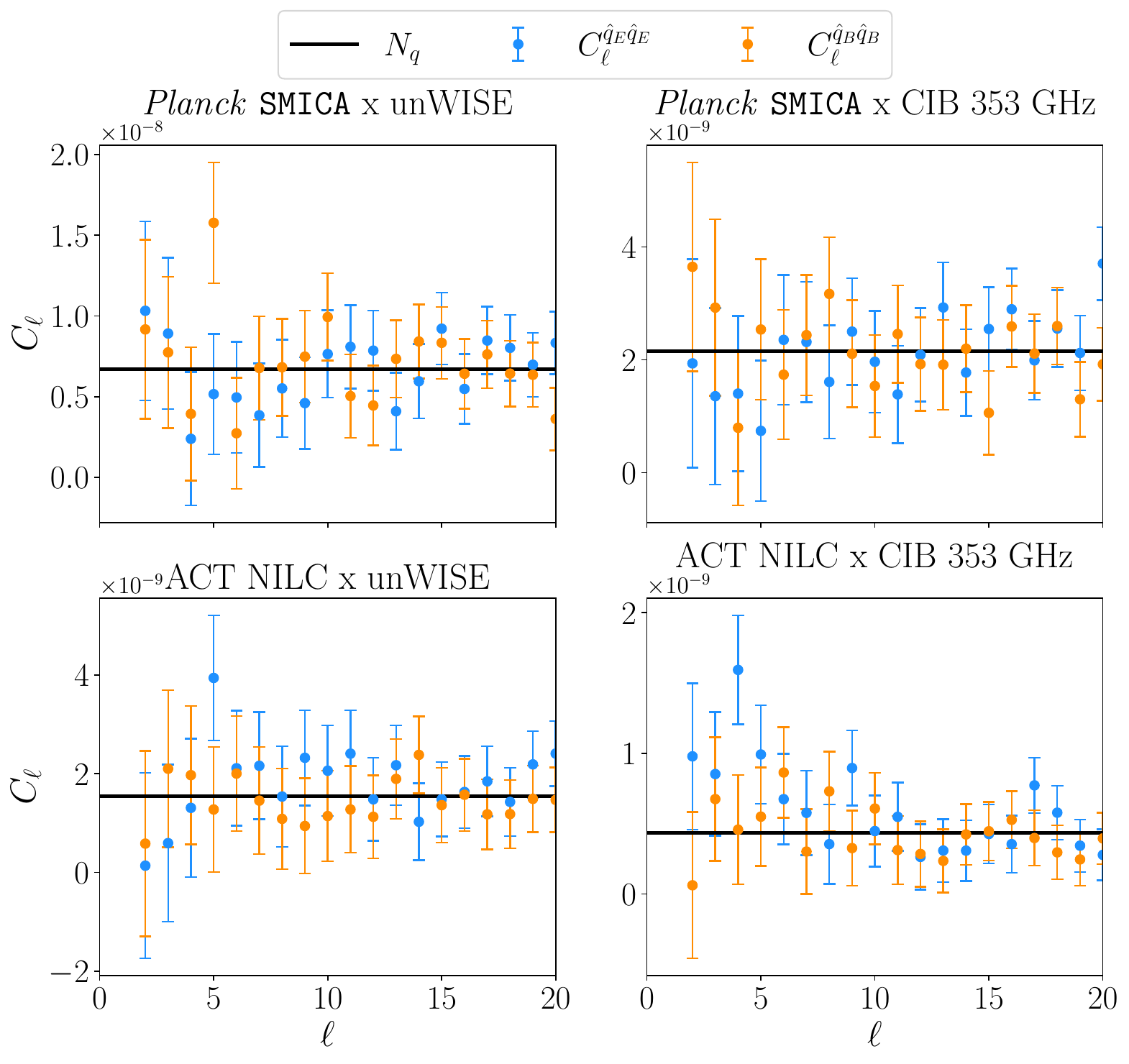}
    \caption{The same as figure~(\ref{fig:ACTinvidunwise_specs}) except for \textit{Planck} \texttt{SMICA} (top row), and ACT NILC (bottom row), $\times$ unWISE (left) and $\times$ CIB 353 GHz (right).}
    \label{fig:PlkSMICA_ACTNILC_specs}
\end{figure}

The component-separated CMB maps provide lower noise than the individual frequency channels and attempt to isolate the blackbody component. We see in figure~(\ref{fig:PlkSMICA_ACTNILC_specs}) the \textit{Planck} \texttt{SMICA} $\times$ unWISE and \texttt{SMICA} $\times$ CIB 353 GHz reconstruction results. Similar to the individual frequency results, the reconstruction is fully consistent with noise when using unWISE as the tracer. There is no evidence of foreground residuals at low-$\ell$ in \texttt{SMICA} $\times$ CIB 353 GHz which demonstrates that \texttt{SMICA} effectively removes the foreground contamination observed in reconstructions based on the individual frequency maps.

In figure~(\ref{fig:PlkSMICA_ACTNILC_specs}) is also the ACT NILC reconstruction. When cross-correlated with unWISE, the spectra are completely consistent with the reconstruction noise. For CIB 353 GHz as the tracer, there is some of the dust foreground contamination that we saw with the ACT 150 and 220 GHz maps. Although NILC is designed to suppress foreground contamination, ILC-based methods minimize total output variance under a CMB-preservation constraint, but the maps we use have not explicitly deprojected individual foreground components. So since the NILC map is a linear combination of all the individual frequency maps, it includes residuals. It is reduced compared to ACT 220 GHz though because it is primarily made up of 90 and 150 GHz since they have the lowest noise. Table~(\ref{tab:Nrec}) shows that NILC $\times$ 353 GHz achieves $N_q = 4.36 \times 10^{-10}$ - the lowest of all data combinations. 

It is important to note that the ACT maps we use are coadds with \textit{Planck}. This removes the atmospheric dust contamination on large angular scales from the ACT maps. This allows us to use multipoles as low as 25 in the quadratic estimator, while still retaining ACT's low noise on small scales. Therefore, the reconstruction noise is better with ACT+\textit{Planck} co-added maps than it would be for ACT alone.

These results demonstrate that component-separated maps are promising for pSZ, as they have lower reconstruction noises compared to their respective survey's individual frequency maps. \texttt{SMICA} is also more successful at cleaning than NILC, due to the larger number of \textit{Planck} frequency channels available for component-separation. This suggests that future pSZ analyses will benefit most from CMB surveys with broad frequency coverage (such as LiteBIRD) which enables more aggressive foreground cleaning and reduces contamination when cross-correlating with CIB tracers. 

\section{Constraints on $b_q$, $\tau_{rei}$, and $r$}\label{sec:constraints}

We now present constraints on $b_v$, $\tau_{\rm rei}$, and $r$. We constrain the optical depth bias using the estimator defined in Eq.~(\ref{eq:bqest}). These results are shown in table~(\ref{tab:bq_results}) for all ACT and tracer map combinations, as well as \textit{Planck} \texttt{SMICA}. All results are consistent with zero within $1$--$2\sigma$; there is no statistically significant detection of the pSZ signal with the current data. Since ACT 90 GHz is the cleanest from foregrounds and has the lowest noise of the individual frequency maps, we take it as our benchmark. The best constraint is therefore ACT 90 GHz $\times$ CIB 353 GHz with $\hat{b}_q=1.02 \pm 2.64$. If our gas-halo connection model accurately describes the data we expect $b_q=1$, so our benchmark uncertainty is close to the threshold needed for a meaningful test of the model, which is encouraging for future analyses.

The comparative performance across all data combinations reflects the reconstruction noise levels in table~(\ref{tab:Nrec}). The CIB maps consistently outperform unWISE as a tracer, with CIB 353 GHz delivering $\sim 1.5$--$2\times$ tighter constraints across all CMB maps due to its higher cross-correlation signal-to-noise. ACT 220 GHz yields substantially larger uncertainties than the other frequency maps ($\sigma_{b_q} \sim 3$--$4\times$ larger) because of its higher noise level and larger foreground contamination. Among component-separated maps, ACT NILC achieves the smallest uncertainty ($\sigma_{b_q} = 2.35$ with CIB 353 GHz), but the mean is biased due to residual dust contamination (shown in Sec.~\ref{sec:component_separated_recons}). \textit{Planck} \texttt{SMICA} is less constraining than ACT NILC despite \textit{Planck}'s $\sim 2\times$ larger sky coverage. This is explained by the estimator uncertainty scaling as $\sigma_{b_q} \propto \sqrt{N_q/f_{\rm sky}}$: ACT NILC's $\sim 4\times$ lower $N_q$ outweighs the smaller sky coverage. 

\begin{table}[h]
\centering
\renewcommand{\arraystretch}{1.3}
\begin{tabular}{l|c|c|c|c}  
  \toprule 
      \multicolumn{5}{c}{$\bm{\hat{b}_q \pm \sigma_{b_q}}$} \\
      \toprule
  \textbf{CMB Map} & \textbf{unWISE} & \textbf{CIB 353 GHz} & \textbf{CIB 545 GHz} & \textbf{CIB 857 GHz} \\
  \midrule
  ACT 90 GHz  & $4.08 \pm 4.81$   & $1.02 \pm 2.64$   & $0.51 \pm 2.98$   & $0.42 \pm 3.18$ \\
  \hline
  ACT 150 GHz & $-5.12 \pm 4.97$  & $-2.79 \pm 2.72$  & $-2.95 \pm 3.07$  & $-4.13 \pm 3.29$ \\
  \hline
  ACT 220 GHz & $-8.50 \pm 14.2$  & $0.13 \pm 8.30$   & $-2.70 \pm 9.32$  & $-7.46 \pm 9.85$ \\
  \hline
  ACT NILC    & $0.21 \pm 4.19$   & $-3.81 \pm 2.35$  & $-4.36 \pm 2.64$  & $-3.97 \pm 2.89$ \\
  \hline
  \textit{Planck} \texttt{SMICA} & $10.65 \pm 6.90$ & $0.85 \pm 4.12$ & $0.68 \pm 4.59$ & $1.03 \pm 4.59$ \\
  \bottomrule
\end{tabular}
\caption{Measured quadrupole amplitude bias, $\hat{b}_q \pm \sigma_{b_q}$, for each CMB map and large-scale
structure tracer combination. }
\label{tab:bq_results}
\end{table}

Using the estimator defined in Eq.~(\ref{eq:tauest}), we constrain the optical depth to reionization; the results are shown in table~(\ref{tab:tau_results}). These results are again consistent with zero within 1--2$\sigma$. Our benchmark of ACT 90 GHz provides the best constraints when cross-correlated with CIB 353 GHz: $\hat{\tau}_{\rm rei} = -0.01 \pm 0.14$. The ACT NILC $\times$ CIB 353 GHz also places tight constraints with $\hat{\tau}_{\rm rei} = -0.20 \pm 0.13$, apart from the slight bias from foregrounds. The \textit{Planck} fiducial value of $\tau_{\rm rei} = 0.054$ lies within the $2\sigma$ uncertainty of all combinations. 
As shown in Sec.~\ref{sec:snr}, with upcoming CMB experiments pSZ will be able to validate existing $\tau_{\rm rei}$ constraints and our bounds are a promising first step in this direction. 

\begin{table}[h]
    \centering 
    \renewcommand{\arraystretch}{1.3}
    \begin{tabular}{l|c|c|c|c}
      \toprule 
      \multicolumn{5}{c}{$\bm{\hat{\tau}_{\rm rei} \pm \sigma_{\tau_{\rm rei}}}$} \\
      \toprule
      \textbf{CMB Map} & \textbf{unWISE} & \textbf{CIB 353 GHz} & \textbf{CIB 545 GHz} & \textbf{CIB 857 GHz} \\
      \midrule
      ACT 90 GHz  & $0.02 \pm 0.27$  & $-0.01 \pm 0.14$ & $-0.05 \pm 0.16$ & $-0.04 \pm 0.18$ \\
      \hline
      ACT 150 GHz & $-0.20 \pm 0.28$ & $-0.10 \pm 0.15$ & $-0.09 \pm 0.17$ & $-0.13 \pm 0.18$ \\
      \hline
      ACT 220 GHz & $-0.14 \pm 0.78$ & $0.40 \pm 0.45$  & $0.34 \pm 0.51$  & $0.26 \pm 0.54$  \\
      \hline
      ACT NILC    & $-0.01 \pm 0.23$ & $-0.20 \pm 0.13$ & $-0.23 \pm 0.15$ & $-0.19 \pm 0.16$ \\
      \hline
      \textit{Planck} \texttt{SMICA} & $0.23 \pm 0.37$ & $0.00 \pm 0.22$ & $-0.01 \pm 0.25$ & $0.02 \pm 0.24$ \\
      \bottomrule
    \end{tabular}
\caption{Measured reionization optical depth, $\hat{\tau}_{\rm rei} \pm \sigma_{\tau_{\rm rei}}$, for each CMB map and
large-scale structure tracer combination, assuming a fiducial $\tau_{\rm rei} = 0.054$.}
\label{tab:tau_results}
\end{table}

To assess the robustness of our results to the choice of large-scale remote quadrupole template, we repeat the analysis using all four \textit{Planck} component-separated maps as templates for the remote quadrupole field. The $\hat{b}_q$ and $\hat{\tau}_{\rm rei}$ results for our benchmark data product (ACT 90 GHz $\times$ CIB 353 GHz) are compared in Appendix~\ref{app:compsep}. \texttt{Commander} and \texttt{SEVEM} agree with each other and are both consistent with zero. \texttt{SMICA} and \texttt{NILC} yield larger central values ($1.5\sigma$ and $2.2\sigma$ from zero in $\hat{b}_q$ respectively). As \texttt{Commander} fits a model and \texttt{SEVEM} subtracts templates, they are better suited to removing the galactic foreground while preserving the CMB on large scales. Since \texttt{SMICA} and \texttt{NILC} minimize variance they can be less effective on large scales where there are fewer available modes. This motivates our use of \texttt{Commander}, with \texttt{SEVEM} serving as an independent cross-check.

Finally, we constrain the tensor-to-scalar ratio using Eq.~(\ref{eq:rest}). In the absence of an empirical measurement of the primordial $B$-mode signal, we replace the ensemble average over simulated $B_{\ell m}$ in Eq.~\eqref{eq:QUBmodetemplate} with a sum over the fixed orthonormal basis spanning the full space of $\ell=2$ $B$-mode quadrupole orientations as defined in Eq.~\eqref{eq:tensorbasis}. We obtain a separate constraint on $r$ for each orientation. The ACT 90 GHz $\times$ CIB 353 GHz results are given in table~(\ref{tab:r_constraints_short}) for $n_t=0$ and for $n_t=-1$. The full table of all data combinations is included in the appendix in table~(\ref{tab:r_constraints}). The constraints tighten significantly for the more negative tilt, since a red spectrum ($n_t < 0$) enhances tensor power on large angular scales and therefore increases the pSZ signal at low $\ell$. This sensitivity to $n_t$ demonstrates the importance of the   modeling assumptions when interpreting these bounds. While the constraints on $r$ are not competitive with those from primary CMB $B$-mode polarization, they are expected to improve substantially with upcoming surveys. At that time, pSZ can provide a complementary and independent probe of primordial gravitational waves. CMB-HD, for instance, identifies this as one of its science objectives~\cite{CMB-HD:2022bsz}, and our results constitute the first proof-of-principle bounds on $r$ from the pSZ effect.

 \begin{table}[h]
    \centering 
    \renewcommand{\arraystretch}{1.3}
    \begin{tabular}{llcc}
    \toprule 
      \multicolumn{4}{c}{$\bm{\hat{r}^{(\alpha)} \pm \sigma_{r^{(\alpha)}}}$} \\
      \toprule
    \textbf{Data Combination} & \textbf{Direction} & \textbf{$n_t=0$} & \textbf{$n_t=-1$} \\
    \midrule
    \multirow{5}{*}{ACT 90\,GHz $\times$ CIB 353\,GHz}
      & $e^{(0)}$ & $-11 \pm 215$ & $0 \pm 5$ \\
      & $e^{(1)}$ & $156 \pm 173$ & $3 \pm 4$ \\
      & $e^{(2)}$ & $20  \pm 162$ & $0 \pm 3$ \\
      & $e^{(3)}$ & $90  \pm 119$ & $2 \pm 2$ \\
      & $e^{(4)}$ & $-9  \pm 118$ & $0 \pm 2$ \\
    \bottomrule
    \end{tabular}
    \caption{ACT 90 GHz $\times$ CIB 353 GHz constraints on the tensor-to-scalar ratio $r$, decomposed
    into the five orthonormal basis directions of the $\ell=2$ $B$-mode quadrupole. Values are
    $\hat{r}^{(\alpha)} \pm \sigma_{r^{(\alpha)}}$ for each direction, given for fiducial model of $n_t=0$ and $n_t=-1$.}
    \label{tab:r_constraints_short}
  \end{table}

\section{Conclusions}\label{sec:conclusions}

In this paper, we have presented the first constraints on the remote quadrupole signal using CMB polarization data from \textit{Planck} and ACT cross-correlated with galaxy and CIB tracers from unWISE and \textit{Planck} CIB maps. We developed a pixel-space quadratic estimator for the remote quadrupole field, and showed that the bispectrum amplitude estimator provides a unified framework for constraining the optical depth bias $b_q$, the optical depth to reionization $\tau_{\rm rei}$, and the tensor-to-scalar ratio $r$. 

All measured amplitudes are consistent with zero within $1$--$2\sigma$, indicating no statistically significant detection of the pSZ signal with current data, as expected from our forecasts. Our benchmark result, ACT 90 GHz $\times$ CIB 353 GHz, yields $\hat{b}_q=1.02 \pm 2.64$, $\hat{\tau}_{\rm rei} = -0.01 \pm 0.14$, $\sigma_r \sim 150$ for $n_t = 0$ and $\sigma_r \sim 3$ for $n_t = -1$. 
Across all combinations, CIB 353 GHz consistently outperforms unWISE as a tracer, delivering $\sim 1.5$--$2 \times$ tighter constraints. This is because the CIB redshift distribution peaks at a higher redshift than unWISE where the remote quadrupole is larger, as seen in figure~(\ref{fig:consrealsl2to4}).

We additionally checked the sensitivity of the pSZ reconstruction to foregrounds. We found that residual galactic dust contamination in the CIB correlates with dust in the higher frequency CMB maps. Either using low frequency CMB channels or component-separated CMB maps avoids this systematic. This suggests that it will be advantageous for pSZ to have CMB surveys with many frequency channels to better clean the map.

An important advantage of pSZ over kSZ is the insensitivity of the reconstruction noise to baryonic physics. Because the dominant signal-to-noise in the pSZ bispectrum comes from angular scales $\ell \sim \mathcal{O}(100)$, which probe the two-halo regime of the LSS--optical depth cross-correlation, the estimator is largely insensitive to the distribution of gas within individual halos. This means that, unlike kSZ, the pSZ signal is not suppressed by baryonic feedback, and the optical depth bias remains close to unity even in scenarios with strong feedback. 

Looking ahead, SO and in particular LiteBIRD, will improve the constraints presented above at an interesting level, potentially leading to a statistically significant detection. Our signal-to-noise forecasts show that LiteBIRD $\times$ CIB 353 GHz is the first combination projected to achieve $\sigma_{b_q} < 1$, reaching $\sigma_{b_q} = 0.769$. PSZ also has the ability to rule out models with large $\tau_{\rm rei}$, as recently proposed to address various cosmological tensions~\cite{Sailer:2025lxj,Sullivan:2026tas}. With LiteBIRD $\times$ CIB 353 GHz we can achieve $\sigma_{\tau_{\rm rei}} = 0.041$. We can also improve our constraining power by using new tracers. For instance, the Spectro-Photometer for the History of the Universe, Epoch of Reionization and Ices Explorer (SPHEREx) survey will provide all-sky galaxy catalog with more than 1 billion galaxies, which will substantially reduce the shot noise and extend redshift coverage relative to unWISE \cite{SPHEREx:2018xfm}. CCAT-prime is also expected to improve measurements of the CIB and overlaps well with SO's footprint \cite{CCAT-Prime_2023}.

Beyond $b_q$ and $\tau_{\rm rei}$, the pSZ effect provides a novel and independent probe of primordial gravitational waves through the remote $B$-mode quadrupole; pSZ can become a complementary probe to primary CMB $B$-mode experiments at next-generation sensitivity levels~\cite{CMB-HD:2022bsz}. Additionally, a measurement of the remote quadrupole will allow us to test $\Lambda$CDM. As it probes inside our past light cone, the remote quadrupole field can probe the homogeneity of the Universe, and since it measures the largest observable scales it can help investigate if the low-$\ell$ CMB anomalies are physical or a statistical fluke~\cite{Cayuso:2019hen}. 

It will also be interesting to assess the sensitivity of current data to the kpSZ effect, and forecast constraining power with SO and LiteBIRD . Another important direction is comparing cluster stacking and optical depth reconstruction with pSZ quadrupole reconstruction. In analogy with kSZ~\cite{Smith:2018bpn} and as shown in Sec.~\ref{sec:bqtaurestimators}, these approaches should be equivalent. While theoretically equivalent, comparing results in practice would provide a valuable consistency check on the reconstruction methodology and help identify any sensitivity to observational systematics. 

The current work demonstrates that existing data can achieve the expected statistical error bar on measurements of the pSZ effect. We can therefore hope to significantly improve on these results with the dramatic gains in sensitivity expected with future CMB experiments. 

\section*{Acknowledgements}
    We thank A. Krolewski for access to unWISE data and W. Coulton for access to the ACT $B$-mode NILC map. We thank M. Alvarez, N. Battaglia, A. Challinor, L. Colombo, G. Fabbian, J. C. Hill, G. Holder, S. Hotinli, F. McCarthy, D. Meerburg, K. Smith, and A. Van Engelen for useful discussions. MCJ is supported by the National Science and Engineering Research Council through a Discovery grant. JK acknowledges support from the Natural Sciences and Engineering Research Council of Canada (NSERC) through the Vanier Canada Graduate Scholarship. EP was supported by NASA grant 80NSSC23K0747 during the preparation of this work. This research was supported in part by grant NSF PHY-2309135 to the Kavli Institute for Theoretical Physics (KITP). This research was supported in part by Perimeter Institute for Theoretical Physics. Research at Perimeter Institute is supported by the Government of Canada through the Department of Innovation, Science and Economic Development Canada and by the Province of Ontario through the Ministry of Colleges and Universities. This project made use of the software tools {\sc ReCCO} ~\cite{Cayuso:2021ljq}, {\sc CAMB}~\cite{Lewis:1999bs,Howlett:2012mh}, {\sc NaMaster}~\cite{Alonso:2018jzx}, {\sc Numpy}~\cite{numpy}, {\sc healpy}~\cite{healpy}, {\sc Matplotlib}~\cite{matplotlib}, and {\sc SciPy}~\cite{Scipy}. Some of the calculations were performed on the Symmetry computing cluster at Perimeter Institute.

\appendix

\section{Remote quadrupole field}\label{app:quadrupolefield}

The contributions to the CMB temperature quadrupole introduced in Eq.~(\ref{eq:quadcontributions}) are given by:
\begin{eqnarray}
    \Theta_{\rm SW}(\hat{n},\chi) &=& 2\left(D_\Psi(\chi_{\rm dec}) - \frac{3}{2}\right)\Psi_i(\chi_e\hat{n}_e+\Delta\chi(a_{\rm dec})\hat{n}) \\
    \Theta_{\rm ISW}(\hat{n},\chi) &=& 2 \int_{a_{\rm dec}}^{a_e} \frac{dD_\Psi}{da} \Psi_i(\chi_e\hat{n}_e+\Delta\chi(a)\hat{n}) da \\
    \Theta_{\rm Doppler}(\hat{n},\chi) &=& \hat{n}\cdot \left[ D_v(\chi_{\rm dec}) \nabla \Psi_i(\chi_e\hat{n}_e+\Delta\chi(a_{\rm dec})\hat{n}) - D_v(\chi_e)\nabla \Psi_i(\chi_e\hat{n}_e+\Delta\chi(a_e)\hat{n}) \right].
\end{eqnarray}
Here, $\chi_{\rm dec}=-\int_1^{a_{\rm dec}}da[H(a)a^2]^{-1}$,  $\Delta\chi(a)=-\int_{a_e}^a da'[H(a')a'^2]^{-1}$, $\Psi_i$ is the primordial Newtonian gravitational potential, $D_\Psi(a)$ is the growth function, and $D_v(a)$ is the velocity growth function.

The scalar $a_{\ell m}$'s are defined as
\begin{equation}
    a_{\ell m}^{X}(\chi) = \int \frac{d^3k}{(2\pi)^3} \ \Delta_\ell^{X}(k,\chi) \ \mathcal{R} (\mathbf{k})\  Y_{\ell m}^*(\mathbf{\hat{k}})
    \end{equation}
where $X$=\{$T$, $E$, $q_E$\} and $\mathcal{R}(\mathbf{k})$ is the primordial curvature perturbation. Then the tensor $a_{\ell m}$'s for $X=$\{$E$, $q_E$\} and $Y$=\{$B$, $q_B$\} are
\begin{align}
    a_{\ell m}^{X;h}(\chi) &= \int \frac{d^3k}{(2\pi)^3} \Delta_{\ell}^{X;h}(k,\chi) \nonumber\\
    &\times \{h_{i,+}(k)
    [{}_2Y_{\ell m}^*(\hat{\mathbf{k}})+ {}_{-2}Y_{\ell m}^*(\hat{\mathbf{k}})] +i
    h_{i,\times}(k)[{}_2Y_{\ell m}^*(\hat{\mathbf{k}})- {}_{-2}Y_{\ell m}^*(\hat{\mathbf{k}})]\} \\
    a_{\ell m}^{Y;h}(\chi) &= \int \frac{d^3k}{(2\pi)^3} \Delta_{\ell}^{Y;h}(k,\chi) \nonumber\\
    &\times \{-h_{i,+}(k) [{}_2Y_{\ell m}^*(\hat{\mathbf{k}})- {}_{-2}Y_{\ell m}^*(\hat{\mathbf{k}})]-i
    h_{i,\times}(k)[{}_2Y_{\ell m}^*(\hat{\mathbf{k}})+ {}_{-2}Y_{\ell m}^*(\hat{\mathbf{k}})]\} \ .
\end{align}
where $h_{i,(+,\times)}(k)$ is the primordial tensor perturbation.

The scalar quadrupole field transfer function is defined as 
\begin{equation}
\begin{split}
    \Delta_\ell^{q_E}(k,\chi)&=-5i^\ell \sqrt{\frac{3}{8}}\sqrt{\frac{(\ell+2)!}{(\ell-2)!}} \frac{j_\ell(k\chi)}{(k\chi)^2} \Theta_2 (\chi,k)
\end{split}
\end{equation}
which is zero for $\ell=0,1$ and $\Theta_2 (\chi,k)$ is 
\begin{eqnarray}
    \Theta_2 (k, \chi) = \int_0^{\eta_0-\chi} d\eta \left[ \mathcal{S}_{\rm SW} (k, \eta) + \mathcal{S}_{\rm ISW} (k, \eta) + \mathcal{S}_{\rm Doppler} (k, \eta) \right] j_2 \left[k(\eta_0 - \chi - \eta )\right] \ .
\end{eqnarray}
We use \texttt{CAMB} to calculate the source functions $\mathcal{S}_{\rm SW} (k, \eta)$, $ \mathcal{S}_{\rm ISW} (k, \eta)$, and $\mathcal{S}_{\rm Doppler} (k, \eta)$.

The tensor quadrupole field contribution was first considered in Ref.~\cite{Alizadeh_2012} and the below results follow their paper as well as Ref.~\cite{Deutsch:2017ybc}. The transfer functions are
\begin{eqnarray}
    \Delta_{\ell}^{q_E;h}(k,\chi) &=& 5 i^\ell E_{\ell}(k,\chi)\mathcal{I}_T^q(k,\chi)\\
\Delta_{\ell}^{q_B;h}(k,\chi) &=& 5 i^\ell F_{\ell}(k,\chi)\mathcal{I}_T^q(k,\chi) \ .
\end{eqnarray}  
The $E_\ell(k,\chi), F_\ell(k,\chi)$ account for projection effects and are defined as
\begin{align}
    E_\ell(k,\chi) &\equiv -\frac{1}{4k^2}\frac{d^2j_\ell(k\chi)}{d\chi^2} - \frac{1}{k^2\chi}\frac{dj_\ell(k\chi)}{d\chi} +j_\ell(k\chi)\left(\frac{1}{4}-\frac{1}{2(k\chi)^2}\right)\\
    F_\ell(k,\chi) &\equiv \frac{1}{2} \left( \frac{2j_\ell(k\chi)}{k\chi} + \frac{1}{k}\frac{d}{d\chi}j_\ell(k\chi)\right).
\end{align}
Then the kernel, $\mathcal{I}_T^q(k,\chi)$, is
\begin{align}
\mathcal{I}_T^q(k,\chi)&=2\pi\sqrt{6}   \int_{a_e}^{a_{\rm dec}} da \frac{dD^T(k,a)}{da} \frac{j_2(k\Delta\chi(a))}{(k\Delta\chi(a))^2} 
\end{align}
where $D^T$ is the tensor growth function 
\begin{equation}
    h_{(+,\times)}(k,a)\equiv D^T(k,a)h_{i,(+,\times)}(k) \ .
\end{equation}

Finally, the $T$, $E$, and $B$ transfer functions are given by
\begin{eqnarray}
    \Delta_\ell^T(k,\chi) &=& \Delta_\ell^{T;\texttt{CAMB}}(k,\chi) \\
    \Delta_\ell^E(k,\chi) &=&  \sqrt{\ell(\ell^2+1)(\ell+2)} \Delta_\ell^{E;\texttt{CAMB}}(k,\chi) \\
    \Delta_\ell^B(k,\chi) &=& \Delta_\ell^{B;\texttt{CAMB}}(k,\chi)\\
    \Delta_\ell^{T;h}(k,\chi) &=& \frac{\pi^3}{3} \sqrt{\ell(\ell^2+1)(\ell+2)} \Delta_\ell^{T;h;\texttt{CAMB}}(k,\chi) \\
    \Delta_\ell^{E;h}(k,\chi) &=& \frac{\pi^3}{3}  \Delta_\ell^{E;h;\texttt{CAMB}}(k,\chi) \\
    \Delta_\ell^{B;h}(k,\chi) &=& \frac{\pi^3}{3} \Delta_\ell^{B;h;\texttt{CAMB}}(k,\chi) \ .
\end{eqnarray}
The $\Delta_\ell^{X;\texttt{CAMB}}(k,\chi)$ are the transfer functions given by \texttt{CAMB}. The $\ell$-dependent factors are used to match the power spectra calculated using \texttt{CAMB}'s transfer functions (using \texttt{get\_cmb\_transfer\_data}), to the power spectra outputted directly by \texttt{CAMB}'s \texttt{get\_cmb\_power\_spectra}. The $\pi^3/3$ terms are to match with how we defined the tensor power spectra in Eq.~(\ref{eq:tensor_spectra}).

\section{Constrained realizations of the remote quadrupole field} \label{app:constrained}

Consider the joint Gaussian distribution over the locally observed CMB temperature and $E$-mode quadrupoles $(T_{\ell m}^0,\, E_{\ell m}^0)$ and the remote quadrupole $q_{\ell m}^E(\chi_i)$ across a set of $i=1\ldots N$ redshift slices. Statistical isotropy implies all fields are uncorrelated across $\{\ell, m\}$, so the full likelihood is
\begin{equation}
  P\!\left(T_{\ell m}^0,\, E_{\ell m}^0,\, q_{\ell m}^E(\chi_i)\right) \propto
  \exp\!\left[-\frac{1}{2}\, \mathbf{a}_{\ell m}^T \, C_\ell^{-1}\, \mathbf{a}_{\ell m}\right],
\end{equation}
where $\mathbf{a}_{\ell m} = (T_{\ell m}^0,\, E_{\ell m}^0,\, q_{\ell m}^E(\chi_1), \ldots, q_{\ell m}^E(\chi_N))^T$ and the covariance matrix is
\begin{equation}
  C_\ell = \begin{pmatrix} C_\ell^{TT} & C_\ell^{TE} & C_\ell^{Tq_E}(\chi_1) & \cdots \\
  C_\ell^{TE} & C_\ell^{EE} & C_\ell^{Eq_E}(\chi_1) & \cdots \\
  C_\ell^{Tq_E}(\chi_1) & C_\ell^{Eq_E}(\chi_1) & C_\ell^{q_Eq_E}(\chi_1,\chi_1) & \cdots \\
  \vdots & \vdots & \vdots & \ddots \end{pmatrix}.
\end{equation}
We would like to make a realization of the remote quadrupole field conditioned on the locally observed $(T_{\ell m}^0,\, E_{\ell m}^0)$. Following
e.g.\ Ref.~\cite{deOliveira-Costa:2006sst} or Ref.~\cite{bucher2012filling}, the conditional distribution is a multivariate Gaussian,
\begin{equation}
  P\!\left(q_{\ell m}^E(\chi_i)\,\Big|\, T_{\ell m}^0,\, E_{\ell m}^0\right) \propto
  \exp\!\left[-\frac{1}{2}\!\sum_{i,j}\!\left(q_{\ell m}^E(\chi_i) -
  \tilde{q}_{\ell m}^E(\chi_i)\right)^T
  \left[C_\ell^{\delta q_E \delta q_E}\right]^{-1}\!\!(\chi_i,\chi_j)
  \left(q_{\ell m}^E(\chi_j) - \tilde{q}_{\ell m}^E(\chi_j)\right)\right],
\end{equation}
where the maximum-likelihood template at each slice is
\begin{equation}
  \tilde{q}_{\ell m}^E(\chi_i) = \begin{pmatrix} C_\ell^{Tq_E}(\chi_i) &
  C_\ell^{Eq_E}(\chi_i) \end{pmatrix} \begin{pmatrix} C_\ell^{TT} & C_\ell^{TE} \\
  C_\ell^{TE} & C_\ell^{EE} \end{pmatrix}^{-1} \begin{pmatrix} T_{\ell m}^0 \\
  E_{\ell m}^0 \end{pmatrix},
\end{equation}
and the covariance is
\begin{equation}
  C_\ell^{\delta q_E \delta q_E}(\chi_i,\chi_j) = C_\ell^{q_Eq_E}(\chi_i,\chi_j)
  - C_\ell^{\tilde{q}_E\tilde{q}_E}(\chi_i,\chi_j).
\end{equation}
To make a constrained realization we add random draws $\delta q_{\ell m}^E(\chi_i)$ from this multivariate Gaussian to the maximum-likelihood template,
\begin{equation}
  q_{\ell m}^{E}(\chi_i) = \tilde{q}_{\ell m}^E(\chi_i) +
  \delta q_{\ell m}^E(\chi_i).
\end{equation}

\section{CMB noise modeling} \label{app:noisemodel}

We model the CMB noise only for the forecasts in Sec.~\ref{sec:snr}. The CMB instrumental noise power spectrum for white noise is $$N^{\rm inst}_\ell = \frac{\Delta_{\rm inst}^2}{B_\ell^2}$$ with $\Delta_{\rm inst}$ the white-noise level and $B_\ell$ a Gaussian beam. For ground-based telescopes we additionally include a large-scale atmospheric noise component, 
$$N^{\rm inst}_\ell = \left[1 + \left(\frac{\ell}{\ell_{\rm knee}}\right)^{-\alpha}\right]\frac{\Delta_{\rm inst}^2}{B_\ell^2}\,.$$ 
The experimental parameters are summarized in table~(\ref{tab:noise_params}). For ACT, since we use ACT+\textit{Planck} coadds, we take the ACT 90 GHz noise model and inverse variance coadd it with the \textit{Planck} 100 and 143 GHz noise:
$$N_\ell^{\rm inst, coadd} = \frac{1}{\frac{1}{N_\ell^{\rm inst,90}}+\frac{1}{(N_\ell^{\rm inst,100}+N_\ell^{\rm inst,143})}}\,.$$ We additionally account for any foreground contamination 
in the maps by taking the spectra of the data $C_\ell^{\rm data}$ and subtracting off the theory spectra and noise model $C_\ell^{\rm th}+N^{\rm inst}_\ell$:
$$N_\ell^{\rm FG} = C_\ell^{\rm data} - C_\ell^{\rm th} - N^{\rm inst}_\ell$$
so the total noise is:
$$N_\ell = N_\ell^{\rm FG} + N_\ell^{\rm inst}. $$
We assume the same foreground contamination for SO as ACT, and since LiteBIRD has many frequency channels we assume that it is cleaned.

  \begin{table}[h]  
  \centering 
  \renewcommand{\arraystretch}{1.3}
  \begin{tabular}{l|c|c|c|c|c} 
  \toprule                                
  \textbf{Experiment} & \textbf{Frequency} & \textbf{FWHM [arcmin]} & $\bm{\Delta_{\rm inst}}$ \textbf{[$\bm{\mu}$K$\cdot$arcmin]} & $\bm{\ell_{\rm knee}}$ & $\bm{\alpha}$ \\
  \midrule   
  \textit{Planck}   & 100 GHz & 9.7  & 83.4 & -- & --   \\
  \textit{Planck}   & 143 GHz & 7.2  & 62.2 & -- & --   \\
  ACT (not coadded)               & 90 GHz  & 5.0  & 21.0 & 300 & 3  \\ 
  SO SAT            & 145 GHz & 30   & 3.7 & 50 & 2.5 \\  
  SO LAT            & 93 GHz  & 2.2  & 11.3 & 50 & 2.5 \\
  LiteBIRD          & --       & 30   & 2.2 & -- & --  \\   
  \bottomrule 
  \end{tabular} 
  \caption{Experimental noise parameters used in the signal-to-noise forecasts. The \textit{Planck} values are from Ref.~\cite{Bloch_2024}, ACT from Ref.~\cite{ACT:2025xdm}, SO from Ref.~\cite{Ade2019,SimonsObservatory:2025wwn}, and LiteBIRD from Ref.~\cite{LiteBIRD:2022cnt}. The ACT beam follows what we use in the reconstruction (explained in Sec.~\ref{sec:recpipeline}) not the true ACT beam.}            
  \label{tab:noise_params}                 
  \end{table}

\section{Reconstruction pipeline details}

The masks used throughout this work are defined in table~(\ref{tab:specmasks}). The masking choices used to calculate the $E$- and $B$-mode power spectra from the data are given in table~(\ref{tab:EEBBspectramasks}). We low and high pass filter the input tracer and CMB maps in the reconstruction pipeline, with the $\ell_{\rm min}$ and $\ell_{\rm max}$ choices listed in table~(\ref{tab:filterlowhigh}). Finally, in table~(\ref{tab:recmasks}) we provide the masks applied to the reconstructed remote quadrupole maps.

\begin{table}[h]
    \centering
    \renewcommand{\arraystretch}{1.3} 
    \begin{tabularx}{\textwidth}{>{\centering\arraybackslash}m{3cm}|>{\raggedright\arraybackslash}X|>{\centering\arraybackslash}m{1.5cm}}
    \toprule
        \textbf{Mask} & \centering \textbf{Definition} & $\bm{f_{\rm sky}}$ \\
        \midrule
        unWISE\_mask & Galactic plane cut retaining 70\% and masking stars, planetary nebulae, and bright sources & 59\% \\
        \hline
        CIB\_mask\_apo & CIB coverage mask apodized with $C_1$ kernel, $2^\circ$ scale & 61\% \\
        \hline
        CIB\_bright\_src & CIB frequency-specific bright-source masks (thresholds: $<$~0.035/0.1/0.2 at 353/545/857 GHz) & 99\% \\
        \hline
        pt\_src\_T\_freq\_mask & \textit{Planck} temperature point source masks (100–353 GHz) & $\sim$100\% \\
        \hline
        pt\_src\_P\_freq\_mask & \textit{Planck} polarization point source masks (100–353 GHz) & $\sim$100\% \\
        \hline
        pt\_src\_T\_joint\_mask & Joint \textit{Planck} temperature point source mask (100–353 GHz combined) & 99.8\% \\
        \hline
        pt\_src\_P\_joint\_mask & Joint \textit{Planck} polarization point source mask (100–353 GHz combined) & 99.8\% \\
        \hline
        galactic\_mask & Galactic plane mask & 60\% \\
        \hline
        galactic\_mask\_apo & Apodized galactic plane mask & 52\% \\
        \hline
        SMICA\_apo\_mask & \texttt{SMICA} confidence mask apodized with $C_1$ kernel, $2^\circ$ scale & 69\% \\
        \hline
        act\_mask\_apo & ACT footprint mask apodized with $C_1$ kernel, $2^\circ$ scale & 49\% \\
        \bottomrule
    \end{tabularx}
    \caption{Definitions of masks used on the galaxy, CIB, and CMB maps.}
    \label{tab:specmasks}
\end{table}

 \begin{table}[]
 \renewcommand{\arraystretch}{1.3} 
     \centering
     \begin{tabular}{c|>{\raggedright\arraybackslash}m{8.5cm}|c}
     \toprule
        \textbf{CMB Map} & \textbf{Mask} & $\bm{f_{\rm sky}}$ \\
        \midrule
         \textit{Planck} 100, 143, 217, 353 GHz & pt\_src\_P\_freq\_mask $\times$ galactic\_mask\_apo & 52\% \\
         \hline
         \texttt{SMICA} & SMICA\_apo\_mask & 69\% \\
         \hline
         ACT 90, 150, 220 GHz and NILC & act\_mask\_apo $\times$ CIB\_mask\_apo $\times$ pt\_src\_P\_joint\_mask & 29\% \\
         \bottomrule
     \end{tabular}
     \caption{Mask choices for calculating the $C_\ell^{EE}$ and $C_\ell^{BB}$ power spectra. The masks are explained in table~(\ref{tab:specmasks}).}
     \label{tab:EEBBspectramasks}
 \end{table}

 \begin{table}[h]
    \centering
    \renewcommand{\arraystretch}{1.3} 
    \begin{tabular}{>{\centering\arraybackslash}m{3.5cm}|>{\centering\arraybackslash}m{2.5cm}|>{\centering\arraybackslash}m{1.5cm}|>{\centering\arraybackslash}m{1.5cm}}
        \toprule
        \textbf{CMB Map} & \textbf{Tracer} & \textbf{$\bm{\ell_{\rm min}}$} & \textbf{$\bm{\ell_{\rm max}}$} \\
        \midrule 
        \multirow{4}{*}{ACT 90~GHz}  & unWISE      & 25  & 4000 \\
                                     & CIB 353~GHz & 25  & 4000 \\
                                     & CIB 545~GHz & 25  & 4000 \\
                                     & CIB 857~GHz & 100 & 4000 \\
        \hline
        \multirow{4}{*}{ACT 150~GHz} & unWISE      & 25  & 4000 \\
                                     & CIB 353~GHz & 25  & 4000 \\
                                     & CIB 545~GHz &  25  & 4000 \\
                                     & CIB 857~GHz &  100 & 4000 \\
        \hline
        \multirow{4}{*}{ACT 220~GHz} & unWISE      & 25  & 4000 \\
                                     & CIB 353~GHz &  200 & 4000 \\
                                     & CIB 545~GHz &  200 & 4000 \\
                                     & CIB 857~GHz &  200 & 4000 \\
        \hline
        \multirow{4}{*}{ACT NILC}    & unWISE      & 25  & 4000 \\
                                     & CIB 353~GHz & 100 & 4000 \\
                                     & CIB 545~GHz & 100 & 4000 \\
                                     & CIB 857~GHz & 200 & 4000 \\
       \hline
        \multirow{4}{*}{\textit{Planck} 100 GHz}    & unWISE      & 25  & 4000 \\
                                     & CIB 353~GHz & 50 & 4000 \\
                                     & CIB 545~GHz & 50 & 4000 \\
                                     & CIB 857~GHz & 150 & 4000 \\
    \hline
        \multirow{4}{*}{\textit{Planck} 143 GHz}    & unWISE      & 100  & 4000 \\
                                     & CIB 353~GHz & 25 & 4000 \\
                                     & CIB 545~GHz & 25 & 4000 \\
                                     & CIB 857~GHz & 250 & 4000 \\
        \hline
        \multirow{4}{*}{\textit{Planck} 217 GHz}    & unWISE      & 350  & 4000 \\
                                     & CIB 353~GHz & 100 & 4000 \\
                                     & CIB 545~GHz & 100 & 4000 \\
                                     & CIB 857~GHz & 300 & 4000 \\
        \hline
        \multirow{4}{*}{\textit{Planck} 353 GHz}    & unWISE      & 450  & 4000 \\
                                     & CIB 353~GHz & 200 & 4000 \\
                                     & CIB 545~GHz & 250 & 4000 \\
                                     & CIB 857~GHz & 375 & 4000 \\
     \hline
        \multirow{4}{*}{\textit{Planck} \texttt{SMICA}}    & unWISE      & 25  & 4000 \\
                                     & CIB 353~GHz & 25 & 4000 \\
                                     & CIB 545~GHz & 25 & 4000 \\
                                     & CIB 857~GHz & 25 & 4000 \\
        \bottomrule
    \end{tabular}
    \caption{CMB map, tracer, and multipole filter range used for each pSZ quadrupole reconstruction.}
    \label{tab:filterlowhigh}
\end{table}

\begin{table}[]
 \renewcommand{\arraystretch}{1.3} 
     \centering
     \begin{tabular}{>{\centering\arraybackslash}m{3.5cm}|>{\centering\arraybackslash}m{2cm}|>{\centering\arraybackslash}m{8.5cm}|>
     {\centering\arraybackslash}m{1cm}}
        \toprule
        \textbf{CMB Map} & \textbf{Tracer} & \textbf{Mask} & \textbf{$\bm{f_{\rm sky}}$} \\
        \midrule 
        \multirow{2}{*}{\textit{Planck} 100, 143 GHz}  & unWISE      & unWISE\_mask $\times$ pt\_src\_P\_freq\_mask &  59\% \\
                                     & CIB & CIB\_mask\_apo $\times$ CIB\_bright\_src $\times$ pt\_src\_P\_freq\_mask & 60\%   \\
    \hline
    \multirow{2}{*}{\textit{Planck} 217, 353 GHz}  & unWISE      & unWISE\_mask $\times$ pt\_src\_P\_freq\_mask  $\times$ galactic\_mask & 51\% \\
                                     & CIB & CIB\_mask\_apo $\times$ CIB\_bright\_src $\times$ pt\_src\_P\_freq\_mask &  60\% \\
    \hline
    \multirow{2}{*}{\textit{Planck} \texttt{SMICA}}  & unWISE      & unWISE\_mask & 59\%  \\
                                     & CIB & SMICA\_apo\_mask $\times$ CIB\_mask\_apo $\times$ CIB\_bright\_src  & 54\%\\
    \hline
        \multirow{2}{*}{ACT}  & unWISE      & act\_mask\_apo $\times$ unWISE\_mask $\times$ pt\_src\_P\_joint\_mask & 27\%  \\
                                     & CIB &  act\_mask\_apo $\times$ CIB\_mask\_apo $\times$ pt\_src\_T\_joint\_mask $\times$ pt\_src\_P\_joint\_mask & 28\% \\
     \bottomrule
     \end{tabular}
     \caption{Masks applied to the reconstructed quadrupole maps. The masks are explained in table~(\ref{tab:specmasks}).}
     \label{tab:recmasks}
 \end{table}

\section{Remote quadrupole reconstruction with \textit{Planck} frequency maps} \label{app:QEPlanck}

In this appendix, we present our remote quadrupole reconstruction for \textit{Planck} individual frequency maps. Although the reconstruction noise from an analysis with \textit{Planck} is far higher than with ACT DR6, the availability of high frequency maps from \textit{Planck} allow us to assess the impact of foregrounds and systematics more carefully than possible with publicly available ACT data products. 

\subsection{Pre-processing and spectra}

We perform the remote quadrupole reconstruction for \textit{Planck} 100-353 GHz maps. This follows the procedure outlined in Sec.~\ref{sec:recpipeline}. For the \textit{Planck} maps we pre-compute $C_\ell^{EE}$ and $C_\ell^{BB}$ with the masks in table~(\ref{tab:specmasks}). For the filters, we use the $\ell_{\rm min}$ and $\ell_{\rm max}$ values listed in table~(\ref{tab:filterlowhigh}), with the choice of $\ell_{\rm min}$ tuned to make $N_q$ agree with the empirical estimator variance at low-$\ell$. We then mask the reconstructed remote quadrupole maps using the masks described in table~(\ref{tab:recmasks}).

\begin{figure}
    \centering
    \includegraphics[width=0.8\linewidth]{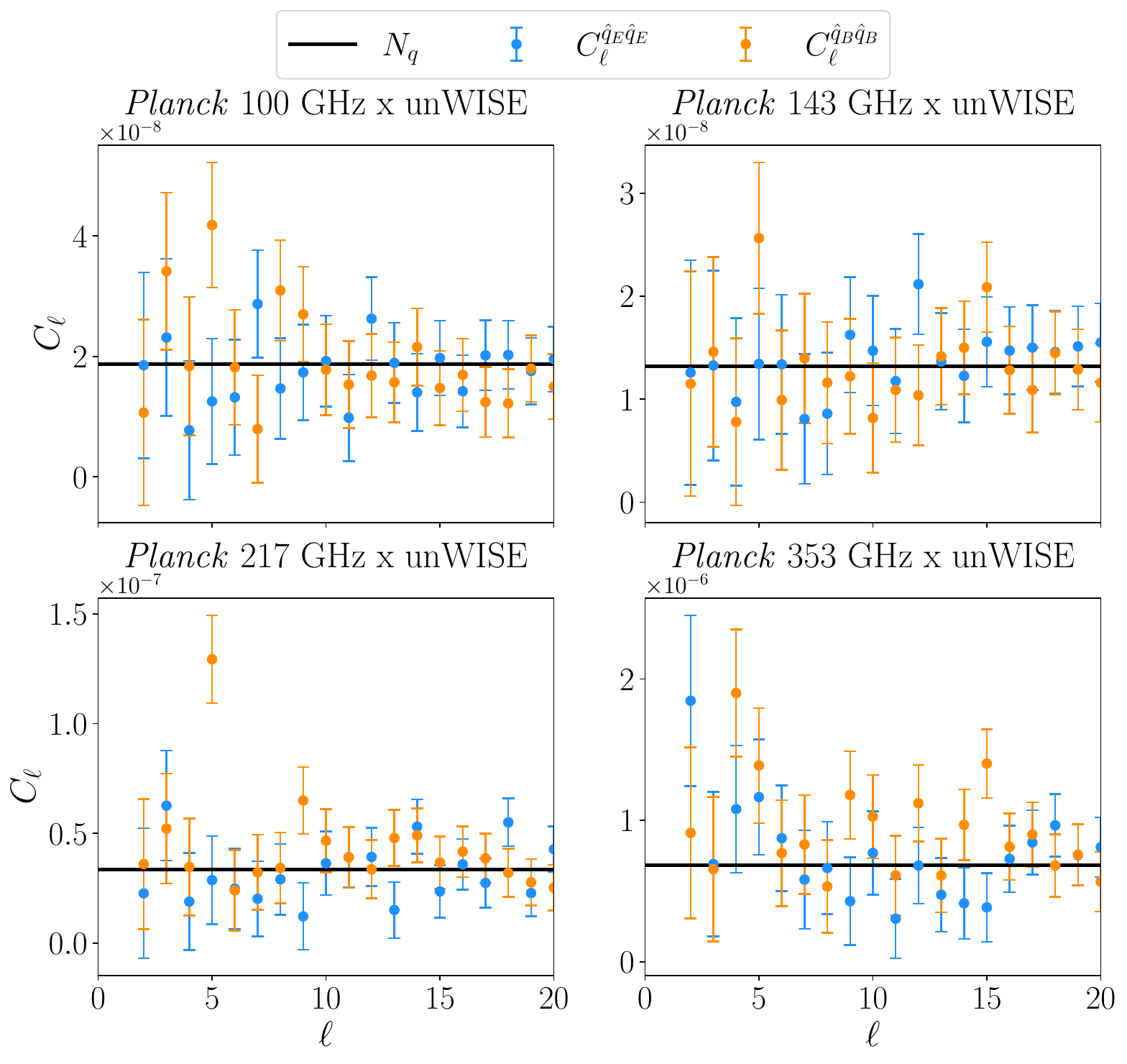}
    \caption{Power spectra of the pSZ remote quadrupole reconstruction, $C_\ell^{\hat{q}_E\hat{q}_E}$ (blue) and $C_\ell^{\hat{q}_B\hat{q}_B}$ (orange), for each \textit{Planck} CMB polarization map cross-correlated with unWISE. The horizontal black line shows the reconstruction noise level $N_q$. Error bars are the 1$\sigma$ cosmic variance error on the cut-sky reconstruction noise     
$\sigma_\ell = \sqrt{2/((2\ell+1)f_{\rm sky})}\, N_q$.}
    \label{fig:Planck_unwise_recspec}
\end{figure}

\begin{figure}
    \centering
    \includegraphics[width=0.8\linewidth]{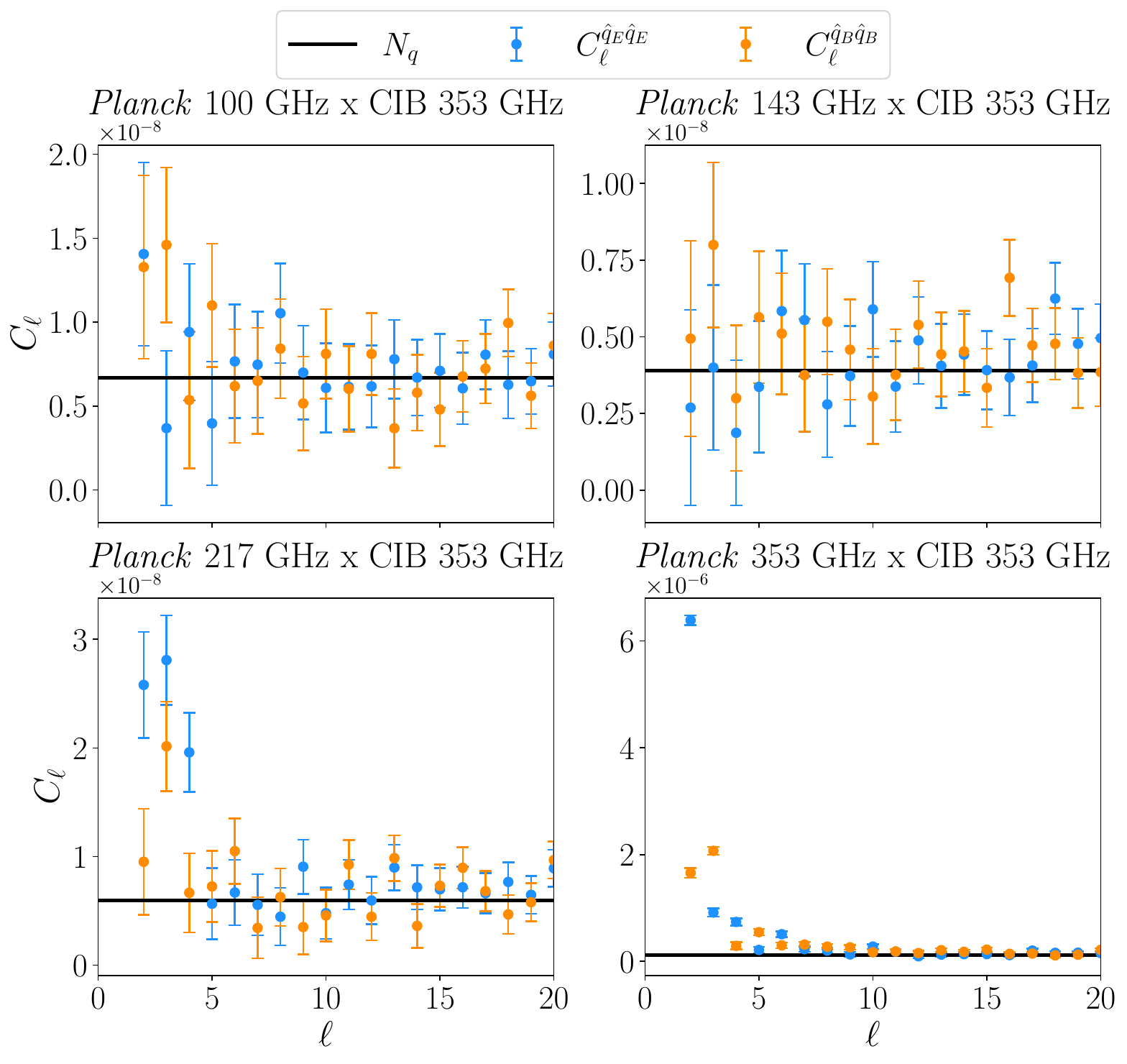}
    \caption{The same as figure~(\ref{fig:Planck_unwise_recspec}), but with \textit{Planck} cross-correlated with CIB 353 GHz.}
    \label{fig:Planck_CIB353_recspec}
\end{figure}

\begin{figure}
    \centering
    \includegraphics[width=0.8\linewidth]{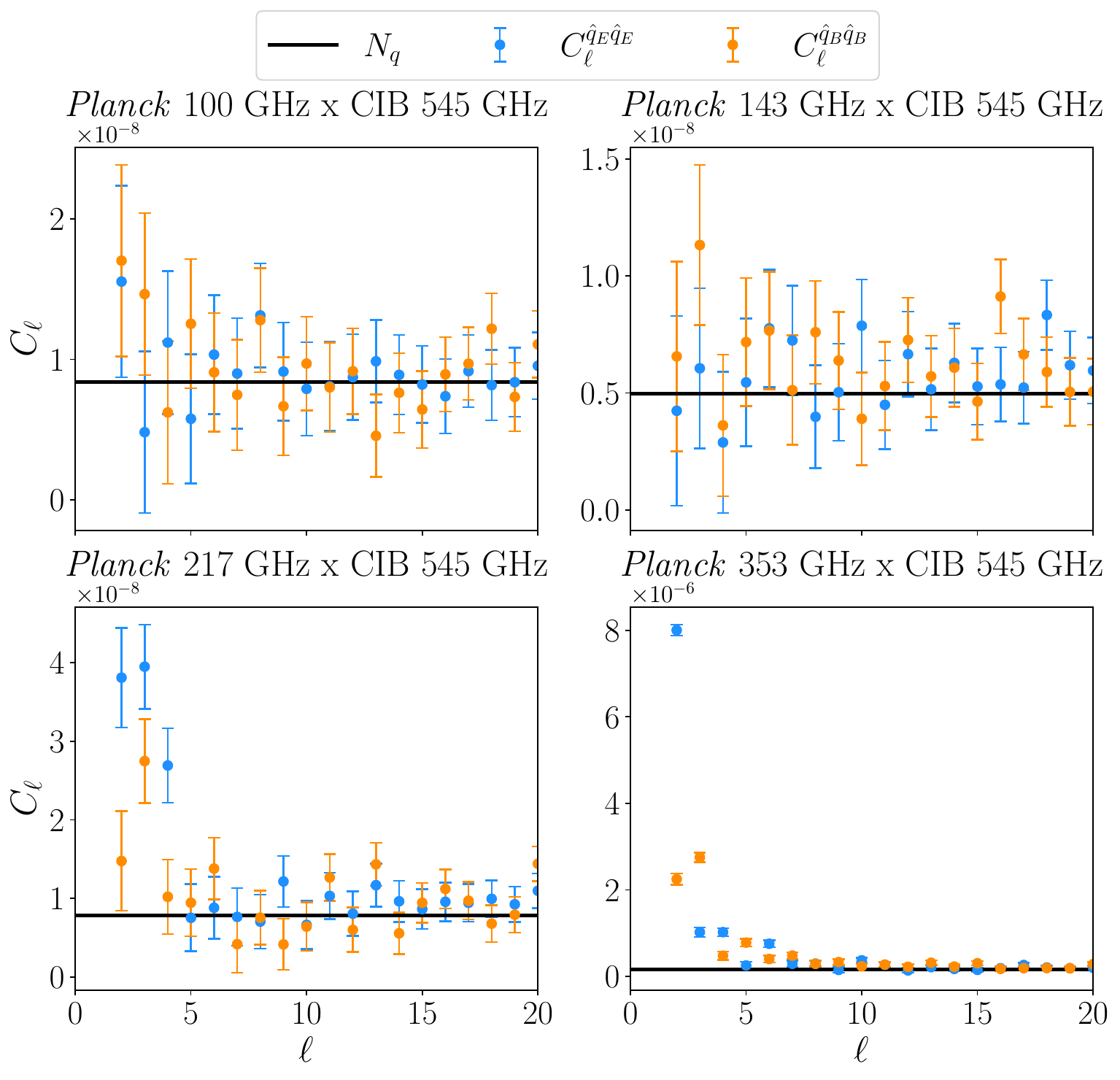}
    \caption{The same as figure~(\ref{fig:Planck_unwise_recspec}), but with \textit{Planck} cross-correlated with CIB 545 GHz.}
    \label{fig:Planck_CIB545_recspec}
\end{figure}

\begin{figure}
    \centering
    \includegraphics[width=0.8\linewidth]{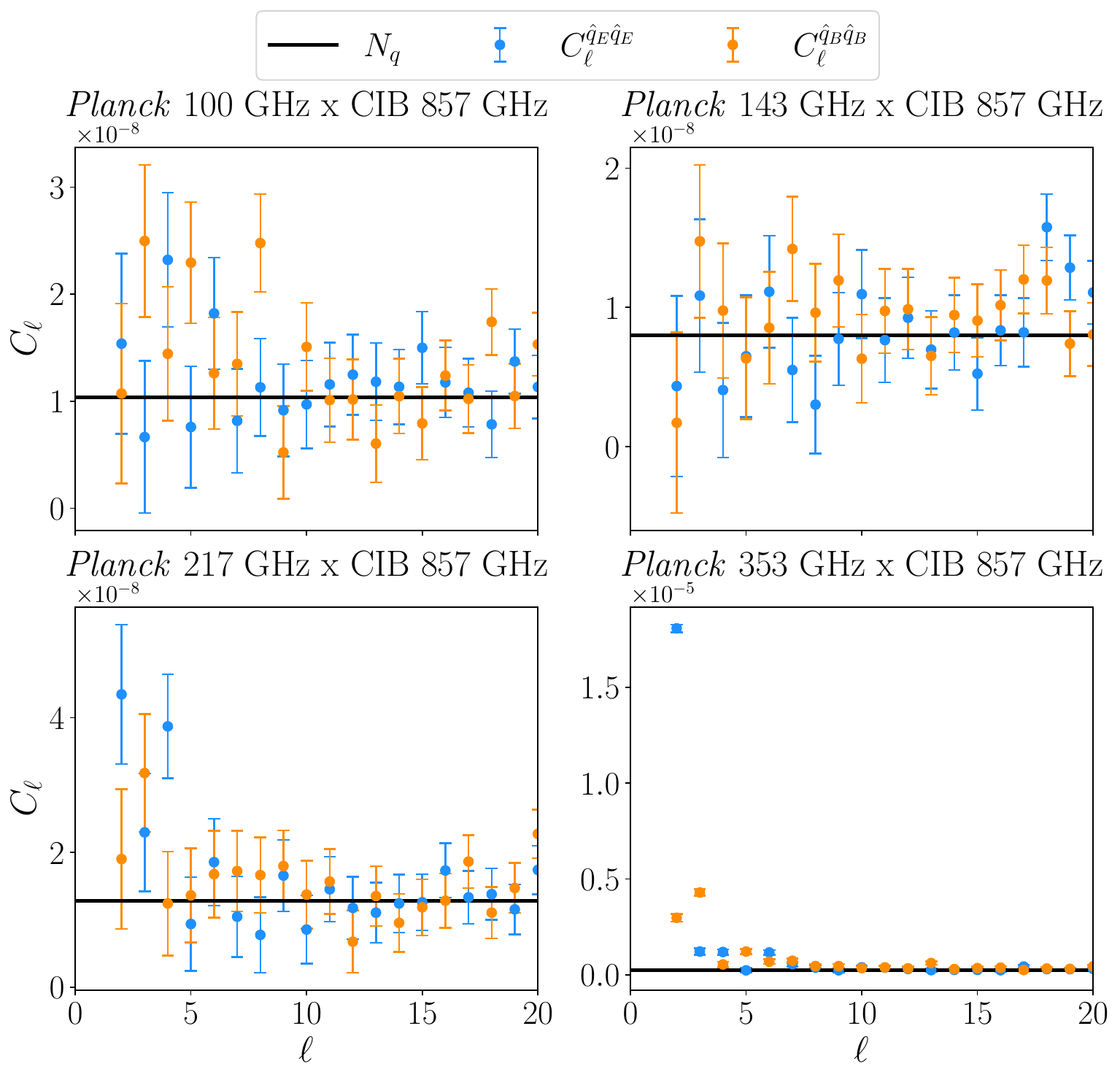}
    \caption{The same as figure~(\ref{fig:Planck_unwise_recspec}), but with \textit{Planck} cross-correlated with CIB 857 GHz.}
    \label{fig:Planck_CIB857_recspec}
\end{figure}

For \textit{Planck} $\times$ unWISE, the estimator spectra are consistent with reconstruction noise, as seen in figure~(\ref{fig:Planck_unwise_recspec}). On the other hand, \textit{Planck} $\times$ CIB (figures~(\ref{fig:Planck_CIB353_recspec}-\ref{fig:Planck_CIB857_recspec})) shows an excess of power at low-$\ell$ in the reconstructions for high frequency \textit{Planck} channels. This low-$\ell$, frequency-dependent, power is a sign of foreground contamination. The CMB polarization maps at low frequency are contaminated by synchotron emission, whereas at high frequency dust dominates. On the other hand, the CIB maps are contaminated by galactic dust. Therefore, it is plausible that we are seeing residual galactic dust in the CIB maps correlating with dust emission in the high-frequency \textit{Planck} channels. We did not perform reconstructions on the LFI \textit{Planck} maps, but we predict that the synchrotron emission in the CMB maps is less correlated with the CIB galactic dust residuals, and it would be interesting to test this low frequency behaviour.

The $N_q$ values for each data combination are listed in table~(\ref{tab:Nrec_planck}). The \textit{Planck} 143 GHz channel has the lowest noise across all tracers. The \textit{Planck} 100 GHz and 217 GHz reconstruction noise levels are close, with the relative performance depending on the tracer field. The \textit{Planck} 353 GHz reconstructions have substantially higher noise, in part due to the higher noise of this frequency channel, and in part due to the much larger $\ell_{\rm min}$ cuts needed in the filtering to account for larger foreground contamination. 

\begin{table}[h]                         
  \centering    
  \renewcommand{\arraystretch}{1.3}  
  \begin{tabular}{l|c|c|c|c}    
      \toprule 
      \multicolumn{5}{c}{$\bm{N^q\times 10^{9}}$} \\
      \toprule 
      \textbf{CMB Map} & \textbf{unWISE} & \textbf{CIB 353 GHz} & \textbf{CIB 545 GHz} & \textbf{CIB 857 GHz} \\  
      \midrule   
      \textit{Planck} 100 GHz & $18.7$ & $6.69$ & $8.38$ & $10.4$ \\   
      \hline   
      \textit{Planck} 143 GHz & $13.2$ & $3.90$ & $4.97$ & $8.01$ \\ 
      \hline      
      \textit{Planck} 217 GHz & $33.6$ & $5.97$ & $7.78$ & $12.8$ \\        
      \hline    
      \textit{Planck} 353 GHz & $685$ & $111$ & $157$ & $243$ \\         
      \hline      
      \textit{Planck} \texttt{SMICA} & $6.71$ & $2.16$ & $2.76$ & $4.45$ \\
      \bottomrule           
  \end{tabular} 
  \caption{Reconstruction noise $N_q$ for each large-scale structure tracer combination with \textit{Planck} 
  CMB polarization maps.}                   \label{tab:Nrec_planck}                   \end{table}

\section{Remote quadrupole reconstruction with ACT maps}\label{app:QEACT}

\begin{figure}
    \centering
    \includegraphics[width=0.8\linewidth]{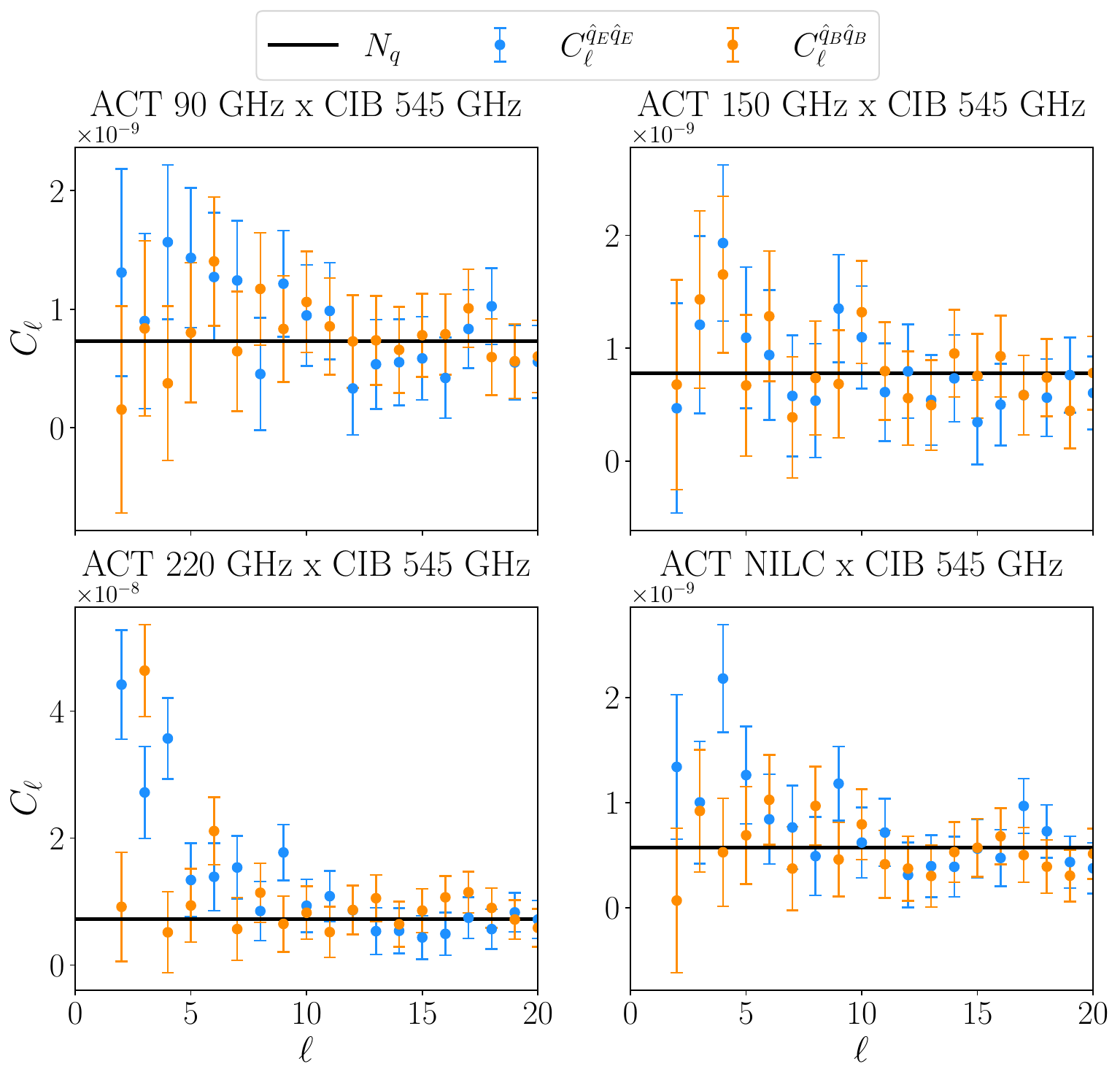}
    \caption{Power spectra of the pSZ remote quadrupole reconstruction, $C_\ell^{\hat{q}_E\hat{q}_E}$ (blue) and $C_\ell^{\hat{q}_B\hat{q}_B}$ (orange), for each ACT CMB polarization map cross-correlated with CIB 545 GHz. The horizontal black line shows the reconstruction noise level $N_q$. Error bars are the 1$\sigma$ cosmic variance error on the cut-sky reconstruction noise     
$\sigma_\ell = \sqrt{2/((2\ell+1)f_{\rm sky})}\, N_q$.}
    \label{fig:ACT_CIB545_recspec}
\end{figure}

\begin{figure}
    \centering
    \includegraphics[width=0.8\linewidth]{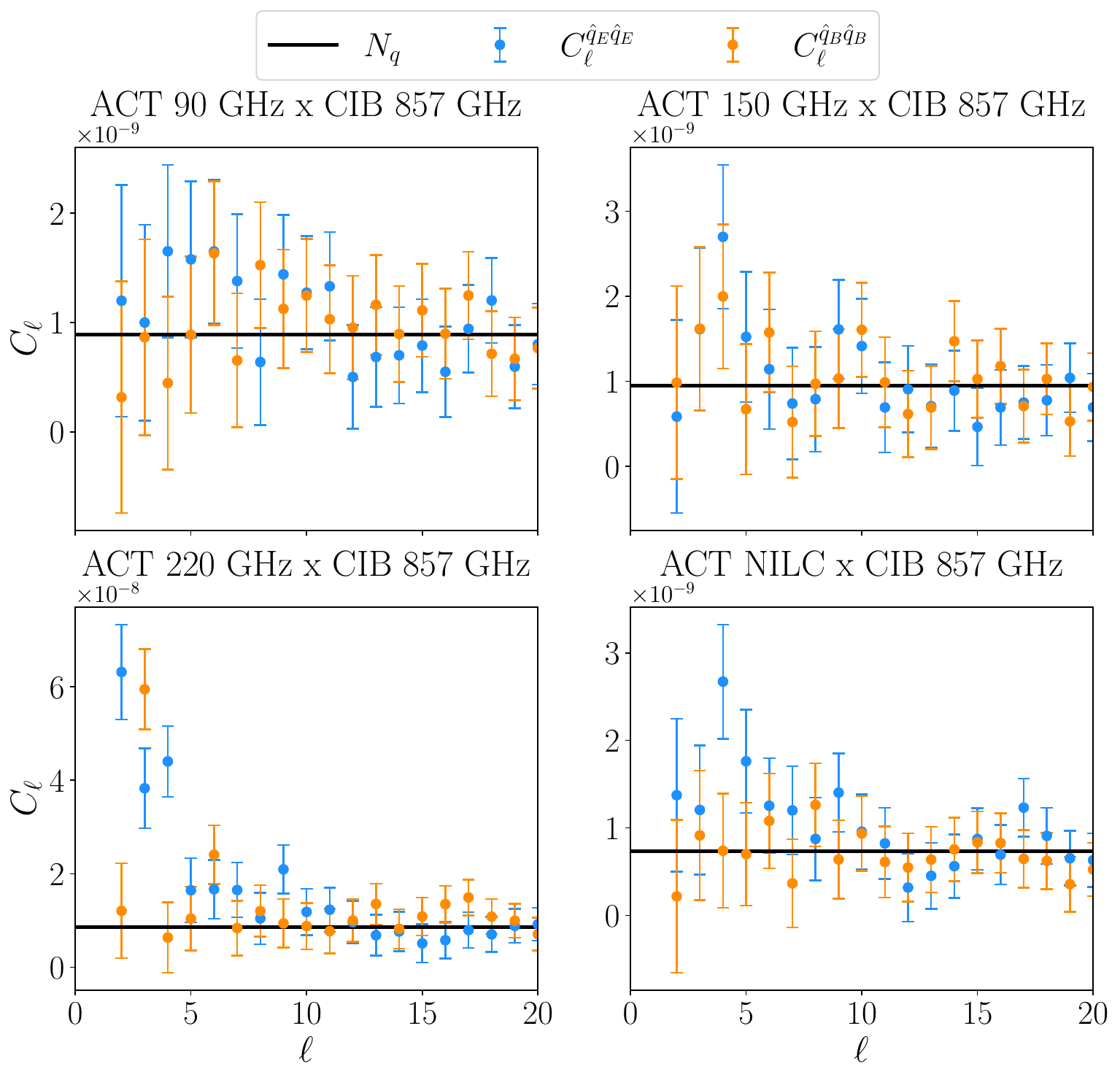}
    \caption{Same as figure~(\ref{fig:ACT_CIB545_recspec}), but with ACT cross-correlated with CIB 857 GHz.}
    \label{fig:ACT_CIB857_recspec}
\end{figure}

Figures~(\ref{fig:ACT_CIB545_recspec}-\ref{fig:ACT_CIB857_recspec}) show the ACT $\times$ CIB 545 GHz and $\times$ CIB 857 GHz remote quadrupole spectra. The ACT 90 GHz channel is the cleanest from dust so the spectra agree with the reconstruction noise. As the ACT frequency increases, the dust foreground increases, which correlates with residual galactic dust in the CIB maps, creating low-$\ell$ excess power. The ACT NILC maps achieve the lowest reconstruction noise, with a minimal amount of foregrounds. 

\section{Comparison of component-separated CMB maps for the quadrupole template} \label{app:compsep}

The pSZ bispectrum estimator requires a template for the remote CMB quadrupole, constructed from the large-scale temperature and polarization of the observed CMB sky. The \textit{Planck} collaboration provides four component-separated full-sky maps: \texttt{Commander}, \texttt{SMICA}, \texttt{SEVEM}, and \texttt{NILC}~\cite{Planck:2018yye}. Each applies a different foreground separation methodology. \texttt{Commander} uses Bayesian parametric fitting in pixel space, \texttt{SMICA} applies spectral matching in harmonic space, \texttt{SEVEM} uses internal template subtraction, and \texttt{NILC} applies a needlet-domain internal linear combination. All four provide inpainted versions in which the masked galactic region is filled via constrained realizations. \textit{Planck} warns against using the inpainted maps for science, but for our application we use scales where they are just Gaussian constrained realizations. We construct the remote quadrupole template (Eq.~(\ref{eq:qE_template})) from each inpainted map independently and compare the results to determine how much the choice of component-separated map affects our results, as well as showing the potential error that can arise from inpainting.

\begin{figure}
    \centering
    \includegraphics[width=0.5\linewidth]{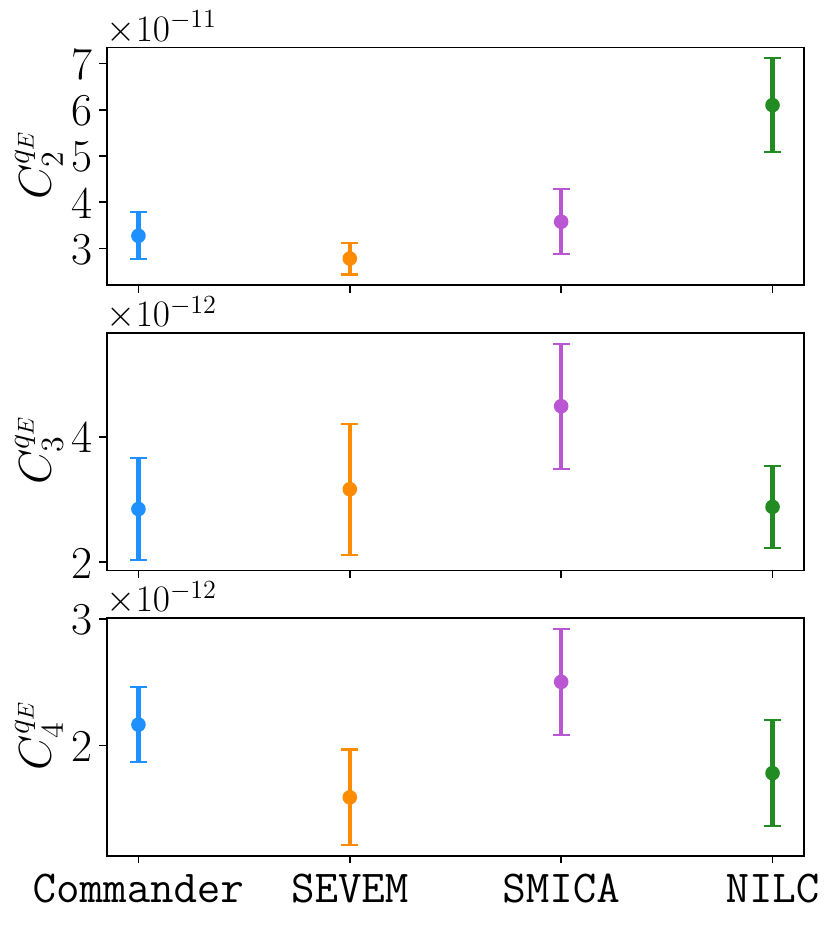}
    \caption{Template power spectra $C_\ell^{q_E}$ for the CIB 353 GHz redshift bin at $\ell = 2$ (top), $\ell = 3$ (middle), and $\ell = 4$ (bottom) for each of the four \textit{Planck} component-separated maps. Error bars show the $1\sigma$ uncertainty from the inpainted galactic region, estimated by generating random Gaussian realizations with the same power spectrum as the inpainted pixels and measuring the resulting variance in $C_\ell^{q_E}$. }
    \label{fig:compsep_clqe}
\end{figure}

Figure~(\ref{fig:compsep_clqe}) shows the resulting template power spectra $C_\ell^{q_E}$ at $\ell = 2$, $3$, and $4$, which are the dominant modes entering the estimator. The error bars represent the uncertainty due to the unknown CMB signal in the galactic masked region, estimated by replacing the inpainted pixels with 100 random Gaussian realizations drawn from the power spectrum of the inpainted map pixels within the galactic mask, corrected by $f_{\rm sky}$. The unmasked pixels are held fixed across all realizations. This method means that the galactic masked region is filled in with realizations that are uncorrelated with the unmasked region, so the resulting variance provides an upper bound on the inpainting uncertainty: because the random realizations are uncorrelated with the unmasked sky, they are statistically less informative than the actual constrained inpainting, and therefore overestimate the sensitivity of the template to the galactic mask. 

The dominant variation arises at $\ell = 2$: \texttt{Commander}, \texttt{SEVEM}, and \texttt{SMICA} give consistent values of $C_2^{q_E} \approx 2.8$--$3.6 \times 10^{-11}$, while \texttt{NILC} yields a value approximately $2\times$ larger ($6.1 \times 10^{-11}$). A pairwise comparison confirms that \texttt{NILC} is inconsistent with each of the other three methods at $\ell = 2$ at the $2$--$3\sigma$ level, while \texttt{Commander}, \texttt{SEVEM}, and \texttt{SMICA} are mutually consistent at $\leq 1\sigma$. For $\ell = 3$ to $20$, all four methods agree within $2\sigma$. This validates that the templates are broadly consistent across all component-separation methods, with the notable exception of \texttt{NILC} at $\ell = 2$. 

\begin{figure}[htbp]
    \centering
    
    \begin{subfigure}{0.48\textwidth}
        \centering
        \includegraphics[width=\linewidth]{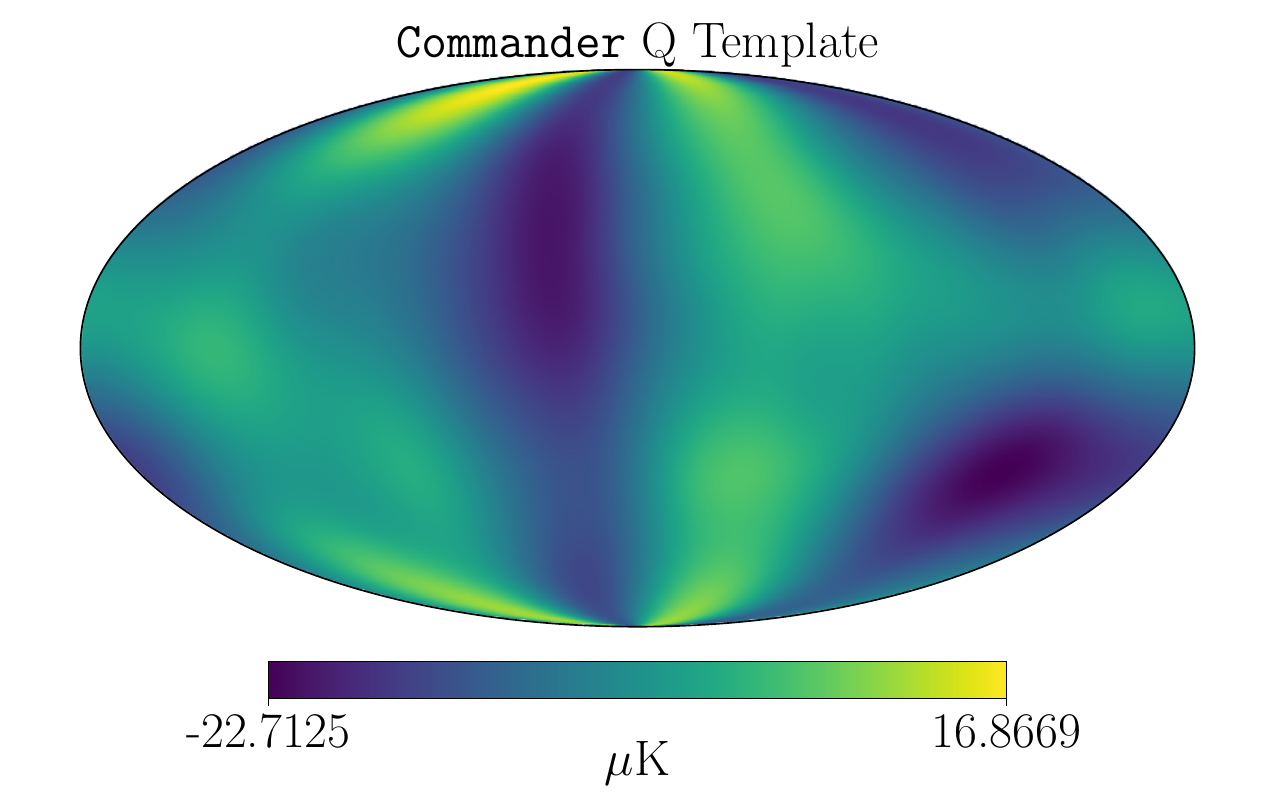}
    \end{subfigure}
    \hfill 
    \begin{subfigure}{0.48\textwidth}
        \centering
        \includegraphics[width=\linewidth]{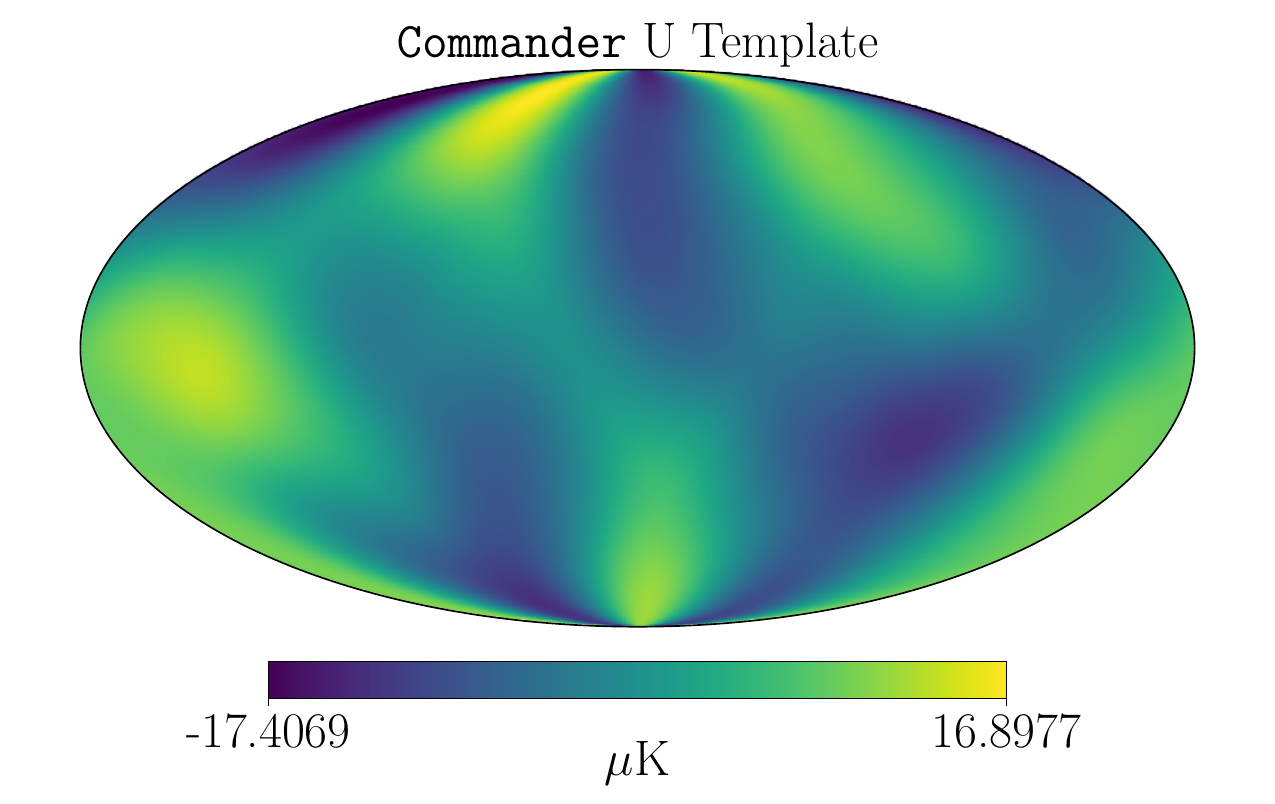}
    \end{subfigure}

    \vspace{0.5cm} 
    \begin{subfigure}{0.48\textwidth}
        \centering
        \includegraphics[width=\linewidth]{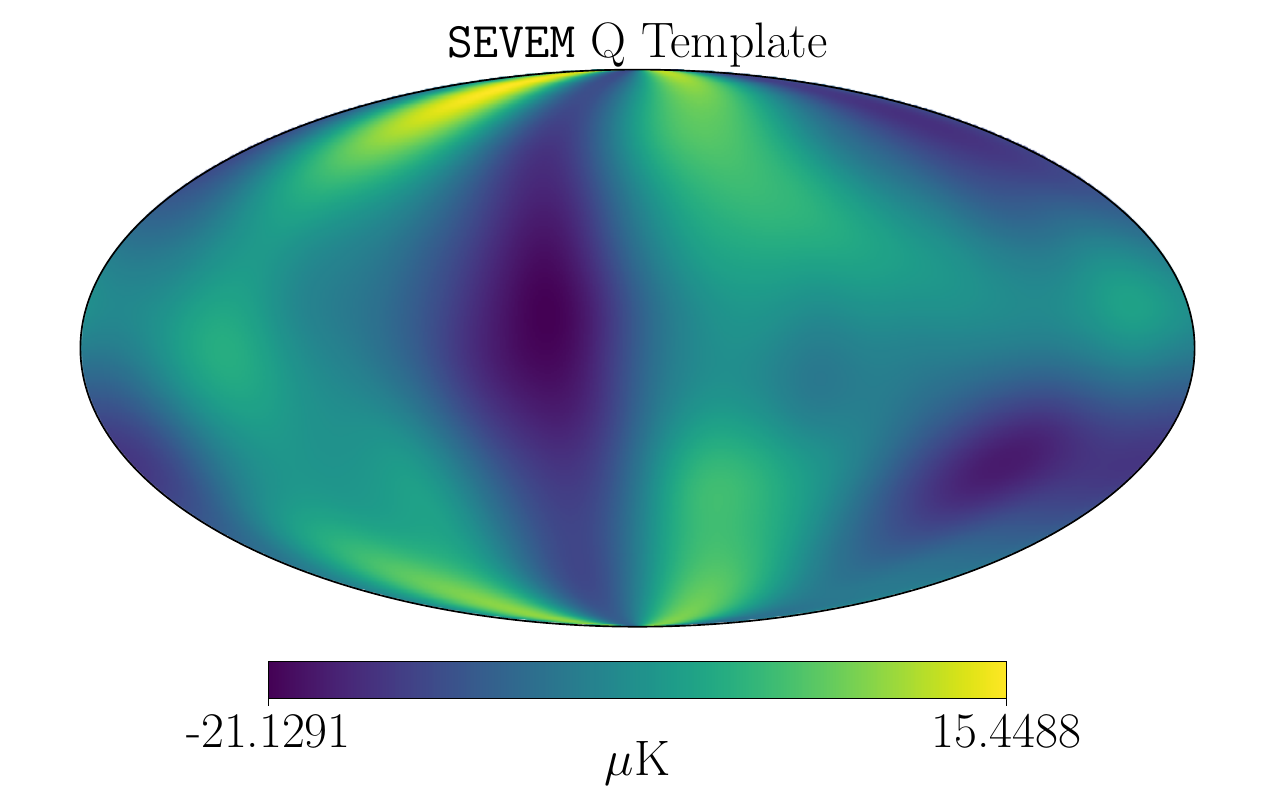}
    \end{subfigure}
    \hfill
    \begin{subfigure}{0.48\textwidth}
        \centering
        \includegraphics[width=\linewidth]{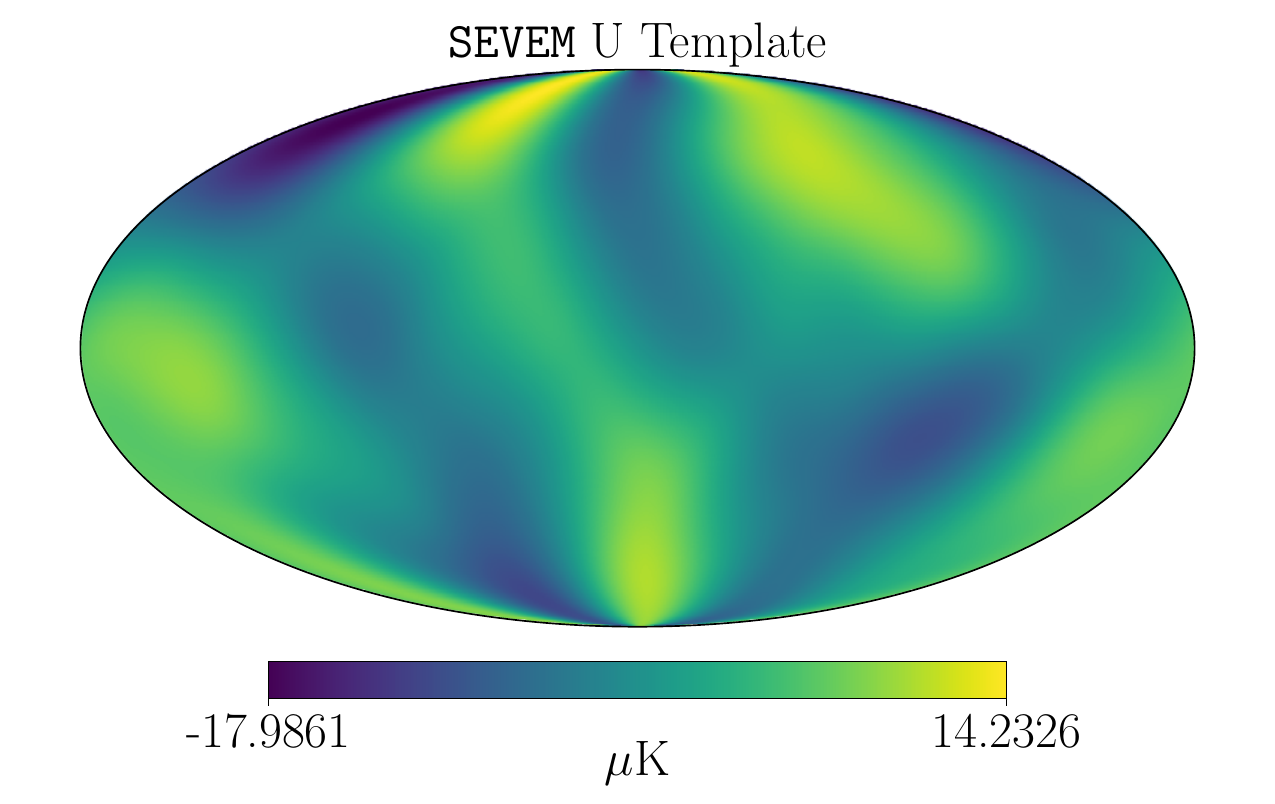}
    \end{subfigure}
    \vspace{0.5cm}

    \begin{subfigure}{0.48\textwidth}
        \centering
        \includegraphics[width=\linewidth]{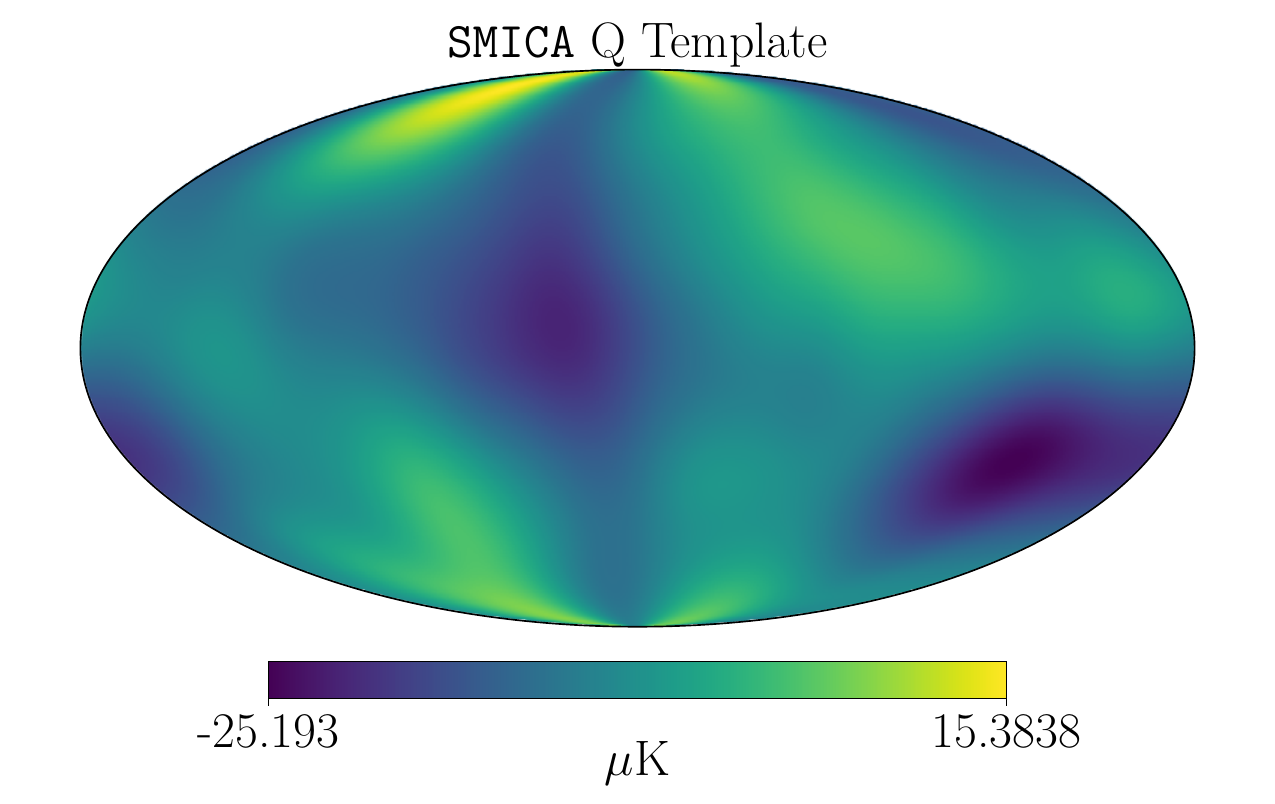}
    \end{subfigure}
    \hfill 
    \begin{subfigure}{0.48\textwidth}
        \centering
        \includegraphics[width=\linewidth]{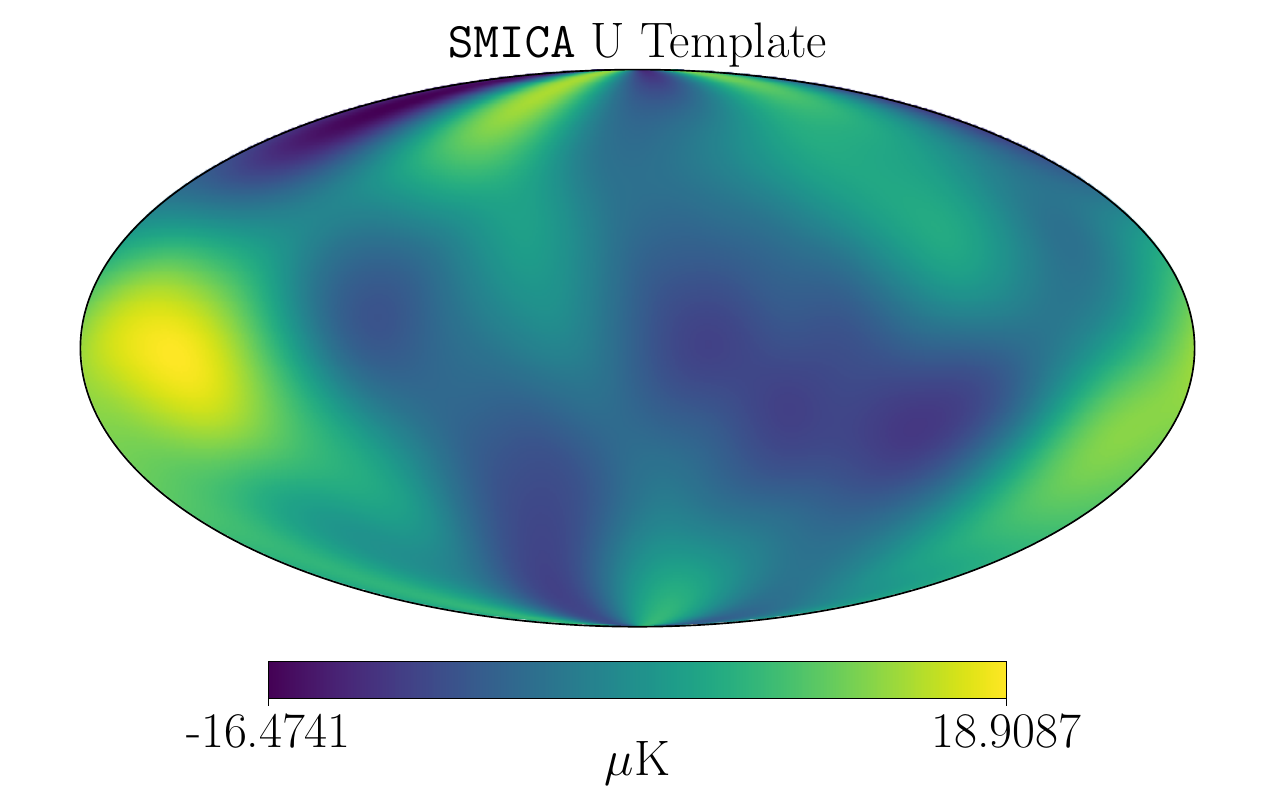}
    \end{subfigure}

    \vspace{0.5cm}

    \begin{subfigure}{0.48\textwidth}
        \centering
        \includegraphics[width=\linewidth]{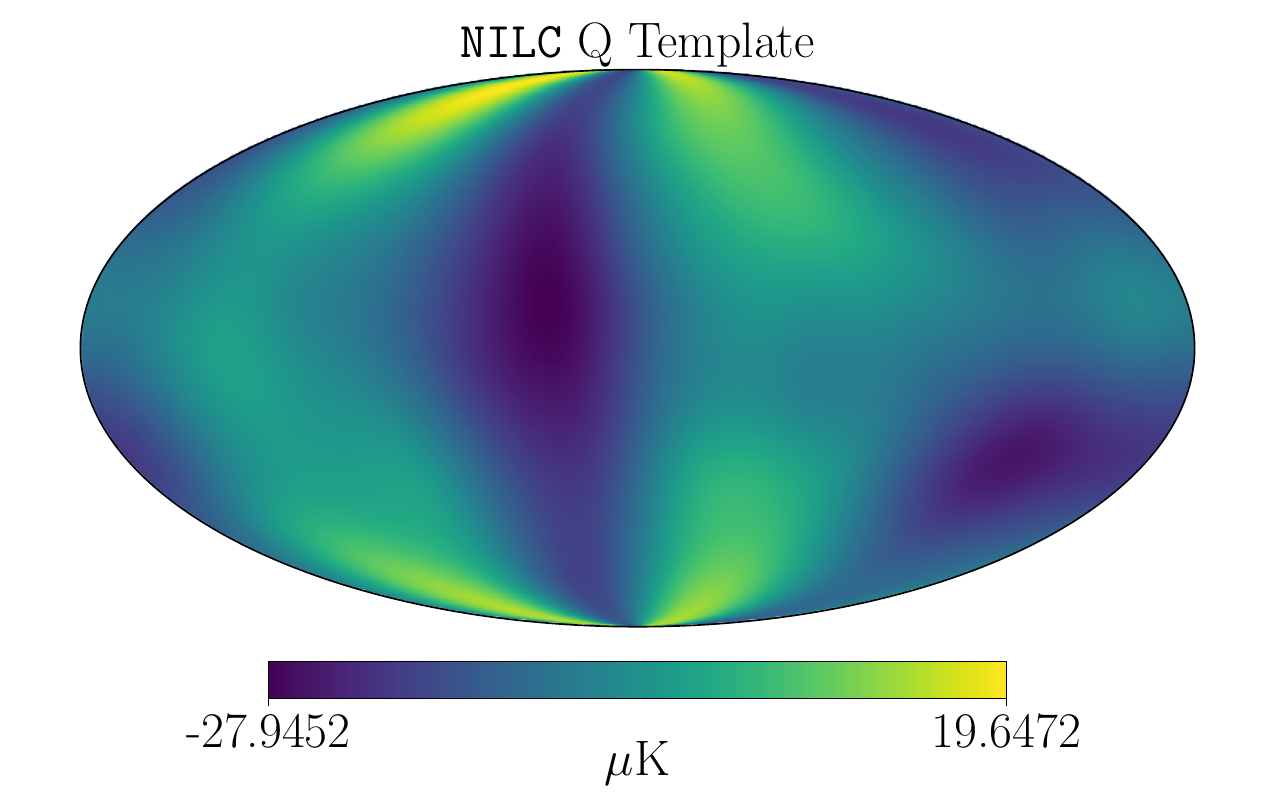}
    \end{subfigure}
    \hfill 
    \begin{subfigure}{0.48\textwidth}
        \centering
        \includegraphics[width=\linewidth]{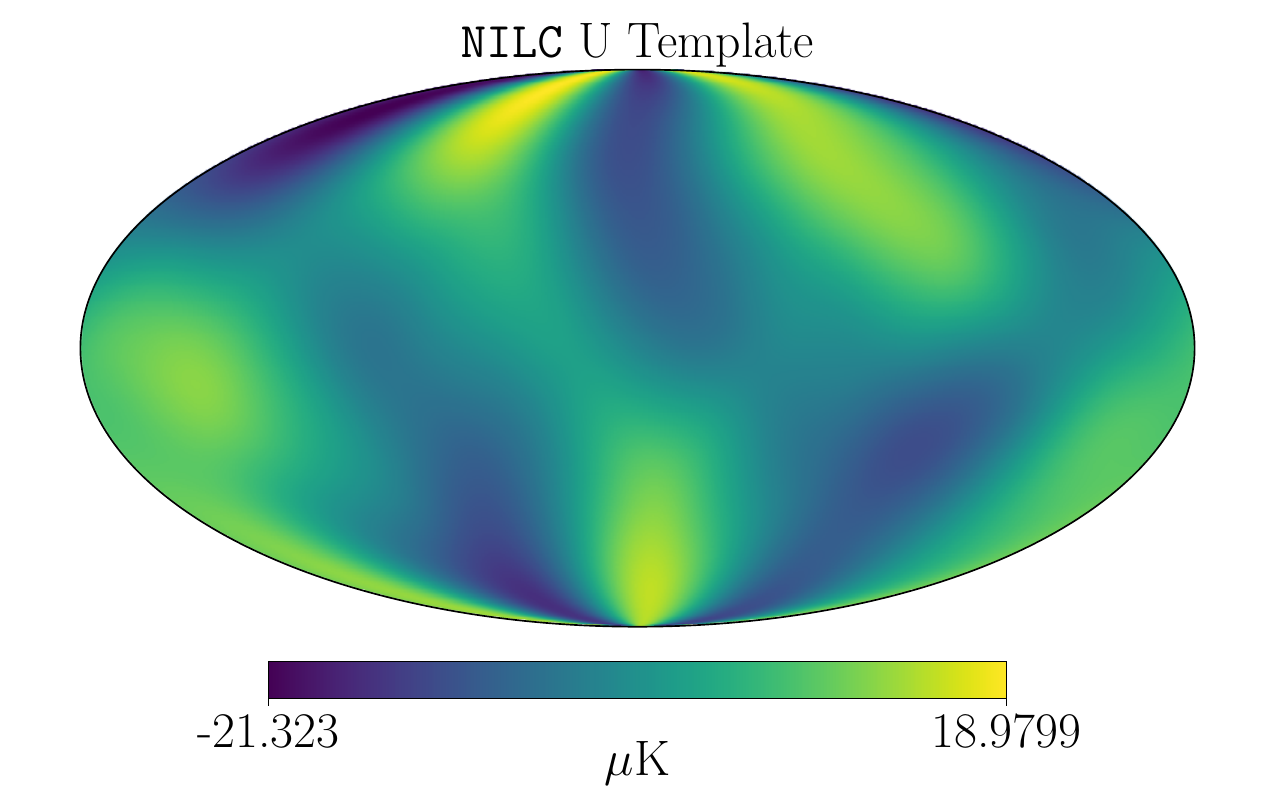}
    \end{subfigure}

    \caption{The $Q$ (left) and $U$ (right) remote quadrupole template maps for CIB 353 GHz from Eq.~\eqref{eq:qE_template} using \texttt{Commander} (first row), \texttt{SEVEM} (second), \texttt{SMICA} (third),  and \texttt{NILC} (last).}
    \label{fig:QUtemplates_compsep}
\end{figure}

Figure~(\ref{fig:QUtemplates_compsep}) compares the $Q$ and $U$ template maps for each component-separated method. All four produce maps with the same large-scale orientation, confirming consistency not only in amplitude but also in direction. The elevated $C_2^{q_E}$ in \texttt{NILC} manifests as a larger overall amplitude in the template map, while the spatial pattern remains qualitatively similar to the other methods.

To show the variation in constraints with different template maps, we calculate the bispectrum estimator for ACT 90 GHz $\times$ CIB 353 GHz for each component-separated map. The resulting $\hat{b}_q$ and $\hat{\tau}_{\rm rei}$ values are shown in table~(\ref{tab:compsep_bqtau}). \texttt{Commander} and \texttt{SEVEM} give mutually consistent results and are both consistent with zero. \texttt{SMICA} and \texttt{NILC} yield larger central values ($1.5\sigma$ and $2.2\sigma$ from zero respectively), with \texttt{NILC} being the most discrepant. This behaviour makes sense, because for a large angular scale template we require a component-separation technique that removes galactic contamination well, but does not project out modes in the primary CMB. Since \texttt{SMICA} and \texttt{NILC} minimize the CMB variance at each angular scale, they are well-suited for the high-$\ell$ CMB power spectrum, but potentially less robust at very large angular scales where galactic foregrounds dominate and the number of available modes is small. In this regime, parametric methods such as \texttt{Commander} and \texttt{SEVEM} that explicitly model and subtract foreground spectral energy distributions may be better suited. We adopt \texttt{Commander} as our primary template and note that the \texttt{SEVEM} results provide an independent confirmation of consistency. 

\begin{table}[h]
\centering
\renewcommand{\arraystretch}{1.3}
\begin{tabular}{lcc}
  \toprule 
  \textbf{Template map} & $\bm{\hat{b}_q \pm \sigma_{b_q}}$ & $\bm{\hat{\tau}_{\rm rei} \pm \sigma_{\tau_{\rm rei}}}$ \\
  \midrule
  \texttt{Commander} & $1.02 \pm 2.64$ & $-0.007 \pm 0.145$ \\
  \texttt{SEVEM}     & $1.40 \pm 3.01$ & $0.012 \pm 0.167$ \\
  \texttt{SMICA}     & $4.46 \pm 2.91$  & $0.181 \pm 0.148$ \\
  \texttt{NILC}      & $6.21 \pm 2.86$ & $0.268 \pm 0.148$ \\
  \bottomrule
\end{tabular}
\caption{Measured $\hat{b}_q$ and $\hat{\tau}_{\rm rei}$ for ACT 90 GHz $\times$ CIB 353 GHz using different \textit{Planck} component-separated maps as the remote quadrupole template.}
\label{tab:compsep_bqtau}
\end{table}

\section{Tensor-to-scalar ratio constraints}

All individual direction constraints in Table~\ref{tab:r_constraints}, $\hat{r}^{(\alpha)}$, are consistent with zero within $1$--$2\sigma$, except \textit{Planck} \texttt{SMICA} which is $\sim 2.6-3\sigma$. This large of a mean value for $e^{(2)}$ is not seen in any of the other data combinations so we do not interpret it as a detection. The tightest individual uncertainties are from ACT NILC $\times$ CIB 353 GHz, with uncertainties ranging from $\sim$100-200. The CIB tracers yield uncertainties roughly $1.5$--$2\times$ smaller than unWISE, consistent with the relative reconstruction noise levels. For ACT 90 GHz, the cleanest CMB map, we can compare the $\hat{r}^{(\alpha)}$ values and uncertainties across the three CIB channels. We find that they are broadly consistent, suggesting the results are not dominated by a single frequency-dependent foreground component.

\begin{table}[h]
    \centering
    \renewcommand{\arraystretch}{1.3}
    \begin{tabular}{llcccc}
    \toprule 
      \multicolumn{6}{c}{$\bm{\hat{r}^{(\alpha)} \pm \sigma_{r^{(\alpha)}}}$} \\
      \toprule
    \textbf{CMB Map} & \textbf{Direction} & \textbf{unWISE} & \textbf{CIB 353\,GHz} & \textbf{CIB 545\,GHz} &
  \textbf{CIB 857\,GHz} \\
    \midrule
    \multirow{5}{*}{ACT 90\,GHz}
      & $e^{(0)}$ & $85  \pm 357$ & $-11 \pm 215$ & $12  \pm 238$ & $-30 \pm 251$ \\
      & $e^{(1)}$ & $19  \pm 301$ & $156 \pm 173$ & $162 \pm 191$ & $186 \pm 201$ \\
      & $e^{(2)}$ & $-152\pm 283$ & $20  \pm 162$ & $19  \pm 179$ & $-4  \pm 188$ \\
      & $e^{(3)}$ & $131 \pm 216$ & $90  \pm 119$ & $90  \pm 132$ & $133 \pm 138$ \\
      & $e^{(4)}$ & $-219\pm 214$ & $-9  \pm 118$ & $-25 \pm 131$ & $-20 \pm 138$ \\
    \midrule
    \multirow{5}{*}{ACT 150\,GHz}
      & $e^{(0)}$ & $129 \pm 369$ & $312 \pm 222$ & $304 \pm 246$ & $228 \pm 259$ \\
      & $e^{(1)}$ & $-261\pm 312$ & $-284\pm 178$ & $-281\pm 197$ & $-299\pm 208$ \\
      & $e^{(2)}$ & $-194\pm 292$ & $-267\pm 167$ & $-278\pm 185$ & $-329\pm 195$ \\
      & $e^{(3)}$ & $-162\pm 223$ & $-13 \pm 123$ & $-15 \pm 136$ & $-6  \pm 143$ \\
      & $e^{(4)}$ & $-49 \pm 222$ & $9   \pm 122$ & $32  \pm 135$ & $28  \pm 143$ \\
    \midrule
    \multirow{5}{*}{ACT 220\,GHz}
      & $e^{(0)}$ & $811 \pm 1053$ & $-227\pm 677$ & $-237\pm 746$ & $-216\pm 775$ \\
      & $e^{(1)}$ & $-454\pm 889$  & $708 \pm 543$ & $863 \pm 599$ & $1027\pm 622$ \\
      & $e^{(2)}$ & $-611\pm 834$  & $-482\pm 509$ & $-516\pm 561$ & $-552\pm 583$ \\
      & $e^{(3)}$ & $-91 \pm 637$  & $278 \pm 374$ & $266 \pm 412$ & $271 \pm 428$ \\
      & $e^{(4)}$ & $30  \pm 632$  & $655 \pm 372$ & $692 \pm 410$ & $738 \pm 426$ \\
    \midrule
    \multirow{5}{*}{ACT NILC}   
      & $e^{(0)}$ & $352 \pm 311$ & $132 \pm 192$ & $140 \pm 211$ & $53  \pm 228$ \\
      & $e^{(1)}$ & $-108\pm 263$ & $-38 \pm 154$ & $-31 \pm 170$ & $-5  \pm 183$ \\
      & $e^{(2)}$ & $47  \pm 247$ & $-38 \pm 144$ & $-36 \pm 159$ & $-65 \pm 171$ \\
      & $e^{(3)}$ & $23  \pm 188$ & $1   \pm 106$ & $-9  \pm 117$ & $10  \pm 126$ \\
      & $e^{(4)}$ & $-207\pm 187$ & $-16 \pm 105$ & $-6  \pm 116$ & $-54 \pm 125$ \\
    \midrule
    \multirow{5}{*}{\textit{Planck} \texttt{SMICA}}
      & $e^{(0)}$ & $199 \pm 394$ & $49  \pm 246$ & $51  \pm 267$ & $30  \pm 261$ \\
      & $e^{(1)}$ & $-17 \pm 334$ & $-58 \pm 209$ & $-29 \pm 226$ & $-65 \pm 221$ \\
      & $e^{(2)}$ & $-951\pm 349$ & $-672\pm 218$ & $-692\pm 236$ & $-589\pm 231$ \\
      & $e^{(3)}$ & $-64 \pm 281$ & $175 \pm 176$ & $170 \pm 191$ & $180 \pm 186$ \\
      & $e^{(4)}$ & $-77 \pm 280$ & $100 \pm 175$ & $83  \pm 190$ & $14  \pm 186$ \\
    \bottomrule
    \end{tabular}
\caption{Constraints on the tensor-to-scalar ratio $r$, decomposed into the five orthonormal basis
directions of the $\ell=2$ $B$-mode quadrupole. Values are $\hat{r}^{(\alpha)} \pm \sigma_{r^{(\alpha)}}$ for
each direction and our model is $n_t=0$.}
\label{tab:r_constraints}
\end{table}

\bibliography{planck_unwise_refs}
\bibliographystyle{minor-changes-JHEP}

\end{document}